%% file: main.tex
\documentclass[final,3p,12pt]{elsarticle}

\usepackage[utf8]{inputenc}
\usepackage{textcomp}
\usepackage{hyperref}
\usepackage{graphicx}
\graphicspath{ {figures/} }
\usepackage{amsmath}
\usepackage{bm}
\usepackage[version=4]{mhchem}
\usepackage{booktabs}
\usepackage{float}
\usepackage{siunitx}
\usepackage{graphicx}
\usepackage{algpseudocode}
\usepackage{algorithm}
\usepackage{enumitem}
\usepackage{optidef}
\usepackage{color}
\usepackage{array}
\usepackage{xcolor}
\usepackage{longtable,tabularx}
\usepackage{subfig}
\usepackage{lineno}
 \usepackage{overpic}
\graphicspath{{figures}/}

\usepackage{amssymb}

\begin{document}

\begin{frontmatter}



\title{H$^3$PC: {\underline H}ypersonic, {\underline H}igh-Order, {\underline H}igh-{\underline P}erformance {\underline C}ode with Adaptive Mesh Refinement and Real Chemistry}

\author{Ahmad Peyvan}
\author{Khemraj Shukla}
\author{George Em Karniadakis}
\affiliation{organization={Division of Applied Mathematics, 182 George Street, Brown University},
    city={Providence},state={RI}, postcode={02912},country={USA}
            }

\input{abstract}





\end{frontmatter}

{\renewcommand\arraystretch{0.75}

\input{introduction}

\input{Governing_equations}

\input{methodology}

\input{results}

\input{conclusion} 
\input{appendix}
\newpage
\section*{Acknowledgments}
This work was supported by the U.S. Army Research Laboratory
W911NF-22-2-0047 and by the MURI-AFOSR FA9550-20-1-0358. The simulations were performed on the Brown University's high performance computing cluster (OSCAR) supported by Center for Computation and Visualization (CCV). We appreciate Prof. Jesse Chan for invaluable assistant and help in developing the Navier-Stokes solver. We would like to thank Prof. Lucas Wilcox from the Naval Postgraduate School for his assistance in porting the \texttt{Mutation++} library to \texttt{Julia}. We gratefully acknowledge Prof. P. Tsoutsanis (Cranfield University) for his support with the UCNS3D code.




\bibliographystyle{elsarticle-num} 
\bibliography{reference}
\end{document}

%% file: abstract.tex
\begin{abstract}
We have developed a hypersonic high-order, high-performance code (H$^3$PC) utilizing the ``Trixi.jl" framework in order to simulate both non-reactive and chemically reactive compressible Euler and Navier-Stokes equations for complex three-dimensional geometries. H$^3$PC is parallel on CPU platforms and can perform exascale parallel computations of hypersonic turbulent flows. The numerical approach is based on the discontinuous Galerkin spectral element method, satisfying the entropy and energy stability conditions for the Euler equations. H$^3$PC  can perform simulations of high-speed flows from subsonic to hypersonic speeds based on  frozen, equilibrium, and non-equilibrium chemistry modeling of the gas mixture, using 
the \texttt{Mutation.jl} , which is a Julia package developed to wrap the C++-based Mutation++ library. H$^3$PC  can also perform parallel adaptive mesh refinement for two- and three-dimensional Euler and Navier-Stokes discretizations with non-conforming elements.  
In this study, we first demonstrate the successful integration of Mutation++ into the H$^3$PC solver, and then verify its accuracy through simulations of Taylor-Green vortex flow,  supersonic flow past a square and circular cylinder, and hypersonic P8-inlet.



\end{abstract}

%% file: introduction.tex
\section{Introduction}

In recent years, there has been an increased demand for designing robust and efficient hypersonic vehicles such as re-entry capsules, commercial low-orbit launchers, and hypersonic aircrafts, a task that requires high-fidelity simulations based on accurate flow dynamics predictions. Experimental data in hypersonic regimes are scarce, while most existing hypersonic codes have low-order accuracy, which could lead to erroneous results in long-time integration and at high Reynolds and Mach numbers. Discontinuous spectral element methods offer high-order accuracy and geometric flexibility to resolve the vast range of turbulent scales specific to flow around complex geometries. However, the high-order accuracy comes with a lack of robustness in resolving the high-gradient features of hypersonic flows. The impact of high-speed flow with the body of a flying object forms high-gradient temperature regions that trigger chemical reactions by dissociation of molecules and recombination of atoms. Therefore, the characteristic time scale will be as short as the chemical reaction time scale. As a result, non-equilibrium chemistry must be considered when modeling hypersonic flows, which significantly  adds to the non-linearity and stiffness of the system of governing equations.

The majority of the existing hypersonic solvers employ low-order discretization in space. US3D \cite{candler2015development} currently is the most popular  hypersonic solver available. It includes most of NASA's DPLR code features but can use an unstructured grid. This solver employs finite volume discretization for which the solution data is stored at the cell center, and fluxes are evaluated at the interfaces of computational cells. Second or third (QUICK) order discretization is used for fluxes, and upwind fluxes are employed if a shock is detected. Eilmer is an open-source, general-purpose compressible flow solver developed at the University of Queensland. It is specifically designed to facilitate research in hypersonics and high-speed aerothermodynamics \cite{gibbons2023eilmer}. Similar to US3D, Eilmer employs finite volume discretization for the spatial domain. The interface fluxes are computed by solving a Riemann problem constructed between two solution states. Godunov's approach is used for flux discretization. The use of Godunov-type flux can be sufficient to capture shock waves. Eilmer can also handle unstructured grids. The Eilmer and US3D frameworks are endowed with a message-passing interface (MPI) library for exascale parallel computations. An open-source hypersonic task-based research (HTR) solver was also developed by Renzo \emph{et al.}~\cite{di2020htr}. HTR employs a low-dissipation sixth-order targeted essentially non-oscillatory (TENO) spatial discretization on structured grids. The numerical scheme is designed to resolve canonical hypersonic flows at high Reynolds numbers, and the solver can be deployed on multiple GPU platforms. 

Antoniadis \emph{et al.} \cite{antoniadis2022ucns3d} have developed the UCNS3D code, which is an open-source computational solver for compressible flows on unstructured meshes. High-order WENO methods are implemented to solve the compressible Navier-Stokes equation on unstructured meshes such as tetrahedral, hexahedral, prisms, etc. UCNS3D is based on three-dimensional WENO schemes for arbitrary shape elements on unstructured meshes with a new stencil construction procedure proposed by Tsoutsanis \emph{et al.}\cite{tsoutsanis2011weno}. UCNS3D is written in Fortran 2008 with MPI and OpenMP functionality and developed for CPU architectures. It utilizes BLAS libraries for several floating-point operations such as matrix-vector and matrix-matrix multiplications. UCNS3D also employs the ParMetis and Metis libraries for partitioning the computational domain discretized by unstructured meshes. In the present study, we will compare the results of H$^3$PC against the UCNS3D for cross-verification.

Another framework that can potentially simulate hypersonic flows is the FLASH code developed by Fryxell \emph{et. al.} \cite{fryxell2000flash}. FLASH solves  the compressible reactive Euler equations. It employs adaptive mesh refinement using the PARMESH library to dynamically adapt the mesh resolution to high-gradient regions of the flow. The spatial discretization in FLASH is based on the piecewise parabolic method (PPM) discretization of Woodward and Colella \cite{woodward1984numerical} and Colella and  Woodward \cite{colella1984piecewise}. Such discretization can provide second-order accuracy in space and time. STAR-CCM+ is a similar commercial software package used to simulate hypersonic flow with simplified chemistry kinetics \cite{cross2019simulation}.

High-order discontinuous spectral element methods (DSEM), such as the Discontinuous Galerkin Spectral element Method (DGSEM) \cite{lomtev1999discontinuous,karniadakis2005spectral}, Spectral Difference (SD) \cite{kopriva1996conservative}, and Flux Reconstruction (FR) \cite{huynh2007flux,vincent2011new,peyvan2021flux} combine the high-order accuracy of spectral methods with the geometric flexibility of finite volume methods. These features make them an excellent choice for solving high-speed turbulent flows around complex geometries \cite{CANTWELL2015205}. Moreover, the DSEM schemes have been proven beneficial for simulating subsonic to supersonic flows~\cite{li2019compressibility, li2021compressibility,komperda2020hybrid}. However, applying these high-order schemes to hypersonic flows has encountered several numerical modeling challenges. For instance, the appearance of strong shock waves induces spurious oscillations, which could lead to non-physical values for species densities, pressure, or internal energies. Negative density and pressure values will cause instant simulation blow-ups. 

This deficiency has been addressed by Peyvan et al. \cite{PEYVAN2023112310} by developing an entropy-stable DGSEM for non-equilibrium thermodynamic models. The non-equilibrium kinetics must be considered at hypersonic speeds since the advection and chemical reaction time scales are of the same order of magnitude. Non-equilibrium chemistry modeling includes production and destruction rates of gas mixture species in the source terms of the transport equations of species \cite{Scoggins2020}. The high fluctuation in temperature values can generate large values in the production source terms, which could lead to numerical instabilities. Computing thermodynamics, chemistry kinetics, and transport properties of gas mixtures under hypersonic conditions requires additional iterative calculations. Iterative calculations must be performed using the least computational effort; otherwise, huge computational overhead renders the non-equilibrium simulations infeasible. Therefore, efficient numerical libraries are required to determine the thermo-chemical properties of the dissociating gas mixture \cite{Scoggins2020}.

In the present study, we develop a hypersonic, high-order, high-performance code (H$^3$PC) with
adaptive mesh refinement (AMR) and real chemistry for simulating reactive and non-reactive Navier-Stokes equations employing equilibrium and non-equilibrium chemistry modeling. The framework is built upon the Trixi.jl code, written in Julia, originally developed by Michael Schlottke-Lakemper,  Gregor Gassner,  Hendrik Ranocha,  Andrew Winters, and Jesse Chan  \cite{schlottkelakemper2020trixi,schlottkelakemper2021purely,ranocha2022adaptive}. The original Trixi.jl code solved hyperbolic conservation equations employing several mesh types such as Tree, Structured, Unstructured, DGMulti, and P4est-based. The P4est-based mesh is an unstructured, curvilinear, non-conforming type for quadrilateral (2D) and hexahedral (3D) cells. It supports quadtree/octree-based adaptive mesh refinement (AMR) via the C library P4est developed by Burstedde, Wilcox, and Ghattas \cite{BursteddeWilcoxGhattas11}. For the development of the H$^3$PC solver, we focused on the P4est mesh type since it supports AMR, complex geometry, and parallel computation capabilities. The development details are described in section \ref{sec:dev}. The H$^3$PC solver can robustly solve conservation equations governing hypersonic flows, and features the following:
\begin{itemize}
    \item High-order accuracy in space and time, kinetic energy-preserving 
 and entropy stability, entropy stable shock capturing, and positivity preservation.
    \item Adaptive mesh refinement for 2D quadrilateral and 3D hexahedral elements with straight and curved sides using the P4est library.
    \item Multi-node parallelization via MPI for Navier-Stokes solvers with static and adaptive meshes.
    \item Flexible chemical kinetics, thermodynamics, and transport coefficients modeling through the Mutation++ library converted to Julia programming language paradigm.
    \item Complex geometry handling with the support of GMesh format. 
\end{itemize}
The rest of the paper is organized as follows:  A brief description of the governing equations is first provided. The methodology of the numerical scheme used in the H$^3$PC solver is then described in detail. Next, the development of the solver and the approaches used for its implementation are described. Five benchmark problems are solved to investigate various aspects of the solver and to validate the accuracy and implementation of the numerical scheme and gas mixture models. Finally, we conclude the paper with a summary.

%% file: Governing_equations.tex
\section{Governing Equations}
We consider  the compressible reacting and non-reacting Navier-Stokes equations that govern hypersonic flows. The dimensional form of the equations can be described using several conservation laws. The system of equations can be described by employing a general matrix form as

\begin{equation}
\frac{\partial \mathbf{U}}{\partial t}+\mathbf{\nabla} . \mathbf{F}^a+\mathbf{\nabla} . \mathbf{F}^v = \mathbf{S}, 
    \label{system_equation}
\end{equation}
where the vector of conserved variables $\mathbf{U}$ is defined as 
\begin{equation}
\mathbf{U}=\left(\rho_1, \cdots, \rho_{Ns}, \rho u, \rho v, \rho w, \rho E\right)^T,
    \label{Uconserv}
\end{equation}
and convective and viscous fluxes vectors consist of three components as $\mathbf{F}^a=\left(\mathcal{F}^a, \mathcal{G}^a, \mathcal{H}^a\right)^T$ and $\mathbf{F}^v=\left(\mathcal{F}^v, \mathcal{G}^v, \mathcal{H}^v\right)^T$ corresponding to three directions of the Cartesian coordinate system. The advective flux components are described as

\begin{equation}
\mathcal{F}^a=\begin{pmatrix}
    \rho_1 u \\ \vdots\\ \rho_{N_s}u\\\rho u^2+p\\ \rho uv\\ \rho uw\\u(\rho E+p)
\end{pmatrix},\quad \mathcal{G}^a=\begin{pmatrix}
    \rho_1 v \\ \vdots\\ \rho_{N_s} v\\\rho vu\\ \rho v^2+p\\ \rho vw\\v(\rho E+p) 
\end{pmatrix},\quad \mathcal{H}^a=\begin{pmatrix}
    \rho_1 w \\ \vdots\\ \rho_{N_s} w\\\rho wu\\ \rho wv\\ \rho w^2+p\\w(\rho E+p) 
\end{pmatrix}.
    \label{adevctive_flux}
\end{equation}
The viscous flux vectors are defined using the following expressions
\begin{equation}
\begin{split}
\mathcal{F}^v=\begin{pmatrix}
    \rho_1 V^1_1 \\ \vdots\\ \rho_{N_s}V^1_{N_s}\\\sigma_{11}\\ \sigma_{21}\\\sigma_{31} \\ q_1-(u\sigma_{11}+v\sigma_{21}+w\sigma_{31})
\end{pmatrix},\quad& \mathcal{G}^v=\begin{pmatrix}
    \rho_1 V^2_1 \\ \vdots\\ \rho_{N_s}V^2_{N_s}\\\sigma_{12}\\ \sigma_{22}\\\sigma_{32} \\ q_2-(u\sigma_{12}+v\sigma_{22}+w\sigma_{32})
\end{pmatrix}\\ \mathcal{H}^v=\begin{pmatrix}
    \rho_1 V^3_1 \\ \vdots\\ \rho_{N_s}V^3_{N_s}\\\sigma_{13}\\ \sigma_{23}\\\sigma_{33} \\ q_1-(u\sigma_{13}+v\sigma_{23}+w\sigma_{33})
\end{pmatrix}.\quad&
\end{split}
    \label{visc_fluxes}
\end{equation}
The source terms $\mathbf{S}$ can be described as
\begin{equation}
\mathbf{S}=\begin{pmatrix}
    \dot{\omega}_1 \\ \vdots\\ \dot{\omega}_{N_s}\\0\\ 0\\0 \\ 0
\end{pmatrix}.
    \label{source_term}
\end{equation}
In Eq.~\eqref{Uconserv}, $\rho_s$ denotes density of $s^\textrm{th}$ species. The terms $u$, $v$, and $w$ are x, y, and z-direction velocities. $E$ indicates the specific total energy of the gas mixture, including internal and kinetic energies defined as
\begin{equation}
E = e+ \frac{1}{2}(u^2+v^2+w^2),
    \label{total_energy}
\end{equation}
with the specific internal energy written as
\begin{equation}
e = \frac{1}{\rho}\sum_{s=1}^{N_s}\rho_s e_s.
    \label{internal_energies}
\end{equation}
The specific internal energy of each species consists of various internal energy modes illustrated in section \ref{termo}. In Eq.~\eqref{adevctive_flux}, $p$ denotes pressure, which is formulated based on Dalton's law as
\begin{equation}
p = \sum_{s=1}^{N_s}\rho_s R_s T.
    \label{eos}
\end{equation}
In Eq.~\eqref{visc_fluxes}, $V^i_s$ defines mass diffusion velocities for $s^\textrm{th}$ in the $i^\textrm{th}$ coordinate system direction, and  $\sigma_{ij}$ describes the stress tensor and $q_i$ denotes heat flux vector component in $i^\textrm{th}$ direction. The species production terms are also shown by $\dot{\omega}_s$ in Eq.~\eqref{source_term}. Equations~\eqref{Uconserv}-\eqref{eos} are derived by considering several assumptions including:

\begin{enumerate}
    \item Ionization of species is neglected. This means that the species do not carry an electric charge.
    \item Free-electron effects are not considered. Therefore, we discard solving the extra transport equation for free-electron energy. 
    \item The effect of radiation is ignored. 
    \item Forces exerted on species due to magnetic and electric fields are neglected.
    \item Total Joule heating is neglected.
    \item All the energy states of every species are in local thermal equilibrium, where we use a single temperature value for all the energy modes. Therefore, we ignore solving extra transport equations for various internal energy modes, such as electronic and vibrational.
    \item Chemical non-equilibrium conditions are considered, which means that the composition of the gas mixture varies due to the occurrence of chemical reactions.
\end{enumerate}
In the appendix, sections \ref{termo}-\ref{chem_kinetics}, we explain thermo-chemical and transport closure models for the system of equations in detail.

%% file: methodology.tex
\section{Methodology}
In this study, we are interested in solving the 3D compressible reacting Navier-Stokes equations \eqref{system_equation} over complex physical domains subjected to particular boundary and initial conditions. In this section, we explain the discretization of the following system of equations

\begin{equation}
\begin{cases}
\frac{\partial \mathbf{U}}{\partial t} +\mathbf{\nabla}. \vec{\mathbf{F}}=\mathbf{S},& \mathbf{x}\in \mathbf{\Omega}\\
\mathbf{U}\left(\mathbf{x},t\right)=\mathbf{\Phi}(t),& \mathbf{x}\in \partial \mathbf{\Omega}\\
\mathbf{U}(\mathbf{x},0) = \mathbf{\Psi}(\mathbf{x}),&\mathbf{x}\in \mathbf{\Omega},
\end{cases}
    \label{system}
\end{equation}
where $\vec{\mathbf{F}}$ denotes the flux vectors that consist of advective and viscous fluxes, and $\mathbf{x}$ denotes the coordinates of points in the 3D physical domain $\mathbf{\Omega}$ with physical boundaries defined as $\partial \mathbf{\Omega}$. We discretize the physical domain into $N$ conforming and non-conforming elements defined such that $\mathbf{\Omega}\approx\cup_{k=1}^N \mathbf{\Omega}_k$. The solution vector is approximated by a union of piecewise local polynomials defined over individual elements as 

\begin{equation}
\mathbf{U}(\mathbf{x},t) = \bigoplus_{k=1}^{N}\mathbf{U}_k^{e,D}(\mathbf{x},t),\quad\mathbf{x}\in \mathbf{\Omega}_k,
    \label{sol_approx}
\end{equation}
and flux vectors are approximated using piecewise continuous local polynomials described as 
\begin{equation}
\vec{\mathbf{F}}(\mathbf{x},t) = \bigoplus_{k=1}^{N}\vec{\mathbf{F}}^e_k(\mathbf{x},t),\quad\mathbf{x}\in \mathbf{\Omega}_k.
    \label{flu_approx}
\end{equation}
In Eq.~\eqref{sol_approx}, superscript $D$ indicates discontinuities at the element interfaces, and $e$ denotes the numerical approximation of the exact quantity. From now on, we omit superscript $D$ and $e$ for brevity. In this study, we employ hexahedron and quadrilateral elements in 3D and 2D configurations, respectively. We then map a physical element into a reference element defined as $\mathbf{X}\in[-1,1]^3$ using the following expression 

\begin{equation}
\mathbf{x}=\sum_{i=1}^{N_c}\mathcal{M}_{i}(\mathbf{X})\mathbf{x}_{i,k},\quad \mathbf{x}\in\mathbf{\Omega}_k
    \label{mapp}
\end{equation}
where $N_c$ denotes the number of corners of the element. According to finite element conventions, for quadrilateral elements $N_c=4$ and for hexahedron elements $N_c=8$. In Eq.~\eqref{mapp}, $\mathcal{M}_{i,k}$ are the shape functions that are defined in \cite{peyvan2022flux}. We now substitute the approximation of solution and flux vector from Eqs.~\eqref{sol_approx} and \eqref{flu_approx} into Eq.~\eqref{system} and employ the mapping defined in Eq.~\eqref{mapp} to derive 
\begin{equation}
\frac{\partial \tilde{\mathbf{U}}_k}{\partial t} +\tilde{\mathbf{\nabla}}. \tilde{\vec{\mathbf{F}}}_k=\tilde{\mathbf{S}}_k,\quad \mathbf{X} \in[-1,1]^3,\quad \textrm{ for }k=1,\cdots,N,
    \label{mapped_equation}
\end{equation}
where $\tilde{\mathbf{U}}_k=|\mathbf{J}_k|\mathbf{U}_k$, $\tilde{\mathbf{\nabla}}=\frac{\partial}{\partial X_i}\vec{e}_i$, $\tilde{\vec{\mathbf{F}}}_k=|\mathbf{J}_k|\frac{\partial X_i}{\partial x_j}\mathbf{F}_{j,k} \vec{e}_i$ and $\tilde{\mathbf{S}}_k=|\mathbf{J}_k|\mathbf{S}_k$ with $|\mathbf{J}_k|$ being the determinant of the transformation Jacobian matrix.  The component of the  Jacobian matrix ($\mathbf{J}_k$) are  defined as
\begin{equation}
J_{ij}=\frac{\partial x_i}{\partial X_j}.
    \label{jacobian}
\end{equation}
After mapping the equations into the reference element, we will solve Eq.~\eqref{mapped_equation} using the discontinuous Galerkin spectral element (DGSEM) approach. First, we must derive a weak form for Eq.~\eqref{mapped_equation} over the reference element. We multiply the equations by the test functions ($L(\mathbf{X})$) and integrate over the reference element to get
\begin{equation}
\int_{\mathbf{\Omega}_k}\frac{\partial \tilde{\mathbf{U}}_k}{\partial t}L(\mathbf{X})d\mathbf{X}+\int_{\mathbf{\Omega}_k}\tilde{\mathbf{\nabla}}. \tilde{\vec{\mathbf{F}}}_k L(\mathbf{X})d\mathbf{X}=\int_{\mathbf{\Omega}_k}\tilde{\mathbf{S}}_k L(\mathbf{X})d\mathbf{X}.
    \label{integration}
\end{equation}
In Eqs.~\eqref{integration}, $L(\mathbf{X})$  denotes Lagrange polynomials defined as 

\begin{equation}
L(\mathbf{X})=l_n(X_1)l_m(X_2)l_r(X_3),
    \label{lagrange_3d}
\end{equation}
where $l_n(X_j)$ is a Lagrange polynomial constructed using the Gauss-Lobatto-Legendre (GLL) points in the interval $[-1,1]$ as
\begin{equation}
l_n(X_j)=\prod_{s=0,s\neq n}^{\mathcal{P}} \frac{X_j-X_{j,s}}{X_{j,n}-X_{j,s}},\quad\textrm{for } j=1,2,3,
\label{lag}
\end{equation}
where $\mathcal{P}$ denotes the polynomial order. Now, we approximate the integrals in Eq.~\eqref{integration} using the GLL quadrature rules with $\mathcal{P}+1$ nodes. After applying the quadrature rule, equation \eqref{integration} reads
\begin{equation}
\tilde{U}_{k,nmr}\omega_n\omega_m\omega_r=-\Big[\mathcal{F}^1_{k,nmr}+\mathcal{F}^2_{k,nmr}+\mathcal{F}^3_{k,nmr}\Big]+\tilde{S}_{k,nmr}\omega_n\omega_m\omega_r,
    \label{divergenc_quadrature}
\end{equation}
where $\omega_n$ are the quadrature weight corresponding to quadrature node $n$. The divergence term (second term on the right-hand side of Eq.~\ref{integration}) is discretized by first performing integration by parts over the corresponding dimension, and then applying the quadrature rule to compute the integrals as

\begin{equation}
\begin{aligned}
    \mathcal{F}^1_{k,nmr} = \left[\tilde{\mathbf{F}}^{*,1}_{k,\mathcal{P}mr}-\tilde{\mathbf{F}}^{*,1}_{0mr}\right]\omega_m \omega_r-\sum_{j=0}^{\mathcal{P}}Q_{nj}\tilde{\mathbf{F}}^1_{k,jmr}\omega_j\omega_m\omega_r\\
    \mathcal{F}^2_{k,nmr} = \left[\tilde{\mathbf{F}}^{*,2}_{k,n\mathcal{P}r}-\tilde{\mathbf{F}}^{*,2}_{k,n0r}\right]\omega_n \omega_r-\sum_{j=0}^{\mathcal{P}}Q_{mj}\tilde{\mathbf{F}}^2_{k,njr}\omega_n\omega_j\omega_r \\
    \mathcal{F}^3_{k,nmr} = \left[\tilde{\mathbf{F}}^{*,3}_{k,nm\mathcal{P}}-\tilde{\mathbf{F}}^{*,3}_{k,nm0}\right]\omega_n \omega_m-\sum_{j=0}^{\mathcal{P}}Q_{rj}\tilde{\mathbf{F}}^3_{k,nmj}\omega_n\omega_m\omega_j,
\end{aligned}
    \label{fluxes_discrete}
\end{equation}
where $Q_{nj}=dl_n(X_{j})/dX$. The summation by parts property of the derivative, mass, and boundary matrix yields the following relation 

\begin{equation}
\begin{aligned}
Q_{nj}\omega_j = B_{nj}-Q_{jn}\omega_n\\
Q_{mj}\omega_j = B_{mj}-Q_{jm}\omega_m\\
Q_{rj}\omega_j = B_{rj}-Q_{jr}\omega_r,
\end{aligned}
    \label{sbp}
\end{equation}
where the boundary matrix is defined as
\begin{equation}
B_{ij}=\begin{cases}
    -1&\textrm{if }i\textrm{ and }j=0\\
1&\textrm{if }i\textrm{ and }j=\mathcal{P}\\
0 & \textrm{Otherwise}.
\end{cases}
    \label{boundary_matrix}
\end{equation}
We combine Eqs.~\eqref{sbp} with Eqs.~\eqref{fluxes_discrete} and derive a strong form of Eq.~\eqref{divergenc_quadrature} as

\begin{equation}
\tilde{U}_{k,nmr}=-\left[ \mathcal{G}^1_{k,nmr}+\mathcal{G}^2_{k,nmr}+\mathcal{G}^3_{k,nmr}\right]+\tilde{S}_{k,nmr},
    \label{strong_form}
\end{equation}
where
\begin{equation}
\begin{aligned}
    \mathcal{G}^1_{k,nmr} = \frac{1}{\omega_\mathcal{P}}\left[\tilde{\mathbf{F}}^{*,1}_{k,\mathcal{P}mr}-\tilde{\mathbf{F}}^{1}_{k,\mathcal{P}mr}\right]-\frac{1}{\omega_0}\left[\tilde{\mathbf{F}}^{*,1}_{k,0mr}-\tilde{\mathbf{F}}^{1}_{k,0mr}\right]+\sum_{j=0}^{\mathcal{P}}D_{jn}\tilde{\mathbf{F}}^1_{k,jmr}\\
    \mathcal{G}^2_{k,nmr} = \frac{1}{\omega_\mathcal{P}}\left[\tilde{\mathbf{F}}^{*,2}_{k,n\mathcal{P}r}-\tilde{\mathbf{F}}^{2}_{k,n\mathcal{P}r}\right]-\frac{1}{\omega_0}\left[\tilde{\mathbf{F}}^{*,2}_{k,n0r}-\tilde{\mathbf{F}}^{2}_{k,n0r}\right]+\sum_{j=0}^{\mathcal{P}}D_{jm}\tilde{\mathbf{F}}^2_{k,njr} \\
    \mathcal{G}^3_{k,nmr} = \frac{1}{\omega_\mathcal{P}}\left[\tilde{\mathbf{F}}^{*,3}_{k,nm\mathcal{P}}-\tilde{\mathbf{F}}^{3}_{k,nm\mathcal{P}}\right]-\frac{1}{\omega_0}\left[\tilde{\mathbf{F}}^{*,3}_{k,nm0}-\tilde{\mathbf{F}}^{3}_{k,nm0}\right]+\sum_{j=0}^{\mathcal{P}}D_{jr}\tilde{\mathbf{F}}^3_{k,nmj},
\end{aligned}
    \label{fluxes_discrete_strong}
\end{equation}
where the derivative matrix is defined as $D_{jn}=dl_n(X_j)/dX$. In Eq.~\eqref{fluxes_discrete_strong}, $\tilde{\mathbf{F}}^{*,1}_{k,nmr}$ for $n=\mathcal{P}$ and $n=0$ represent the common fluxes at the interfaces with normal vectors in the $X_1$ mapped coordinate direction. The term $\tilde{\mathbf{F}}^{*,2}_{k,nmr}$ when $m=\mathcal{P}$ and $m=0$ denotes common fluxes at the interfaces of the elements with normal vector in $X_2$ direction. Similarly, $\tilde{\mathbf{F}}^{*,3}_{k,nmr}$ with $r=\mathcal{P}$ and $r=0$ indicates the common flux at the interfaces of elements with normal vector in $X_3$ direction. The flux vectors consist of advective and viscous fluxes. We treat the advective and viscous fluxes differently for the split form of the DGSEM approach. In the following,  we explain how we use split forms for advective fluxes and how we compute viscous fluxes at the elements' interfaces.

\subsection{Advective Fluxes}
The advective fluxes on the elements' interfaces are computed by solving a Riemann problem with two constant states of solution vectors from adjacent elements. The Riemann problem is solved using approximation methods such as Lax-Friedrichs, HLLC, and HLL. Therefore, the interface fluxes are computed using the following formulas

\begin{equation}
\tilde{\mathbf{F}}^{*,1}_{k,0mr}=F^{Rie}\left(\mathbf{U}_{k,0mr},\mathbf{U}_{k-1,\mathcal{P}mr},\vec{n}_{10}\right),\textrm{ } \tilde{\mathbf{F}}^{*,1}_{k,\mathcal{P}mr}=F^{Rie}\left(\mathbf{U}_{k,\mathcal{P}mr},\mathbf{U}_{k+1,0mr},\vec{n}_{1\mathcal{P}}\right)
    \label{riemann_problem1}
\end{equation}
\begin{equation}
\tilde{\mathbf{F}}^{*,2}_{k,n0r}=F^{Rie}\left(\mathbf{U}_{k,n0r},\mathbf{U}_{k-1,n\mathcal{P}r},\vec{n}_{02}\right),\textrm{ } \tilde{\mathbf{F}}^{*,2}_{k,\mathcal{P}mr}=F^{Rie}\left(\mathbf{U}_{k,n\mathcal{P}r},\mathbf{U}_{k+1,n0r},\vec{n}_{2\mathcal{P}}\right)
    \label{riemann_problem2}
\end{equation}
\begin{equation}
\tilde{\mathbf{F}}^{*,3}_{k,nm0}=F^{Rie}\left(\mathbf{U}_{k,nm0},\mathbf{U}_{k-1,nm\mathcal{P}},\vec{n}_{30}\right),\textrm{ } \tilde{\mathbf{F}}^{*,3}_{k,\mathcal{P}mr}=F^{Rie}\left(\mathbf{U}_{k,nm\mathcal{P}},\mathbf{U}_{k+1,nm0},\vec{n}_{3\mathcal{P}}\right).
    \label{riemann_problem3}
\end{equation}
In Eqs.~\eqref{riemann_problem1} to \eqref{riemann_problem3}, $\vec{n}_{ir}$ denotes the normal vector to the corresponding interface. For the advective volume fluxes, Fisher \emph{et al.}~\cite{fisher2013high} and Carpenter \emph{et al.}~\cite{carpenter2014entropy} proved that the derivative operator with the diagonal SBP property can be reformulated into a sub-cell based finite volume scheme as 

\begin{equation}
\begin{aligned}
\sum_{j=0}^{\mathcal{P}}D_{jn}\tilde{\mathbf{F}}^1_{k,jmr} = \frac{\bar{\mathbf{F}}^1_{k,(n+1)mr}-\bar{\mathbf{F}}^1_{k,(n)mr}}{\omega_n},\quad n,m,r=0,\cdots,\mathcal{P}\\
\sum_{j=0}^{\mathcal{P}}D_{jm}\tilde{\mathbf{F}}^2_{k,njr}=\frac{\bar{\mathbf{F}}^2_{k,n(m+1)r}-\bar{\mathbf{F}}^2_{k,n(m)r}}{\omega_m},\quad n,m,r=0,\cdots,\mathcal{P}\\
\sum_{j=0}^{\mathcal{P}}D_{jr}\tilde{\mathbf{F}}^3_{k,nmj}=\frac{\bar{\mathbf{F}}^3_{k,nm(r+1)}-\bar{\mathbf{F}}^3_{k,nm(r)}}{\omega_r},\quad n,m,r=0,\cdots,\mathcal{P}
\end{aligned}
    \label{differencing}
\end{equation}
The flux differencing from Eq. \eqref{differencing} proves the conservation of the derivative operator. Fisher \emph{et al.}~\cite{fisher2013high} and Carpenter \emph{et al.}~\cite{carpenter2014entropy} showed that the volume flux differencing relation \eqref{differencing} can be written in terms of two-point entropy conserving flux functions $F^{**I}_{EC}$ as

\begin{equation}
\frac{\bar{\mathbf{F}}^1_{k,(n+1)mr}-\bar{\mathbf{F}}^1_{k,(n)mr}}{\omega_n}\approx 2\sum_{j=0}^\mathcal{P}D_{jn}F^{**I}_{k,EC}(\mathbf{U}_{k,nmr},\mathbf{U}_{k,jmr}).
    \label{fluxdiff}
\end{equation}
Fisher \emph{et al.}~\cite{fisher2013high} and Carpenter \emph{et al.}~\cite{carpenter2014entropy} then showed that the two-point entropy conservative flux results in a high-order accurate discretization. Gassner \emph{et al.}~\cite{gassner2016split} later concluded that the high-order accuracy resulted from the symmetry and consistency of the two-point flux function and not from the entropy conservation. Therefore, they rewrote Eq.~\eqref{fluxdiff} such that

\begin{equation}
\frac{\bar{\mathbf{F}}^1_{k,(n+1)mr}-\bar{\mathbf{F}}^1_{k,(n)mr}}{\omega_n}\approx 2\sum_{j=0}^\mathcal{P}D_{jn}F^{**I}_{k}(\mathbf{U}_{k,nmr},\mathbf{U}_{k,jmr}),
    \label{fluxdiffcons}
\end{equation}
where the two-point flux function $F^{**I}_{k}$ satisfies consistency and symmetry conditions as 
\begin{equation}
    F^{**I}_k(\mathbf{U}_{k,i},\mathbf{U}_{k,i})=F^I_k(\mathbf{U}_{k,i})\quad\textrm{and}\quad F^{**I}_k(\mathbf{U}_{k,i},\mathbf{U}_{k,j})=F^{**I}_k(\mathbf{U}_{k,j},\mathbf{U}_{k,i}).
    \label{conssymm}
\end{equation}
The two-point entropy conservative fluxes can be determined using the semi-discrete setting of Tadmor \cite{tadmor1987numerical,tadmor2003entropy}. Peyvan \emph{et al.} \cite{PEYVAN2023112310} used the Tadmor relation to derive new entropy conservative fluxes for the hypersonic equations. Here, we can also employ the split form fluxes presented in the literature such as standard DG \cite{kopriva2009implementing}, Morinishi \cite{morinishi2010skew}, Ducros \emph{et al.}~\cite{ducros2000high}, Kennedy and Gruber \cite{kennedy2008reduced}, and Pirozzoli \cite{pirozzoli2011numerical}. Gassner \emph{et al.}~\cite{gassner2016split} presented the two-point flux formulations of all the split forms. This strategy avoids the tedious calculations of the entropy conservative fluxes for the hypersonic formulations. It also provides a flux differencing framework that can be exploited to construct a conservative hybrid discretization for shock capturing. Substituting Eq.~\eqref{differencing} into \eqref{fluxes_discrete_strong}, we can derive the following for discrete advective fluxes as

\begin{equation}
\begin{aligned}
    \mathcal{G}^1_{k,nmr} = \frac{1}{\omega_\mathcal{P}}\left[\tilde{\mathbf{F}}^{*,1}_{k,\mathcal{P}mr}-\tilde{\mathbf{F}}^{1}_{k,\mathcal{P}mr}\right]-\frac{1}{\omega_0}\left[\tilde{\mathbf{F}}^{*,1}_{k,0mr}-\tilde{\mathbf{F}}^{1}_{k,0mr}\right]+\frac{\bar{\mathbf{F}}^1_{k,(n+1)mr}-\bar{\mathbf{F}}^1_{k,(n)mr}}{\omega_n}\\
    \mathcal{G}^2_{k,nmr} = \frac{1}{\omega_\mathcal{P}}\left[\tilde{\mathbf{F}}^{*,2}_{k,n\mathcal{P}r}-\tilde{\mathbf{F}}^{2}_{k,n\mathcal{P}r}\right]-\frac{1}{\omega_0}\left[\tilde{\mathbf{F}}^{*,2}_{k,n0r}-\tilde{\mathbf{F}}^{2}_{k,n0r}\right]+\frac{\bar{\mathbf{F}}^2_{k,n(m+1)r}-\bar{\mathbf{F}}^2_{k,n(m)r}}{\omega_m} \\
    \mathcal{G}^3_{k,nmr} = \frac{1}{\omega_\mathcal{P}}\left[\tilde{\mathbf{F}}^{*,3}_{k,nm\mathcal{P}}-\tilde{\mathbf{F}}^{3}_{k,nm\mathcal{P}}\right]-\frac{1}{\omega_0}\left[\tilde{\mathbf{F}}^{*,3}_{k,nm0}-\tilde{\mathbf{F}}^{3}_{k,nm0}\right]+\frac{\bar{\mathbf{F}}^3_{k,nm(r+1)}-\bar{\mathbf{F}}^3_{k,nm(r)}}{\omega_r},
\end{aligned}
\label{fluxes_discrete_fluxdifferencing}
\end{equation}
with the high-order flux defined as 

\begin{equation}
\begin{aligned}
\Bar{\mathbf{F}}^1_{k,0mr}=\mathbf{F}^{a,1}_{k,0mr},
\textrm{ }
\Bar{\mathbf{F}}^{a,1}_{k,jmr}=\sum_{p=j}^{\mathcal{P}}\sum_{q=0}^{j-1}2\omega_{q}D_{qp}F^{**,I}_{k}(\mathbf{U}_{qmr},\mathbf{U}_{pmr})\textrm{ } j=1,\cdots,\mathcal{P},
\textrm{ }
\Bar{\mathbf{F}}^{a,1}_{k,(\mathcal{P}+1)mr}=\mathbf{F}^{a,1}_{k,\mathcal{P}mr},\\
\Bar{\mathbf{F}}^2_{k,n0r}=\mathbf{F}^{a,2}_{k,n0r},
\textrm{ }
\Bar{\mathbf{F}}^{a,2}_{k,njr}=\sum_{p=j}^{\mathcal{P}}\sum_{q=0}^{j-1}2\omega_{q}D_{qp}F^{**,I}_{k}(\mathbf{U}_{nqr},\mathbf{U}_{npr})\textrm{ } j=1,\cdots,\mathcal{P},
\textrm{ }
\Bar{\mathbf{F}}^{a,2}_{k,n(\mathcal{P}+1)r}=\mathbf{F}^{a,2}_{k,n\mathcal{P}r},\\
\Bar{\mathbf{F}}^3_{k,nm0}=\mathbf{F}^{a,3}_{k,nm0},
\textrm{ }
\Bar{\mathbf{F}}^{a,3}_{k,nmj}=\sum_{p=j}^{\mathcal{P}}\sum_{q=0}^{j-1}2\omega_{q}D_{qp}F^{**,I}_{k}(\mathbf{U}_{nmq},\mathbf{U}_{nmp})\textrm{ } j=1,\cdots,\mathcal{P},
\textrm{ }
\Bar{\mathbf{F}}^{a,3}_{k,nm(\mathcal{P}+1)}=\mathbf{F}^{a,3}_{k,nm\mathcal{P}},
\end{aligned}
\label{DGvec}
\end{equation}
We can now blend the high-order flux with a first-order finite volume flux constructed on sub-cell interfaces. The sizes of those finite volume sub-cells are given by the quadrature weights of each GLL point \cite{hennemann2021provably}. In the sub-cells, the solution values on the GLL points are taken to be the sub-cell solution averages \cite{hennemann2021provably}. We blend the high-order flux function with the low-order finite volume one to stabilize the high-order scheme. Using a combination of the high and low-order fluxes, we rewrite Eq.~\eqref{fluxes_discrete_fluxdifferencing} as
\begin{equation}
\begin{aligned}
    \mathcal{G}^1_{k,nmr} = \frac{1}{\omega_\mathcal{P}}\left[\tilde{\mathbf{F}}^{*,1}_{k,\mathcal{P}mr}-\tilde{\mathbf{F}}^{1}_{k,\mathcal{P}mr}\right]-\frac{1}{\omega_0}\left[\tilde{\mathbf{F}}^{*,1}_{k,0mr}-\tilde{\mathbf{F}}^{1}_{k,0mr}\right]+\frac{1}{\omega_n}\Delta \left((1-\alpha)\vec{\bar{\mathbf{F}}}^1_{k,mr}+\alpha\vec{\mathbf{F}}^{1,FV}_{k,mr}\right)\\
    \mathcal{G}^2_{k,nmr} = \frac{1}{\omega_\mathcal{P}}\left[\tilde{\mathbf{F}}^{*,2}_{k,n\mathcal{P}r}-\tilde{\mathbf{F}}^{2}_{k,n\mathcal{P}r}\right]-\frac{1}{\omega_0}\left[\tilde{\mathbf{F}}^{*,2}_{k,n0r}-\tilde{\mathbf{F}}^{2}_{k,n0r}\right]+\frac{1}{\omega_m}\Delta \left((1-\alpha)\vec{\bar{\mathbf{F}}}^2_{k,nr}+\alpha\vec{\mathbf{F}}^{2,FV}_{k,nr}\right) \\
    \mathcal{G}^3_{k,nmr} = \frac{1}{\omega_\mathcal{P}}\left[\tilde{\mathbf{F}}^{*,3}_{k,nm\mathcal{P}}-\tilde{\mathbf{F}}^{3}_{k,nm\mathcal{P}}\right]-\frac{1}{\omega_0}\left[\tilde{\mathbf{F}}^{*,3}_{k,nm0}-\tilde{\mathbf{F}}^{3}_{k,nm0}\right]+\frac{1}{\omega_r}\Delta \left((1-\alpha)\vec{\bar{\mathbf{F}}}^3_{k,nm}+\alpha \vec{\bar{\mathbf{F}}}^{3,FV}_{k,nm}\right),
\end{aligned}
\label{fluxes_discrete_with_fv}
\end{equation}
with 
\begin{equation}
\Delta=\begin{pmatrix}-1&1&0&\cdots&\cdots&0\\0&-1&1&0&\cdots&0\\\vdots&\ddots&\ddots&\ddots&\ddots&\vdots\\0&\cdots&0&-1&1&0\\0&\cdots&\cdots&0&-1&1
\end{pmatrix}\in \mathbb{R}^{(\mathcal{P}+1)\times(\mathcal{P}+2)},
    \label{finitevolumeder}
\end{equation}
and with $\alpha$ as the blending coefficient defined as $\alpha \in [0,\alpha_{max}]$. The blending coefficient $\alpha$ is constant over each element. The limit $\alpha_{max}$ is a user-defined value that satisfies $\alpha_{max}\le 1$. The value of $\alpha$ is determined based on the decay rate of the energy stored at high modes on each element \cite{hennemann2021provably}. The finite volume flux vector is expressed as

\begin{equation}
\Vec{F}_k^{I,FV}=\begin{pmatrix}0\\\Tilde{F}_k^{*I,a}\left(\mathbf{U}_{k,0},\mathbf{U}_{k,1}\right)\\\vdots\\ \Tilde{F}_k^{*I,a}\left(\mathbf{U}_{k,\mathcal{P}-1},\mathbf{U}_{k,\mathcal{P}}\right)\\ 0
\end{pmatrix}\in \mathbb{R}^{\mathcal{P}+2}
    \label{finitevolumeflux}
\end{equation}
where the $*$ superscript denotes the Riemann flux calculated from the two solution states at an interface. When $\alpha=1$, the discretization reduces to a finite volume scheme, and when $\alpha=0$, the discretization reverts to the high-order DGSEM scheme. The blending technique is identical to the work of Hennemann \emph{et al.}~\cite{hennemann2021provably}.

\subsection{Viscous Fluxes}
The viscous fluxes are functions of primitive variables such as velocity, temperature, and species molar fractions according to Eqs.~\eqref{diffusion}-\eqref{internal_cond}:

\begin{equation}
\vec{\tilde{\mathbf{F}}}^{v}_{k,j}=F^{I,v}_{k,j}(\mathbf{W}_{k,j}),
    \label{visfluxcalc}
\end{equation}
where $\mathbf{Q}_{k,j}=\mathbf{W}_{x,k,j}$ is the first order derivative of the primitive variable. Bassi and Rebay \cite{bassi1997high} regard the gradient of the conservative variables as auxiliary unknowns of the Navier-Stokes equations. In contrast, in this study, we define the gradient of the primitive variables as the auxiliary unknowns to avoid transforming between derivatives of conservative variables and derivatives of primitive variables. We first define the equation $Q^I=W^I_x$ and map it onto the reference element. We then formulate the weak form similar to Eqs.~\eqref{integration} as

\begin{equation}
\sum_{j=0}^\mathcal{P} \Tilde{Q}_{k,j}^I l_i(\eta_j)\omega_j=-\frac{1}{|J|}\left(\Tilde{W}^{*I}_{r,k}-\Tilde{W}^{*I}_{l,k}-\sum_{j=0}^\mathcal{P}\Tilde{W}^{I}_{j,k}\frac{d l_i(\eta_j)}{d \eta} \omega_j\right).
    \label{auxil}
\end{equation}
The equivalent strong form of Eq.~\eqref{auxil} reads as
\begin{equation}
\Vec{Q}^I_k=-\frac{1}{|J|}\left(\mathbf{M}^{-1}\mathbf{B}\left(\Vec{W}^{*I}_k-\Vec{W}^I_k\right)-\mathbf{D}\Vec{W}^I_k\right),
    \label{auxstrongf}
\end{equation}
with $\Vec{W}^{*I}_k=\left(W^{*I}_{l,k},0,\cdots,0,W^{*I}_{r,k}\right)^T$ where the common interface values are determined using an arithmetic mean such that $W^{*I}_{l,k}=\frac{1}{2}\times(W^{I}_{k-1,\mathcal{P}}+W^{I}_{k,0})$ and $W^{*I}_{r,k}=\frac{1}{2}\times(W^{I}_{k,\mathcal{P}}+W^{I}_{k+1,0})$. After calculating the auxiliary variable $\Vec{Q}^I_k$, we employ the relations \eqref{visc_fluxes} to determine the viscous fluxes at the GLL points. For the viscous interface fluxes, we enforce continuity by averaging the fluxes across the interface.

\section{New Developments}
\label{sec:dev}
The original Trixi.jl framework lacked four main components for the target hypersonic viscous solver. 
\begin{itemize}
    \item A solver for the parabolic equations using the general P4est mesh type.
    \item The MPI implementation of the parabolic solver.
    \item Non-equilibrium chemistry modeling, including non-equilibrium thermodynamics, chemical kinetics, and transport processes.
    \item Existing equation types were unsuitable for adapting to hypersonic viscous flow governing equations.
\end{itemize}
In the following, we describe how we implemented these four main components in Trixi.jl framework to construct the H$^3$PC solver.

\subsection{P4est based Adaptive Parabolic Solver}
In Trixi.jl, the main solver is written based on the Discontinuous Galerkin method that employs DGmulti for simplex elements and DGSEM for hypercube or tensor product elements. We focused on developing a parabolic solver with conforming and non-conforming 2D and 3D quadrilateral and hexahedron elements. Similar to other Trixi.jl solvers, the parabolic solver computes the volume and surface integrals of the viscous (parabolic) part of the Navier-Stokes equations. First, computing the volume integral of the viscous fluxes for P4est mesh requires the computation of the gradient of primitive variables defined in Eq.\eqref{auxil}. In the gradient module, flux patching of the auxiliary equation is performed on the interfaces of elements and the mortars that connect non-conforming elements. The mortars in 2D and 3D configurations employ 1:2 and 1:4 division standards. The solver, L$_2$-projection and interpolation matrices are used to transfer the solution from the large sides to the small sides of the mortars and vice versa. After the gradient computation, the volume integral of the divergence of the viscous fluxes is computed using the definitions presented in Eq.~\eqref{visc_fluxes}. In the next stage, viscous fluxes are patched across interfaces and mortars to calculate the surface integrals. In the end, the viscous flux computed in the parabolic solver are combined with hyperbolic fluxes, and the time stepping scheme updates the solution variables for the next time step. 

\subsection{P4est Parabolic Solver Parallelization}
In Trixi.jl, all the data structures containing information on elements, interfaces, boundaries, and mortars are initialized  and stored in a cache container. This cache container is carried out through the code workflow, and various functions act on the data structures and update the solution variables for all the degrees of freedom in the spatial discretization. A cache existed for hyperbolic discretization, and hence we implement another cache for the parabolic discretization that carries data structures required for the solution update during the execution of the code. The parabolic cache is designed only to include elements, interfaces, boundaries, and viscous container data structures. For all the data structures related to mortars, the parabolic solver uses the already stored data within the hyperbolic cache to avoid extensive memory usage. 

In the Trixi.jl framework, for parallel execution of hyperbolic solver, additional data structures are defined for MPI interfaces, MPI mortars, and an MPI cache container. MPI interfaces and MPI mortars are assigned to interfaces and mortars that exist at the boundary of local partitions on each MPI rank. The MPI cache consists of all the required data for non-blocking communication between MPI ranks, including sending and receiving buffers for MPI mortar and MPI interface data. Since the communication data structure for both hyperbolic and parabolic solvers are identical, we used the same parallel hyperbolic cache for all the communications needed for the parabolic solver. Using this approach, we avoid allocating extra data structures on memory for communications. We implemented new functions for prolongation to interfaces and mortars for parallel execution mode. The calculation of mortar and interface fluxes for parallel execution are redesigned for the parabolic solver. It is worth mentioning that prolongation for gradient and divergence computations are completely different since we require directional information for the flux patching. When AMR functionality is enabled in parallel mode, the parabolic solver resizes all the data structures and then applies load balancing and mesh repartitioning procedures. The parallel parabolic solver can efficiently perform 2D and 3D simulations using AMR and static mesh mode.

\subsection{Thermo-Chemistry and Transport Coefficient Modeling}
\label{sec:mutation}
For the modeling of gas mixture at hypersonic speeds, we need to model thermodynamics, transport processes, and chemical reactions kinetics. We employ the Mutation++ library for such modeling in H$^3$PC solver. The Mutation++ library is developed in C++ using an application binary interface (ABI) that leverages C++ features such as polymorphism and inheritance. The current implementation of Mutation++ provides an ABI for both Fortran and C languages. Since Julia does not natively support polymorphism, we compiled the Mutation++ library as a dynamic Fortran library and then developed all the subroutines in Fortran as Julia native functions, using the ccall() mechanism. In Julia, ccall is a function used to call external functions written in other programming languages (such as C, Fortran, or C++). This is part of Julia's Foreign Function Interface (FFI), which allows Julia code to interoperate with libraries and functions written in other languages. The ccall() function provides a way to interface with shared libraries or dynamic link libraries (DLLs) directly from Julia. 

\subsection{Hypersonic Equations Type}
\begin{figure}[t!]
\centering
\includegraphics[width=0.7\textwidth,height=0.3\textwidth]{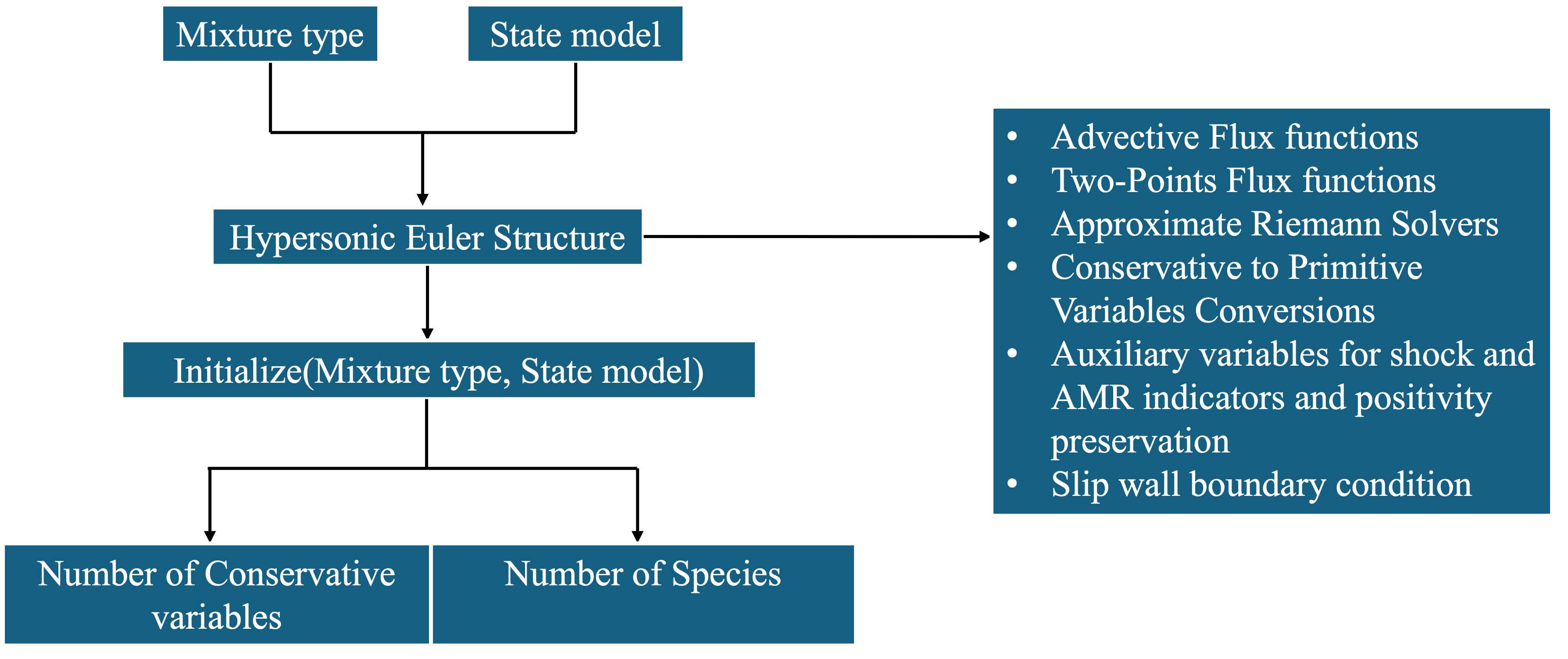}
  \caption{Schematic of Hypersonic-Euler equations type for H$^3$PC solver.}
  \label{fig:equation_type}
\end{figure}

For the H$^3$PC solver, we employ the external Mutation++ library. Therefore, we are required to implement a new equation type within the framework of Trixi.jl that is entirely different from the existing equations. We define two equation types, including Hypersonic-Euler and Hypersonic-Navier-Stokes, for 1D, 2D, and 3D spatial configurations. According to Fig.~\ref{fig:equation_type}, the user can choose a mixture type such as ``air5" and a state model such as ``ChemNonEq1T" and input to the equation structure. The Hypersonic-Euler structure initializes Mutation++ according to the mixture type and state model. After the initialization, all the functionality of the Mutation++ library will be accessible through the Julia ABI described in section \ref{sec:mutation}. The structure stores a variable for the number of conservative variables and one for the number of species computed by the Mutation++ library based on the mixture type. The Hypersonic-Euler type provides auxiliary utilities that various parts of the H$^3$PC solver can use for different tasks. These utilities include individual functions that compute advective fluxes, two-points fluxes, approximate Riemann solvers, variables for shock, AMR  and positivity preservation schemes, and specific boundary conditions such as the slip wall condition. 
The structure of the Hypersonic-Navier-Stokes type is very similar to Hypersonic-Euler type with different auxiliary functions to compute viscous fluxes, boundary conditions for gradient and divergence computations.

%% file: results.tex
\newcommand{\ocircleplus}{\mathrel{\ooalign{$\bigcirc$\cr\hidewidth$+$\hidewidth\cr}}}
\section{Results}
In this section, we evaluate the hypersonic solver through a series of benchmark test cases. Initially, the integration of the Mutation++ library with the $\textrm{H}^3\textrm{PC}$ solver is validated using a blast wave problem. Subsequently, the accuracy of the H$^3$PC solver is verified through the Taylor-Green vortex case. The solver's capability to capture vortex street formations in supersonic flows is then demonstrated for a two-dimensional configuration. We also simulate hypersonic flow inside the P8-inlet geometry. 


To compare the results obtained from the current high-order solver, we use the UCNS3D code developed by Antoniadis et al. \cite {antoniadis2022ucns3d}, which is based on high-order WENO scheme. UCNS3D is an open-source, high-order compressible CFD solver designed for unstructured static meshes. UCNS3D features a comprehensive set of high-resolution numerical schemes, such as Central schemes, Monotone Upstream-centered Scheme for Conservation Laws (MUSCL), Weighted Essentially Non-Oscillatory (WENO) schemes, Central WENO (CWENO) schemes, CWENOZ schemes, and Multidimensional Optimal Order Detection (MOOD) schemes. These schemes offer spatial accuracy from 1st to 7th order and are complemented by a wide range of temporal discretization methods for both steady and transient flow problems.

\subsection{Non-Equilibrium Blast Wave}
This section addresses the 2D form of the reactive Euler equations described in \eqref{system_equation}. Grossman and Cinnella \cite{grossman1990flux} proposed a 1D Sod problem that models air as a gas mixture of five species under chemical non-equilibrium. A verification test is included in the appendix to validate the integration of the Mutation++ library with the H$^3$PC solver. This study uses the gas mixture's driver and driven states to initialize a circular region at the center of a 2D square domain. The central region is set to the driver state, while the surrounding area corresponds to the driven state of Grossman’s test problem. Initially, the driver and driven states of the gas mixture are in equilibrium, which can be fully characterized by the gas mixture’s temperature and pressure. The square domain is defined as $x, y \in [-1, 1]^2$, and the solution is initialized using conservative variables computed based on the equilibrium state defined as
\begin{equation}
\left(T,u,P\right)= 
\begin{cases}
     \left(9000\textrm{ K}, 0\textrm{ }\frac{\textrm{m}}{\textrm{s}}, 195,256 \textrm{ Pa}\right),& \text{if } \sqrt{\left(x^2+y^2\right)}<0.5\\
     \left(300 \textrm{ K}, 0\textrm{ }\frac{\textrm{m}}{\textrm{s}}, 10,000 \textrm{ Pa}\right),              &\text{if } \textrm{Otherwise}
\end{cases},
    \label{eq:grossman}
\end{equation}
which translates to the values for the species densities presented in Table~\ref{tbl:grossman}.
\begin{table}[!t]
\caption{Non-Equilibrium Blast Wave: The equilibrium  total and species densities for driver and driven sections of the non-equilibrium Sod problem. $r=\sqrt{x^2+y^2}$.}  
\centering
\begin{tabular}{lcccccc} 
\hline
&$\rho_N$&$\rho_O$&$\rho_{NO}$&$\rho_{N_2}$&$\rho_{O_2}$&$\rho$\\
\hline
$r<0.5$&0.027912&0.00894 &3.49306e-5 &  0.00158&     5.01766e-7 &  1.3042e-05\\
$r>0.5$&8.83171e-81&4.62783e-42 &  3.11979e-17 & 0.08872   &  0.02694& 0.232918\\
\hline
\end{tabular}
\label{tbl:grossman}
\end{table}

\begin{figure}[t!]
  \begin{center}
    \begin{tabular}{cccc}

\includegraphics[width=0.24\textwidth,height=0.20\textwidth]{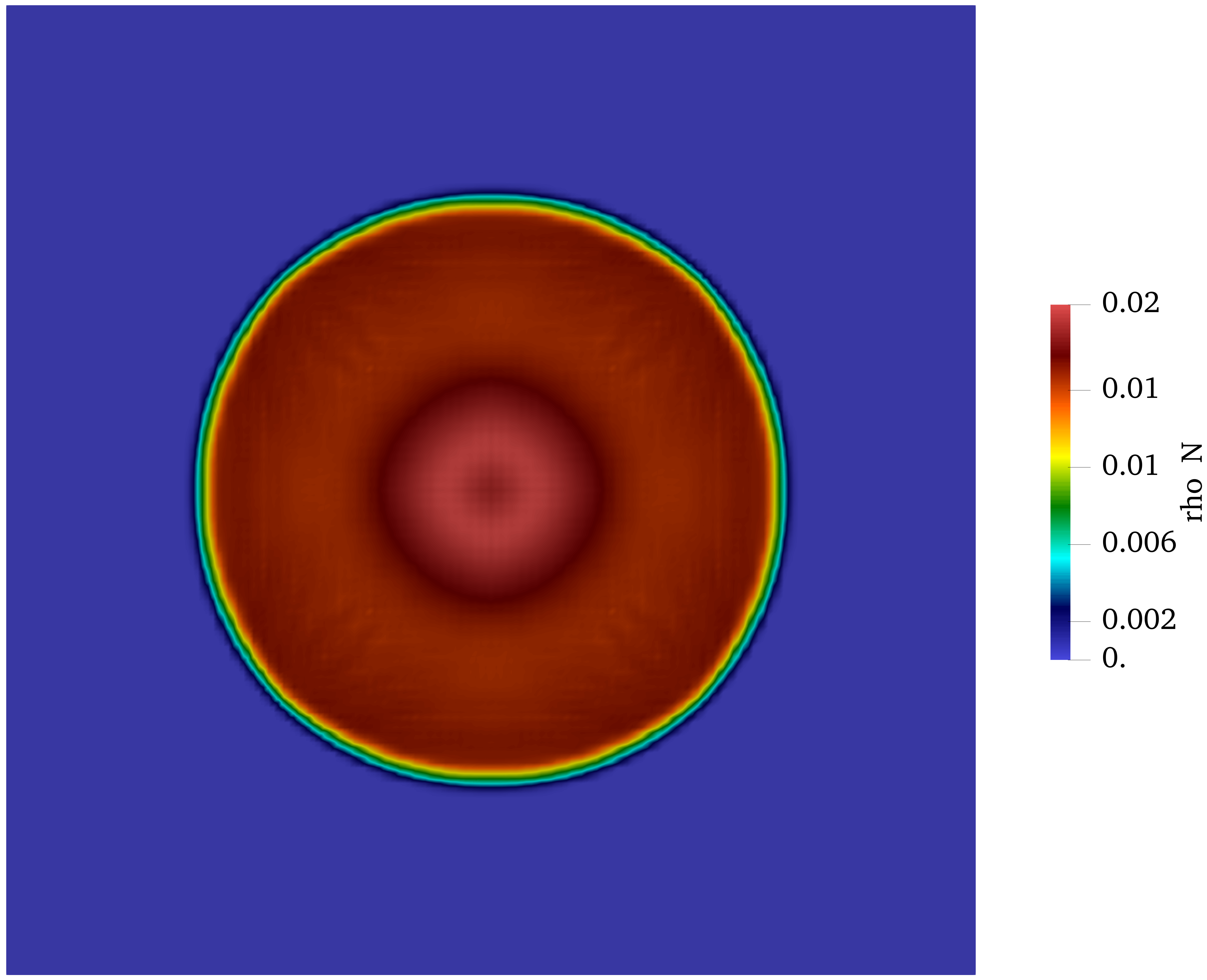}
 &
 \includegraphics[width=0.24\textwidth,height=0.20\textwidth]{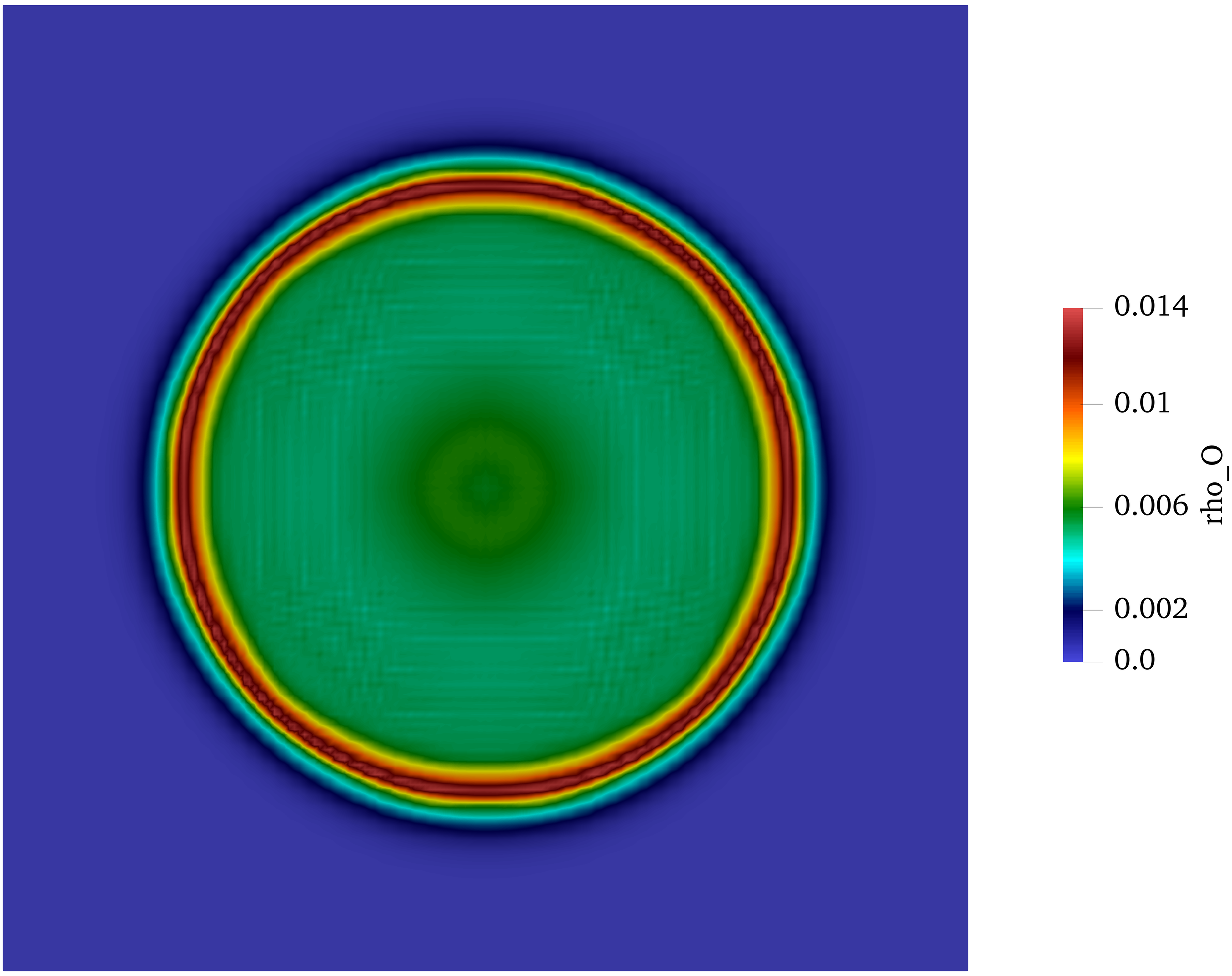}

 &

\includegraphics[width=0.24\textwidth,height=0.20\textwidth]{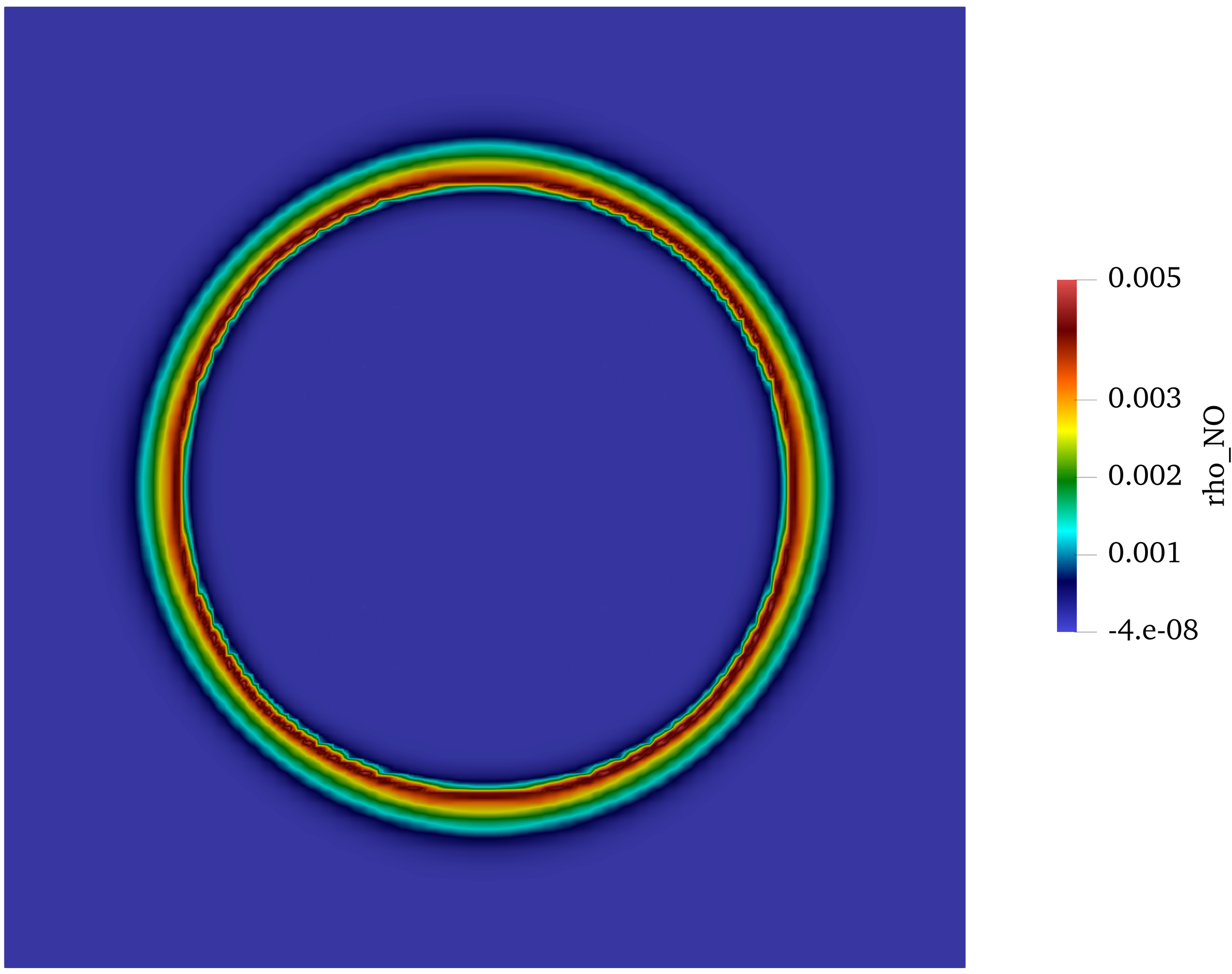}
 &

\includegraphics[width=0.24\textwidth,height=0.20\textwidth]{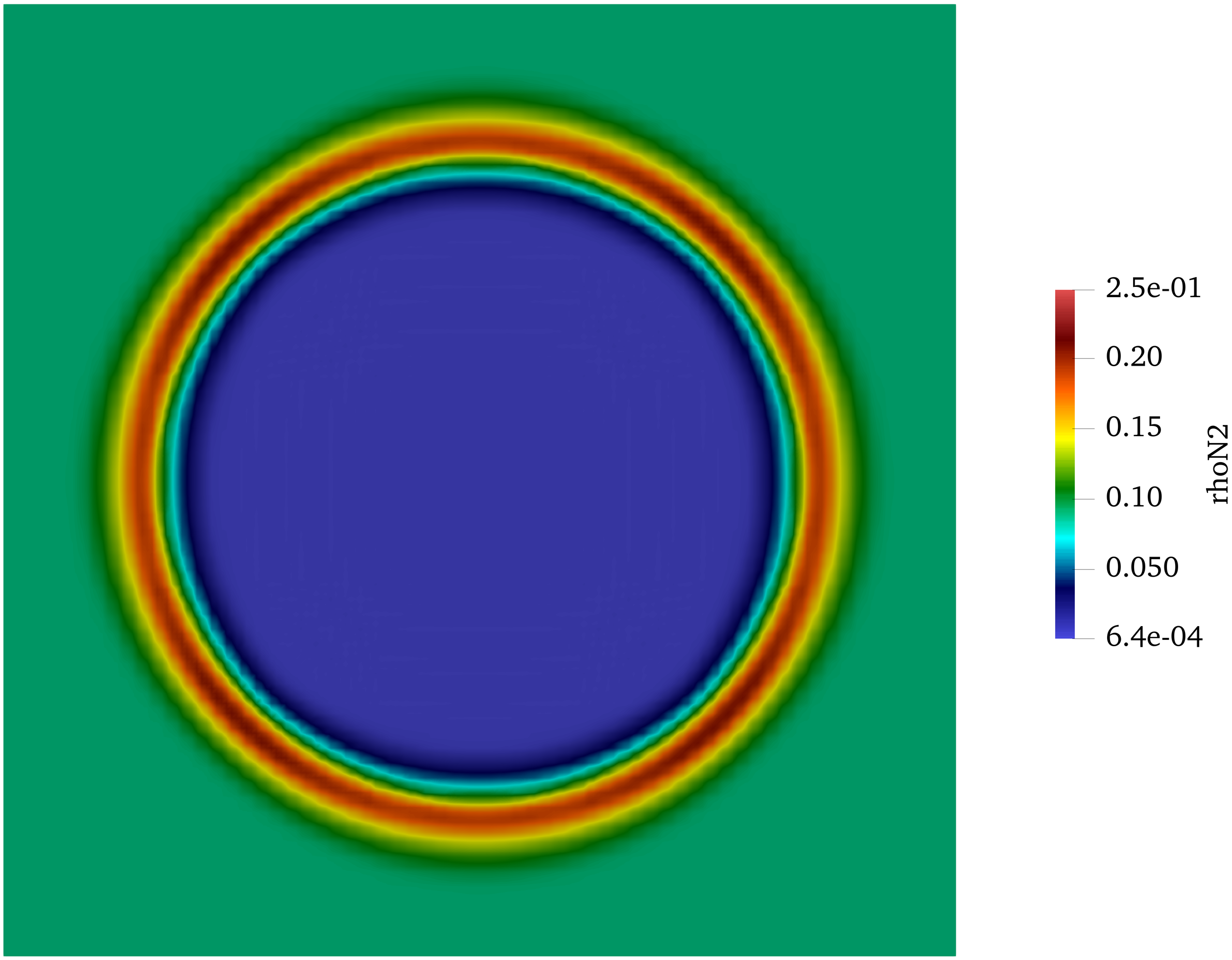}
  \\
  (a) $\rho_N$& (b) $\rho_O$ & (c) $\rho_{NO}$ & (d) $\rho_{N_2}$
  \\
  \includegraphics[width=0.24\textwidth,height=0.20\textwidth]{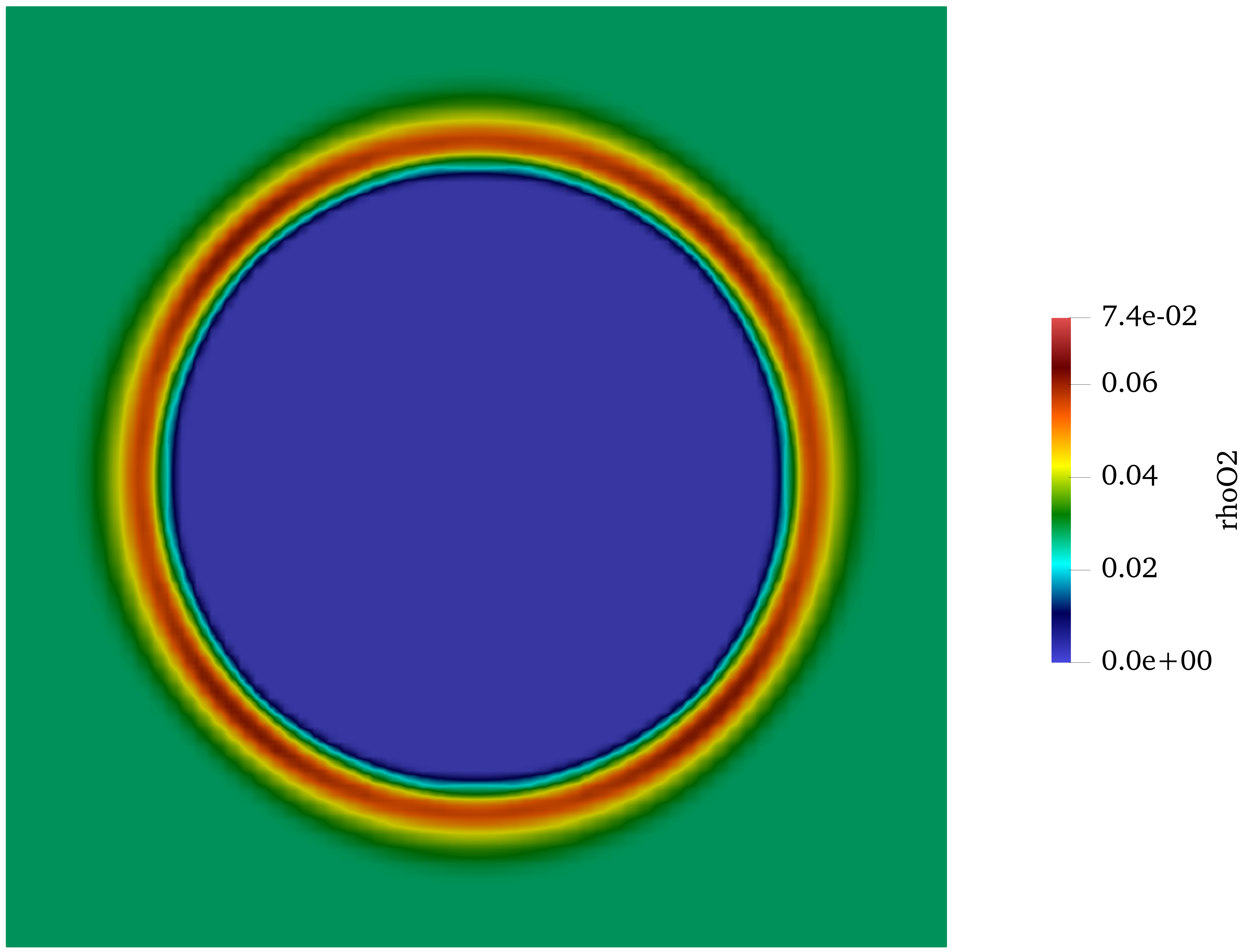}
  &
  \multicolumn{2}{c}{\includegraphics[width=0.24\textwidth,height=0.20\textwidth]{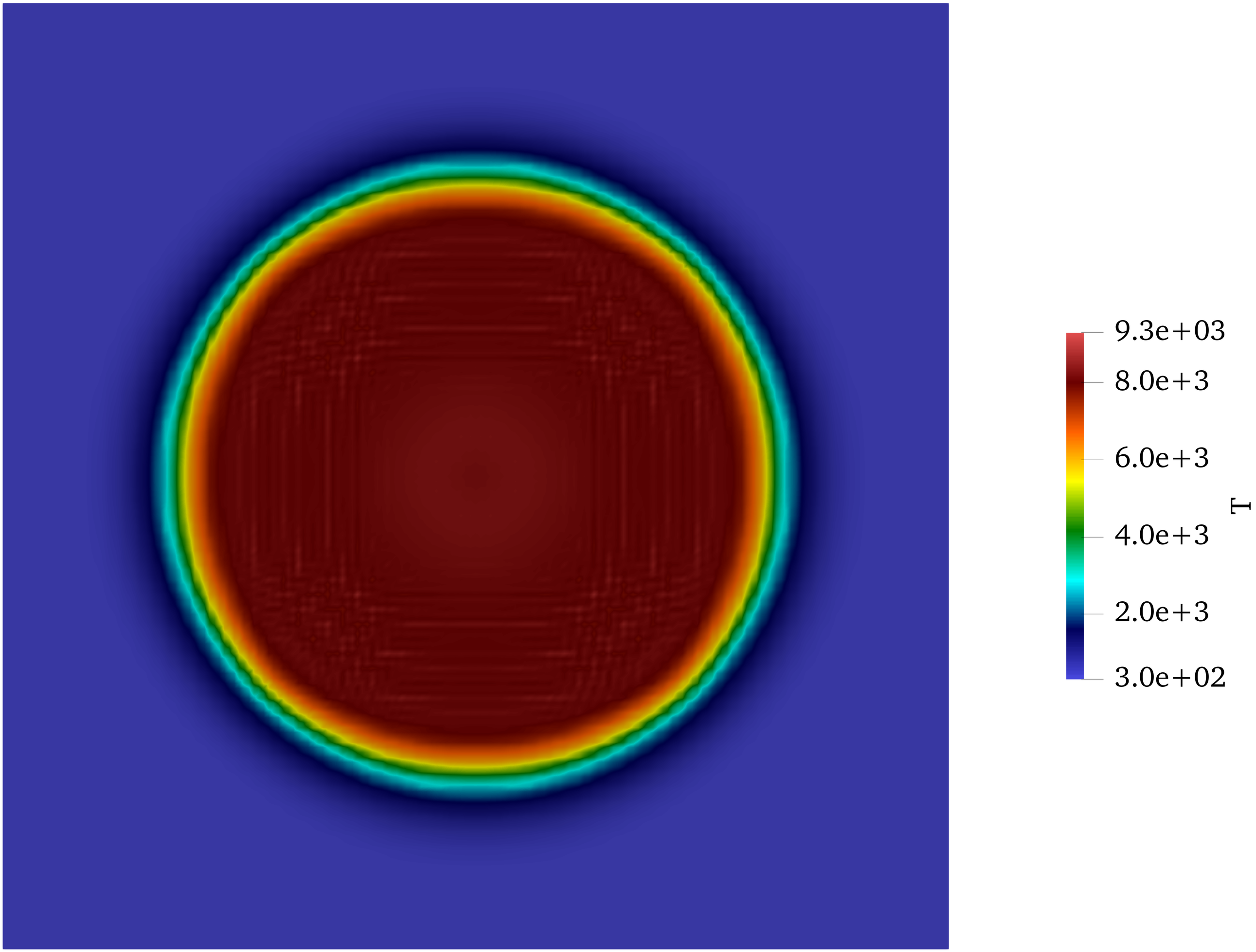}}

  &
\includegraphics[width=0.24\textwidth,height=0.20\textwidth]{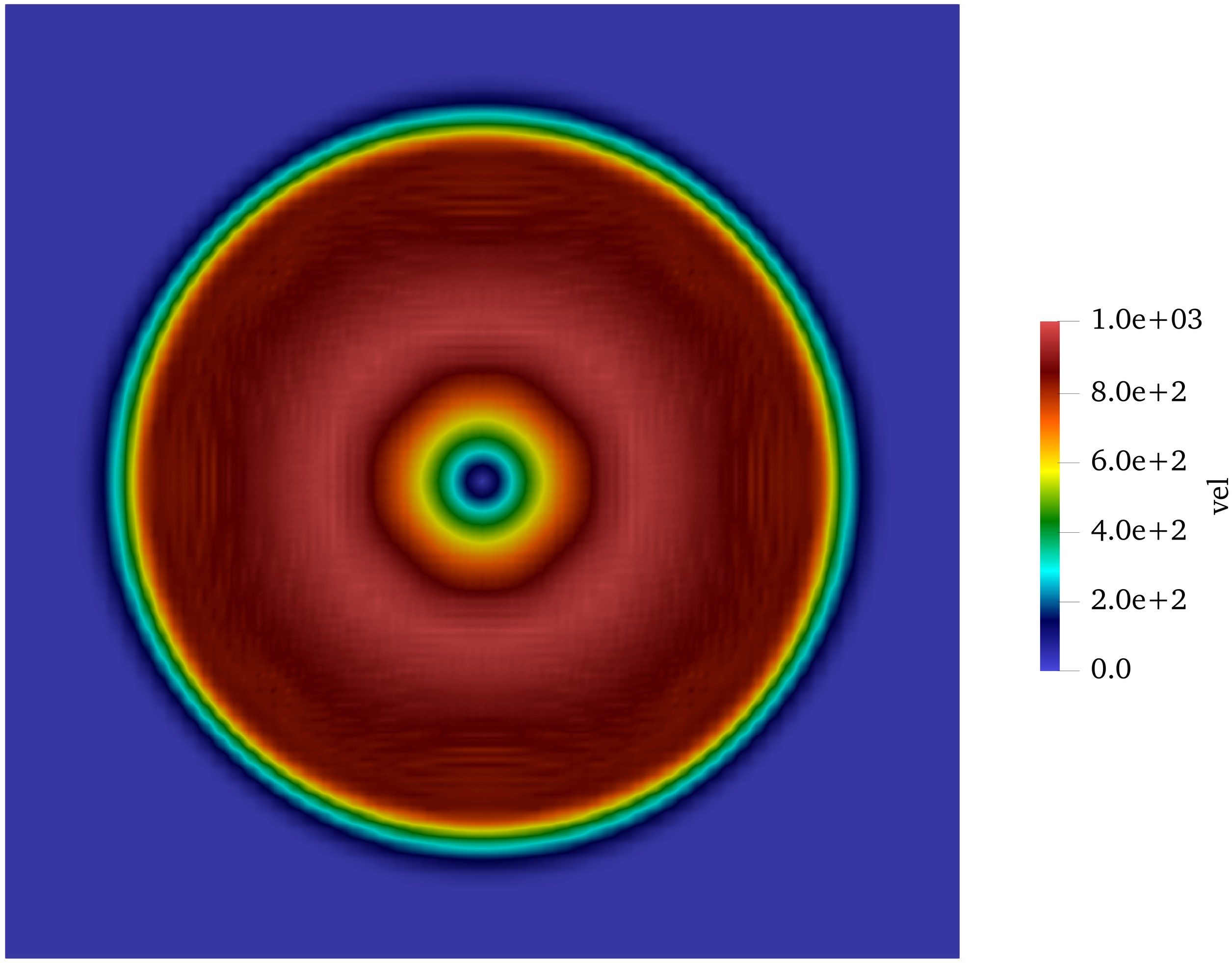}
\\
(e)$\rho_{O_2}$& \multicolumn{2}{c}{(f) Temperature}&(g) Velocity magnitude
\end{tabular} 
\caption{Hypersonic blast wave, (a) Density of $N$, (b) Density of $O$, (c) Density of $NO$, (d) Density of $N_2$, (e) Density of $O_2$ (f) Temperature, (g) Velocity magnitude. The domain is discretized by employing $64\times64$ elements with $\mathcal{P}=3$. The flow is initialized using the modified Sod problem by Grossman \cite{grossman1990flux}. A high temperature region is imposed inside a circle with radius of 0.5 at the center of the square domain with $x,y \in [-1,1]^2$.}
    
    \label{fig:circular_grossman}
  \end{center}
  
\end{figure}
We define free-stream boundary conditions on all the boundaries of the domain. We use 32 elements with order $\mathcal{P}=3$ for each Cartesian direction. Lax-Friedrichs and Ducros approaches are employed for the surface and volume flux functions. Species production terms and thermodynamics of the gas mixture are computed using the Mutation++ library \cite{Scoggins2020}. The gas mixture thermo-chemistry modeling is performed using chemical non-equilibrium with a single temperature model (``ChemNonEq1T") and ``air$_5$" gas mixture. The equations are solved until $t=0.0002 s$. The positivity preserving scheme of Zhang and Shu \cite{zhang2011positivity} enforces positivity of all the species densities and pressure weakly. The time-stepping scheme for the $\textrm{H}^3\textrm{PC}$ solver is the four-stage third-order strong stability preserving Runge-Kutta (SSPRK43)\cite{kraaijevanger1991contractivity,conde2018embedded,ranocha2022optimized} with an adaptive time step size.
\begin{figure}[t!]
  \begin{center}
    \begin{tabular}{ccc}

\includegraphics[width=0.33\textwidth,height=0.28\textwidth]{./figures/T.png}
 &
 \includegraphics[width=0.33\textwidth,height=0.28\textwidth]{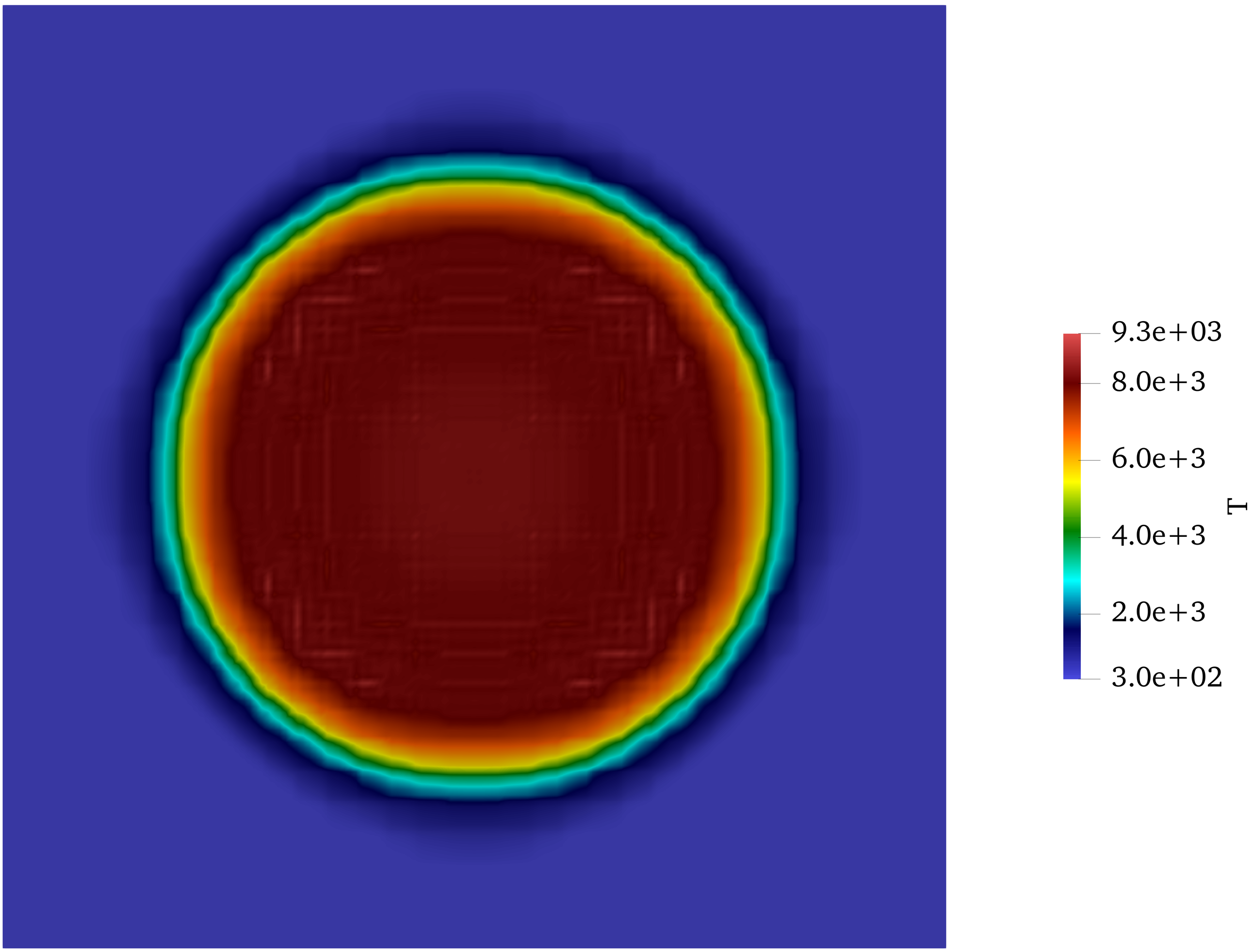}

 &

\includegraphics[width=0.33\textwidth,height=0.28\textwidth]{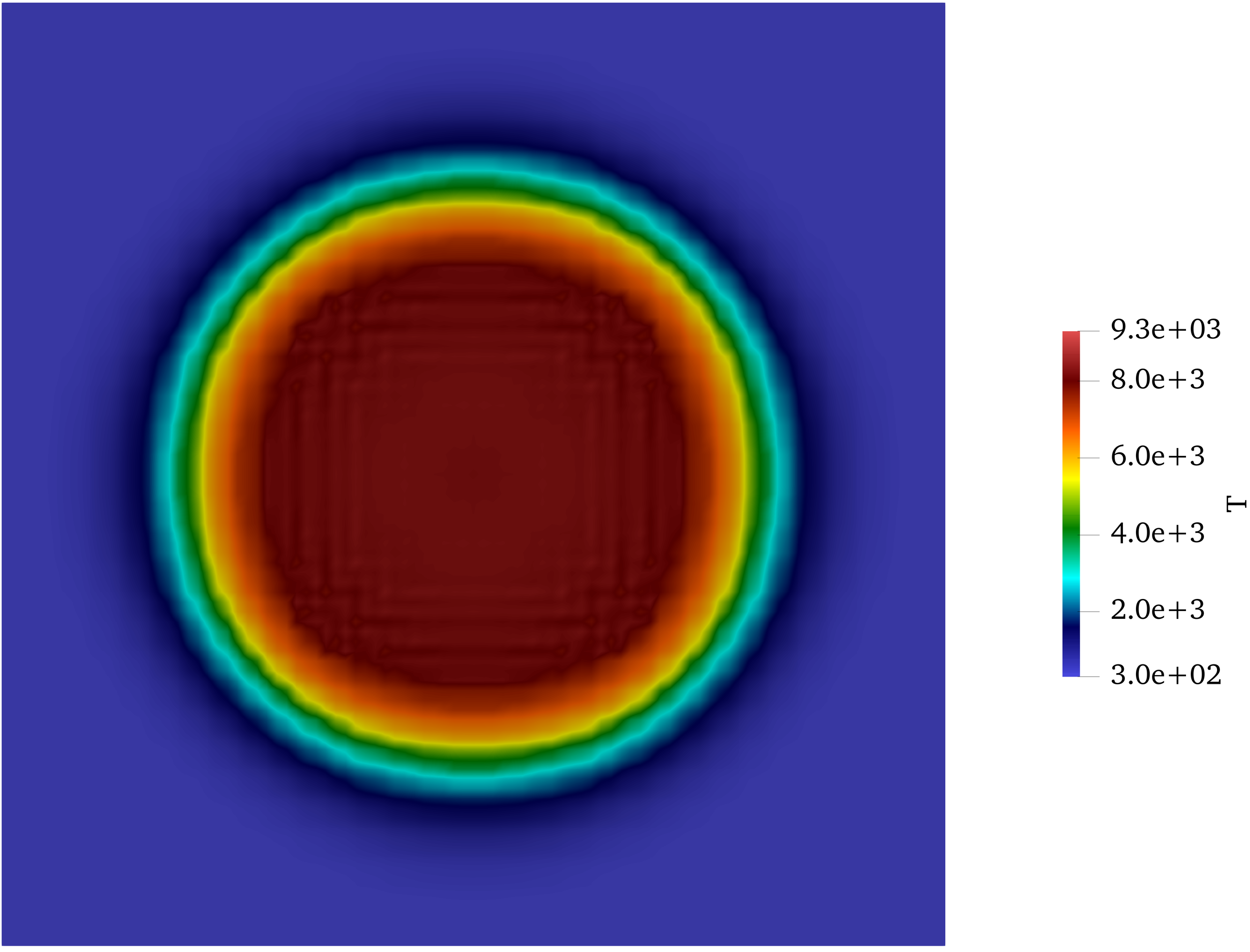}
  \\
  (a) $N=64$ and $\mathcal{P}=3$& (b) $N=32$ and $\mathcal{P}=5$ & (c) $N=32$ and $\mathcal{P}=3$

\end{tabular} 
\caption{Comparison of element size and polynomial order for the non-equilibrium blast wave problem.}
    \label{fig:circular_grossman_compare}
  \end{center}
\end{figure}

Figure~\ref{fig:circular_grossman} depicts all the primitive variables at $t=0.0002$ s for the circular 2D blast wave. The simulation is performed using 64 elements in each Cartesian directions using $\mathcal{P}=3$. The initial discontinuity has a circular shape, which  makes simulating using a Cartesian grid challenging because the grid will not be aligned with the circular shape of the leading shock wave. However, the H$^3$PC solver robustly computes the discontinuous solution without generating significant spurious oscillations. Moreover, the discontinuous solutions for the gas mixture species are also captured without oscillations even though the values of density for $N$, $O$ and $NO$ are small and close to zero. We have also investigated the effect of the spatial resolution on the quality of the results for non-equilibrium problems with real chemistry in Fig.~\ref{fig:circular_grossman_compare}. Comparing Fig.~\ref{fig:circular_grossman_compare}(a) and (b), we can conclude that by just using higher polynomial orders we can accurately compute the discontinuous solution with a comparable sharpness relative to the high resolution case with $N=64^2$ and $\mathcal{P}=3$. We can also observe in Fig.~\ref{fig:circular_grossman_compare} (b) and (c) that using higher polynomial order reduces the numerical diffusion in capturing the circular leading shock wave in the temperature profile. 

\subsection{Subsonic and Supersonic Taylor-Green Vortex Benchmarks} 
\label{TGV_sub}
The Taylor-Green vortex problem is formulated to assess the capability of H$^3$PC in capturing turbulent scales. In this test case, we solve the three-dimensional compressible Navier-Stokes equations under the assumptions of frozen chemistry and a calorically perfect gas. The computations are carried out within a cubic domain defined as $\Omega = [0, 2\pi]^3$, employing periodic boundary conditions. The physical domain is initialized with laminar velocity, density, and pressure fields as follows:
$$
\begin{aligned}
u & =\sin \left(\frac{x}{L}\right) \cos \left(\frac{y}{L}\right) \cos \left(\frac{z}{L}\right) \\
v & =-\cos \left(\frac{x}{L}\right) \sin \left(\frac{y}{L}\right) \cos \left(\frac{z}{L}\right) \\
w & =0 \\
p & =p_0+\frac{1}{16}\left(\cos \left(\frac{2 x}{L}\right)+\cos \left(\frac{2 y}{L}\right)\right)\left(\cos \left(\frac{2 z}{L}\right)+2\right) \\
\rho &= 1 
\end{aligned}
$$

where $L=2\pi$. For this test case, $M=0.1$ and the Reynolds number is $Re=1600$. The initial laminar structures evolve and interact with each other. The simulations have been performed for three mesh resolutions. In the case of DGSEM, we have used $\mathcal{P}=7$, providing $8^\textrm{th}$ order accuracy. The mesh resolutions are designed to have $11$, $16$, and $22$ elements in each Cartesian direction. A low-storage 9-stage $4^\textrm{th}$-order Runge Kutta time stepping scheme advances the solution in time. This scheme was introduced as  $\textrm{RK}4(3)9_F[3S^*_{+}]$ by Ranocha \emph{et al.} \cite{ranocha2022optimized} and is optimized for compressible flows. The time step size is computed using an error-control-based controller, and the error tolerance is selected at $10^{-8}$. 

In the UCNS3D solver, the semi-discretization of the compressible Navier-Stokes equations is achieved using the WENO method \cite{tsoutsanis2011weno} with a polynomial order of 4, which is necessary for the accurate WENO reconstruction. A WENO scheme generally involves two steps. First, a piecewise linear reconstruction of the variables is performed. The variables are then interpolated to the edges of the control volumes. Monotonicity principles are applied at this stage to eliminate over-/undershoots by using limiters. The variables at the edges of the control volumes are treated as initial data for the Riemann solver. In the second step, the variables on either side of the control volume edges are used to compute the numerical flux. Therefore, to ensure convergence and avoid any over-/undershoots, we employ the Venkat limiter \cite{venkatakrishnan1995convergence} in this case. For the time integration, we use a 4th-order explicit Runge-Kutta scheme with a CFL number of 1.3. To obtain dissipation statistics, the simulation runs until \(t = 20\). The UCNS3D solver was not able to resolve the TGV flow employing $8^\textrm{th}$ order accuracy. Therefore, we have used $4^\textrm{th}$ order accuracy with element numbers of $22$, $32$, and $48$ in each Cartesian direction.

\begin{figure}[t]
\includegraphics[width=\textwidth]{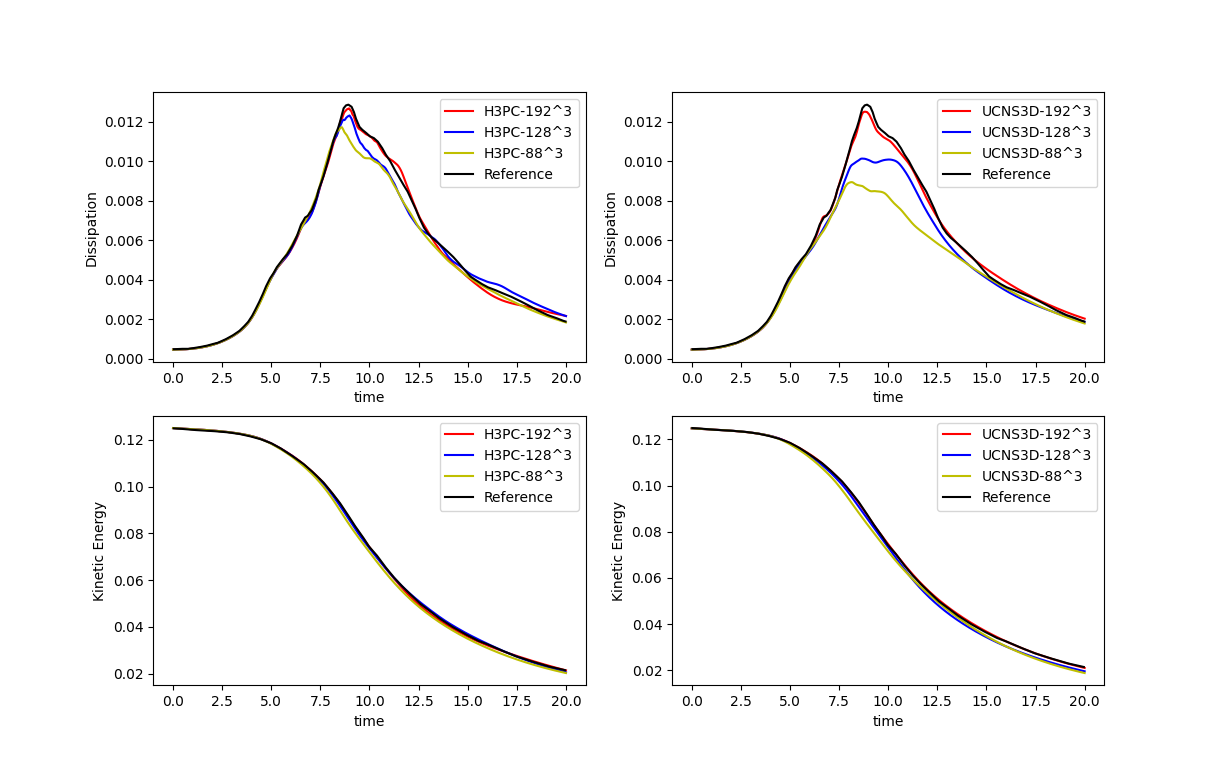}
\caption{Taylor-Green Vortex for $M=0.1$ and $Re=1600$. shown are the spatial dissipation and kinetic energy time evolution. For the $\textrm{H}^3\textrm{PC}$, $\mathcal{P}=7$ and the number of elements for each Cartesian direction is set as 11, 16, and 24. For the UCNS3D solver, we have employed 22, 32, and 48 with accuracy order of 4.}
\label{fig:dissp}
\end{figure}

Figure~\ref{fig:dissp} compares the spatial dissipation and kinetic energy profile versus time obtained by the $\textrm{H}^3\textrm{PC}$ and UCNS3D solvers. The nonlinearity of the NS equations causes the structures to break down into smaller scales and then transition into a fully turbulent flow. As the scales of vortical structures become smaller, the dissipation peaks at $t=8.95$, where we have the largest range of scales in the flow. The total kinetic energy of the flow constantly decreases in time due to the viscosity effects. According to Fig.~\ref{fig:dissp}, both solvers can accurately predict the evolution of the total kinetic energy in time at all mesh resolutions. However, the accuracy of spatial dissipation prediction is drastically different at coarse mesh resolutions. For $88^3$ and $128^3$ resolutions, the $\textrm{H}^3\textrm{PC}$ solver can resolve a larger range of scales compared to the UCNS3D solver due to the $8^\textrm{th}$ accuracy order scheme we have used. Typical spectral element methods at high polynomial orders are unstable and require stabilization techniques such as modal filtering and over-integration to avoid aliasing-triggered solution blow-ups. The TGV results show that the entropy stable DGSEM scheme possesses an anti-aliasing feature that can be employed effectively for the under-resolved simulations of turbulent flows since it can predict the peak dissipation point with $88^3$ and $128^3$ resolutions without the use of stabilization techniques.

\begin{figure}[t!]
  \begin{center}
    \begin{tabular}{ccc}

\includegraphics[width=0.3\textwidth,height=0.3\textwidth]{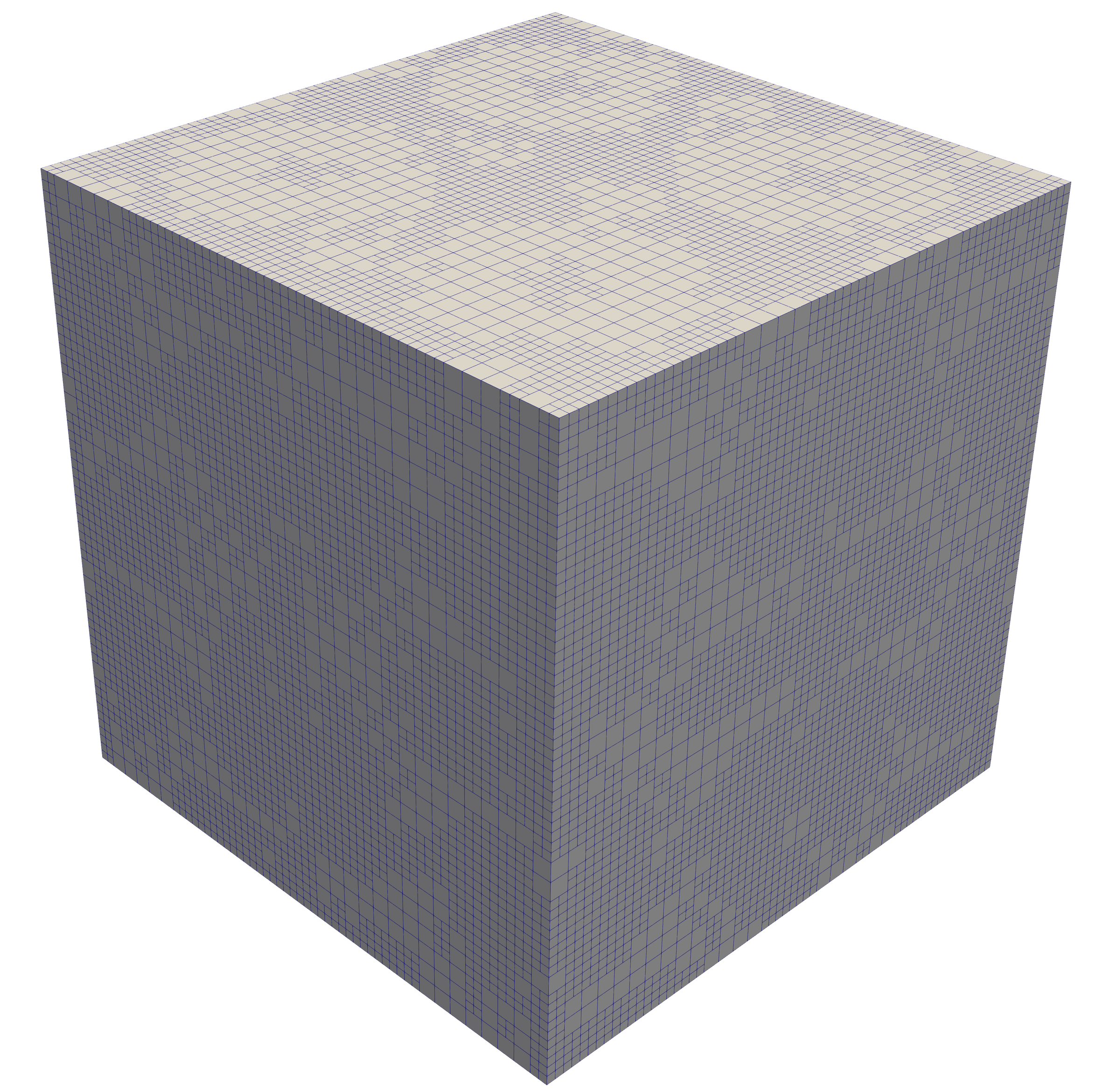}
 &
 \includegraphics[width=0.3\textwidth,height=0.3\textwidth]{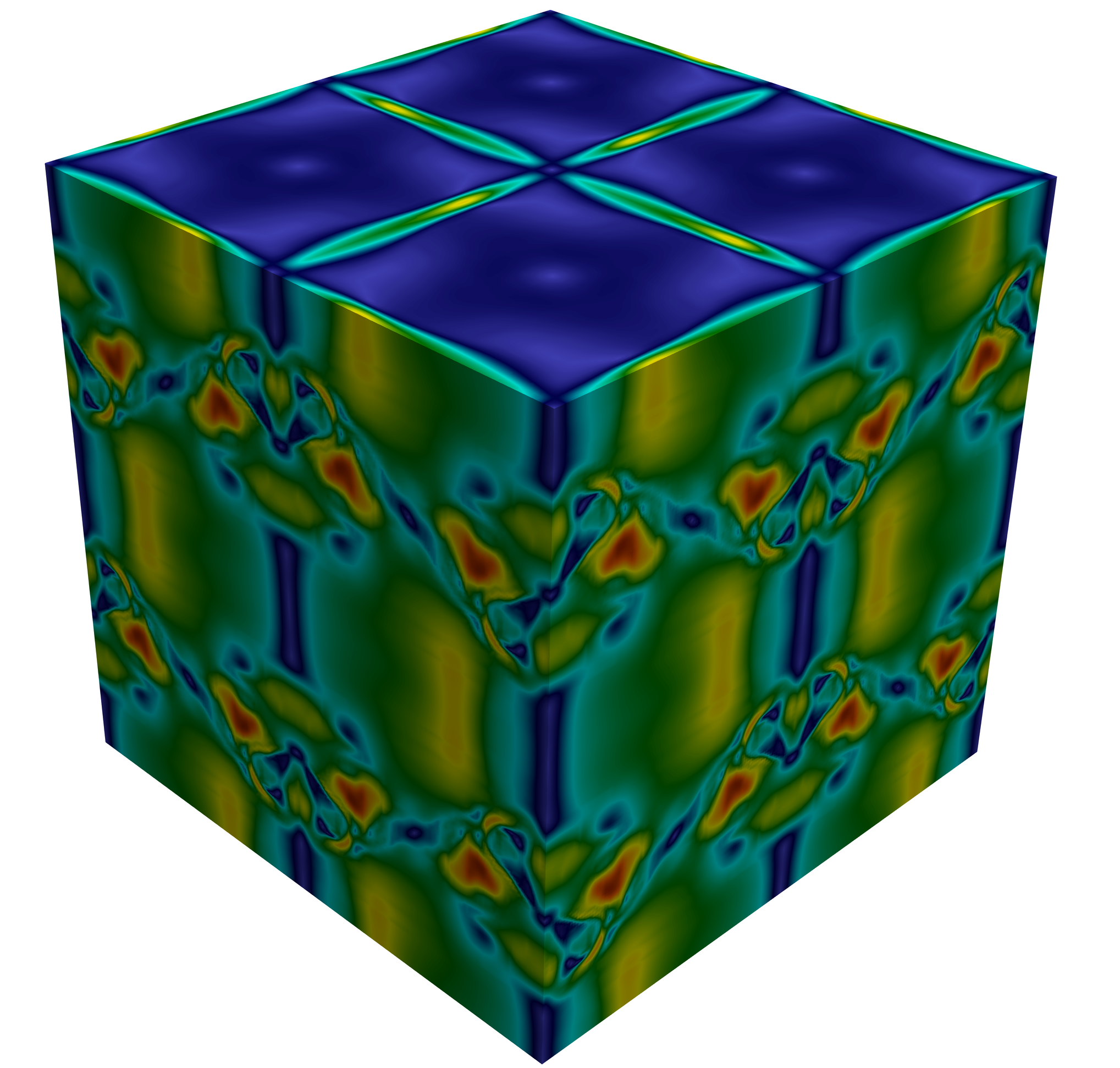}

 &

\includegraphics[width=0.3\textwidth,height=0.3\textwidth]{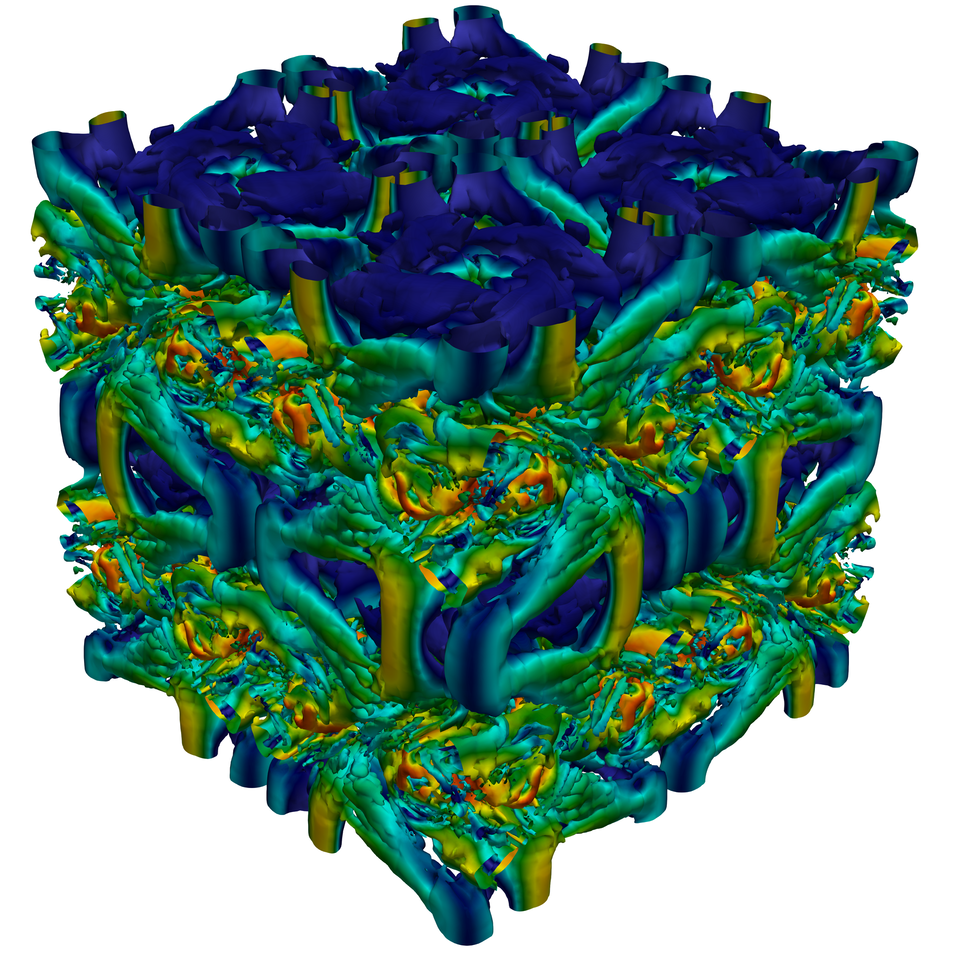}
  \\
  (a) Adaptive Mesh& (b) Velocity magnitude & (c) Iso surfaces 
 
\end{tabular} 

\caption{Taylor-Green Vortex contour plots for $M=0.1$ and $Re=1600$. (a) Adaptive mesh refinement based on velocity magnitude indicator, (b) Velocity magnitude contours for $t=9.13$, (c) Iso-surfaces of Q-criterion=0.01 colored by the velocity magnitude. The color range for the velocity magnitude ranges from dark blue at 2.3e-4 to dark red at $1.2$.}
    
    \label{fig:TGV_qcrit}
  \end{center}
  
\end{figure}

\subsubsection{Adaptive Mesh Refinement}
The $\textrm{H}^3\textrm{PC}$ solver has adapted the P4est mesh refinement library \cite{BursteddeWilcoxGhattas11,IsaacBursteddeWilcoxEtAl15,Burstedde20} to refine mesh resolution according to an indicator function. The indicator functions take in a variable computed by the solution variables, which could be density, pressure, density-pressure, or other derived variables. The indicator function can be set as suggested by Lohner \cite{LOHNER1987323}, which estimates the input variable's weighted second derivative. The other option, the Hennemann-Gassner indicator, is constructed based on a shock-capturing indicator introduced by Hennemann \emph{et al.} \cite{HENNEMANN2021109935}. This indicator calculates a coefficient between zero and one by computing the energy ratio of the highest mode to the summation of all modes' energies calculated using the input variable. We have extended Trixi.jl's AMR capability, which was for Euler equations to Navier-Stokes equations. The AMR is extended for efficient parallel computations in the $\textrm{H}^3\textrm{PC}$ solver. Figure~\ref{fig:TGV_qcrit} depicts the performance of the AMR solver in simulating TGV flow. The Lohner indicator is applied using the velocity magnitude as the AMR input variable. According to Figs.~\ref{fig:TGV_qcrit}(a) and (b), the mesh resolution is accurately adapted to the variation of velocity magnitude executed using 32 processing units. For this case, we have employed five levels of refinement, with the medium and maximum levels set at levels 2 and 4, respectively, using thresholds of 0.1 and 0.2. The AMR is performed by splitting large quadrilateral(2D) and hexahedra (3D) elements into four and eight small elements. The AMR requires the solver to handle non-conforming elements cemented together using mortars. At each mortar, for the patching operation of the advective and viscous fluxes, we use interpolation and projection operations to transfer solution data from small to large elements and from large to small elements, respectively. The projection operation from small to large elements is performed using the exact L2 projection, where we first interpolate to Gauss-Legendre points, then, after the projection, we convert the solution back to Gauss-Lobatto nodes.

\subsubsection{Scaling Study}


\begin{figure}[t!]
  \begin{center}
    \begin{tabular}{ccc}
    \includegraphics[width=0.45\textwidth,height=0.3\textwidth]{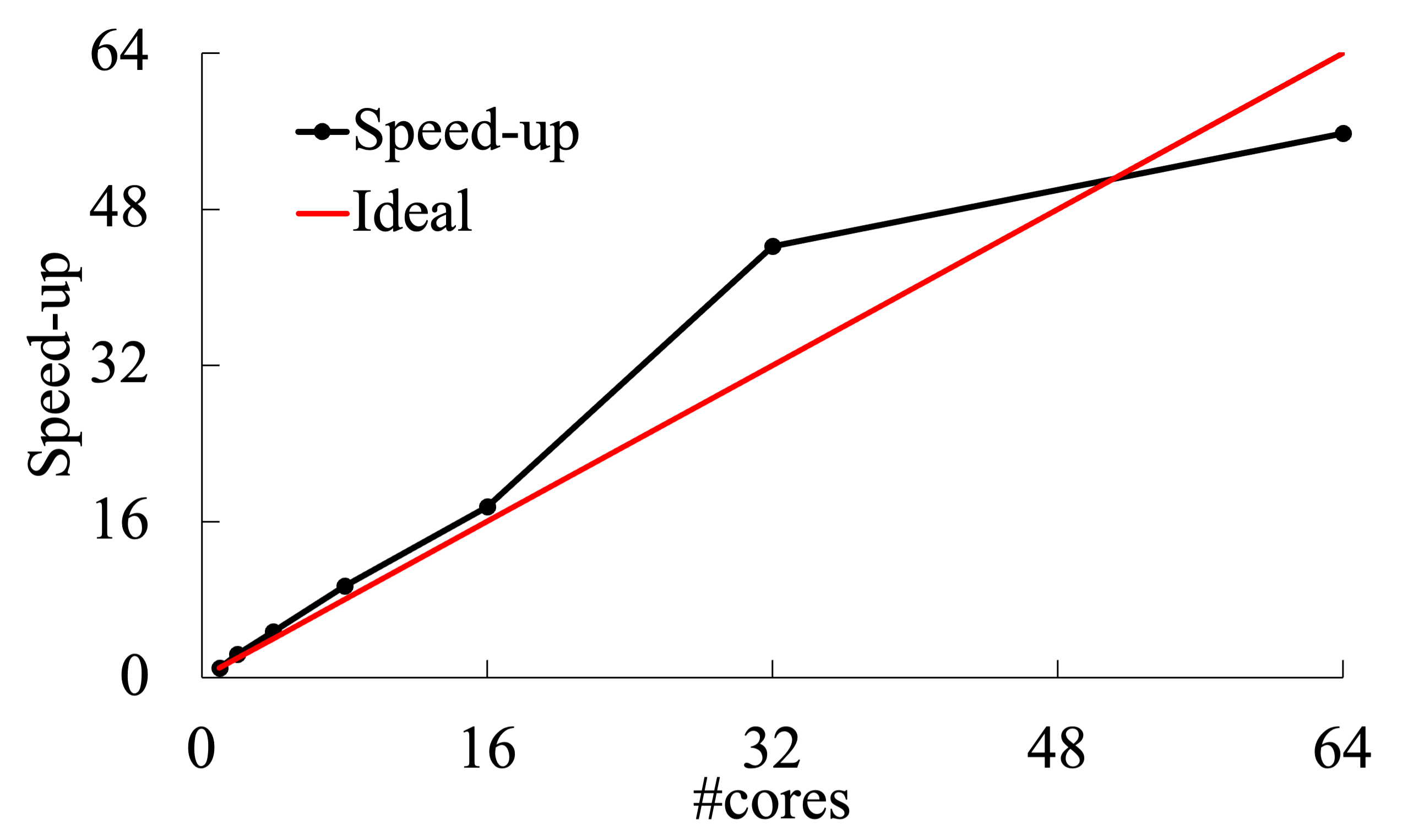}
 &
 \includegraphics[width=0.45\textwidth,height=0.3\textwidth]{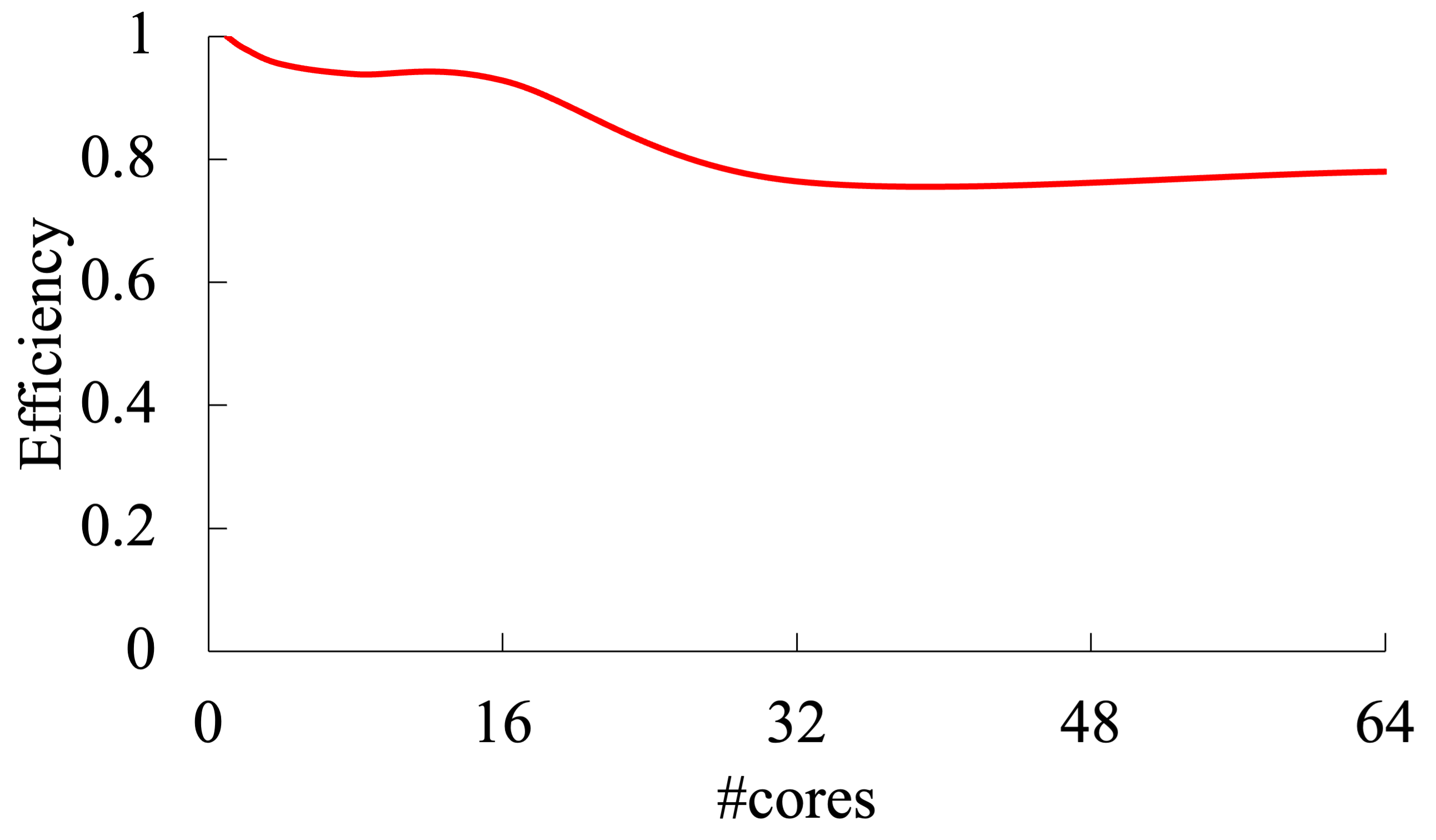}
    
  \\
  (a) Speed-up& (b) Weak parallel efficiency \\

\end{tabular} 
\caption{Scaling studies (a) Strong scaling study, (b) Weak scaling study}
\label{fig:scaling}
\end{center}
\end{figure}

After the extension of parallelism capability to the hyperbolic-parabolic equations such as Navier-Stokes, we want to show that the parallel efficiency of Trixi.jl is carried out through the  $\textrm{H}^3\textrm{PC}$ framework. For this purpose, we designed strong and weak scaling studies for the hyperbolic parabolic solver by employing the viscous TGV case. For the weak scaling, we employed $4^3$ elements per core and a fixed polynomial order $\mathcal{P}=15$. We solve the subsonic TGV problem using the RK45 of Carpenter and Kennedy \cite{CK94a} time stepping scheme for 20 time steps. The weak scaling study was performed for up to 64 MPI ranks. We use $88^3$ degrees of freedom, and we run the solver for 20 time steps using RK45 of Carpenter and Kennedy \cite{CK94a} time stepping scheme. We use a static mesh without any dynamic refinement for this test case. We ran simulations with 1, 2, 4, 8, 16, 32, and 64 cores on the Oscar CCV cluster. Fig.~\ref{fig:scaling} depicts the speed up of the 3D Navier-Stokes solver on a static mesh.

\subsubsection{Supersonic Taylor Green Vortex}
In section \ref{TGV_sub}, we assess the capability of the H$^3$PC solver in under-resolved simulations of TGV in subsonic regimes. In this section, we set up a supersonic TGV problem and investigate the shock capturing schemes in resolving transitional supersonic flows. The simulation setup is identical to the work of Chapellier \emph{et al.} \cite{chapelier2024comparison}. We solve the system of compressible non-reactive Navier-Stokes equations \eqref{system_equation} with calorically perfect gas assumption in a computational domain defined as a cubic box $\Omega=[-\pi L,\pi L]^3$. The initial condition for the supersonic TGV is slightly different than the subsonic TGV such that 
$$
\begin{aligned}
u & =U_o \sin \left(\frac{x}{L}\right) \cos \left(\frac{y}{L}\right) \cos \left(\frac{z}{L}\right) \\
v & =-U_o \cos \left(\frac{x}{L}\right) \sin \left(\frac{y}{L}\right) \cos \left(\frac{z}{L}\right) \\
w & =0 \\
p & =p_o+\frac{\rho_o U_o}{16}\left(\cos \left(\frac{2 x}{L}\right)+\cos \left(\frac{2 y}{L}\right)\right)\left(\cos \left(\frac{2 z}{L}\right)+2\right) \\
T &= T_o 
\end{aligned}
$$
where the density is computed using the ideal gas law along with $p_o$ and $T_o$ values. Here, $T_o$ is the reference temperature of the Sutherland law, $p_o=101325$ Pa, and $U_o=M c_{o}$ where $c_o= \sqrt{\gamma p_o/\rho_o}$. We set $\rho_o=1.225\textrm{ kg}/m^3$, $\gamma=1.4$, Mach number $M=1.25$, and $Re=1600$. Here, the dynamic viscosity is computed using the Sutherland law as 
\begin{equation}
\mu(T)=\frac{1.4042(T/T_o)^{1.5}}{T/T_o+0.4042}*\mu_o
    \label{eq:sutherland}
\end{equation}
where $T_o=p_o/(R\rho_o)$, $R=287.058$ J/(Kg.K) and $\mu_o=\rho_o*L*U_o/Re$. We discretize the cube using 8, 16, 32, and 64 divisions in each Cartesian coordinate and employ $\mathcal{P}=7$ in all cases. We employ $\textrm{RK}4(3)9_F[3S^*_{+}]$ scheme \cite{ranocha2022optimized} with error tolerance of $10^{-8}$ for temporal advancement. The positivity preservation scheme of Zhang and Shu \cite{zhang2011positivity} is activated, and the shock capturing of Hennemann \cite{hennemann2021provably} with shock indicator of pressure is used for stabilization.

In this problem, the quantities of interest are the kinetic energy, solenoidal, and dilatational components of kinetic energy dissipation. The kinetic energy is computed using the following formulation as

\begin{equation}
E = \frac{1}{2\rho_o U_o^2 |\Omega|}\int_\Omega \rho \mathbf{u}.\mathbf{u} d\Omega 
    \label{eq:TGV_KE}
\end{equation}
The solenoidal and dilatational components of the kinetic energy dissipation are defined as
\begin{equation}
\epsilon_s = \frac{L^2}{Re U_o^2 |\Omega|}\int_\Omega \frac{\mu(T)}{\mu_o} \mathbf{\omega}.\mathbf{\omega} d\Omega 
    \label{eq:TGV_SD}
\end{equation}
and 
\begin{equation}
\epsilon_d = \frac{L^2}{3Re U_o^2 |\Omega|}\int_\Omega \frac{\mu(T)}{\mu_o} \left( \nabla . \mathbf{u}\right)^2 d\Omega. 
    \label{eq:TGV_DD}
\end{equation}
In Eq.\eqref{eq:TGV_SD}, $\mathbf{\omega}$ is the vorticity vector.

\begin{figure}[t]
\includegraphics[width=\textwidth]{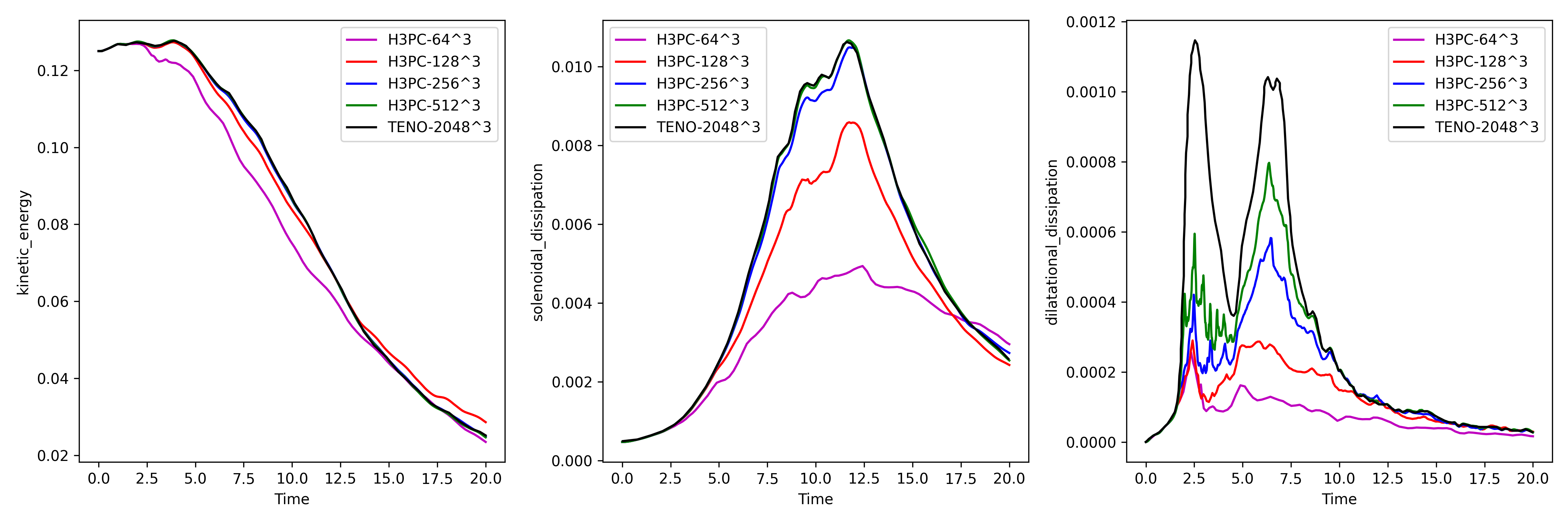}
\caption{Supersonic Taylor-Green Vortex for $M=1.25$ and $Re=1600$. shown are the kinetic energy and solenoidal and dilatational components of kinetic energy dissipation and  time evolution. For the $\textrm{H}^3\textrm{PC}$, $\mathcal{P}=7$ and the number of elements for each Cartesian direction is set as 8, 16, 32, and 64. The TENO results are from \cite{chapelier2024comparison}. }
\label{fig:super_TGV_dissp}
\end{figure}

Figure~\ref{fig:super_TGV_dissp} shows the H$^3$PC solver's performance. It captures large scales (kinetic energy), small scales (solenoidal dissipation), and compressibility effects (dilatational dissipation) from left to right. In the kinetic energy plot, the resolutions $256^3$ and $512^3$ accurately predict the kinetic energy evolution. For solenoidal dissipation, only the $512^3$ resolution matches the reference profile. For dilatational dissipation, higher resolution brings predictions closer to the reference profile, which is computed with $2024^3$ resolution. 

\subsection{Supersonic flow over cylinders}
In this section, we investigate the ability of the H$^3$PC solver in resolving transient supersonic flow past cylinders. We first validate the solver by computing Strouhal number of oscillations in the wake of a 2D circular cylinder in supersonic flow and a moderate Reynolds number. We then simulate the possibility of vortex street behind a square cylinder in a supersonic flow and at various Reynolds numbers.

\subsubsection{Validation of H$^3$PC solver for supersonic flow around circular cylinder}
\label{sec:circ_cyl}
\begin{figure}[t!]
  \centering
\includegraphics[width=0.8\textwidth]{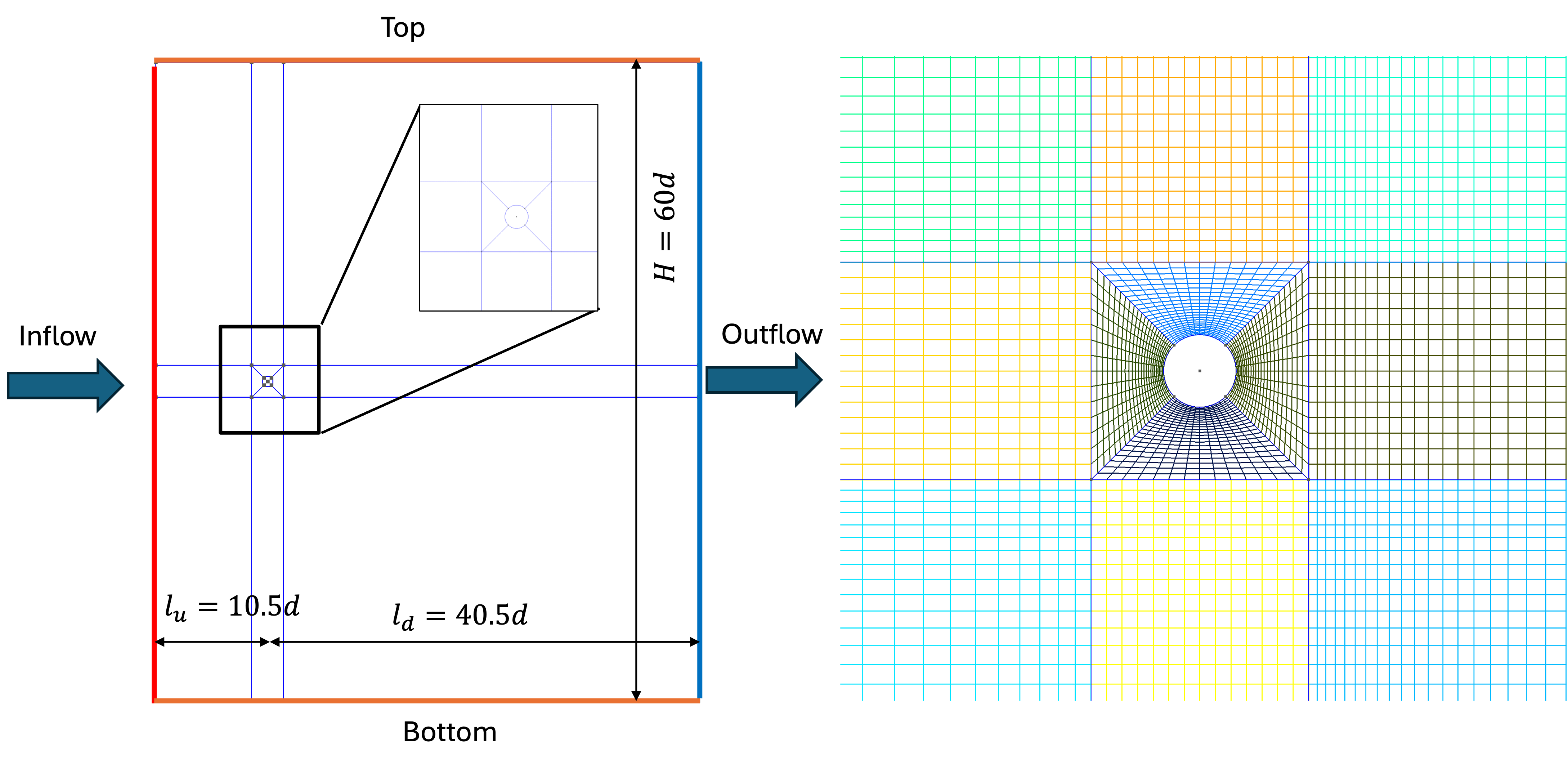} 
\caption{Domain and the grid used for circular cylinder in supersonic flows at $M=4.0$ and Re=$2\times10^4$.}
\label{fig:circ_cyl_domain}  
\end{figure}

To verify the H$^3$PC solver results, we performed a 2D simulation of supersonic flow at Mach=4 around a circular cylinder based on the setup mentioned in \cite{schmidt2015oscillations}. The Reynolds number considered is Re$=2\times10^4$. According to \cite{schmidt2015oscillations}, the Strouhal number computed using the Fourier analysis of oscillations in the wake of a circular cylinder is $0.3\pm0.025$ based on the experimental results. Therefore, we created a computation domain around a cylinder (see Fig.~\ref{fig:circ_cyl_domain}) and used 11,452 quadrilateral elements to discretize the domain in a structured formation. The polynomial order is $\mathcal{P}=5$. We impose supersonic inflow at the red boundary and supersonic outflow at the blue boundary, while the top and bottom boundaries are defined as an inflow boundary condition. The domain height is selected such that the oblique shock lines emanating from the cylinder do not intersect with the top and bottom borders. The free-stream values are taken to be
\begin{equation}
\rho=1, \quad u_1=1, \quad u_2=0, \quad p=\frac{1}{\mathrm{M}^2 \gamma}. 
    \label{init_cond_circ}
\end{equation}
The initial condition and the exterior states on the left, top, and bottom boundaries are set using free-stream values. The center of the coordinate system is at the center of the circle. We place two probe points at $p_1=(2d,0)$ and $p_2=(4d,0)$ to record all the primitive variables time history. The simulation is run for 10 flow-through times (FT), and we record the time history for the entire duration of the simulation.

\begin{figure}[h!]
\begin{center}
   \begin{tabular}{cc}
\includegraphics[width=0.45\textwidth]{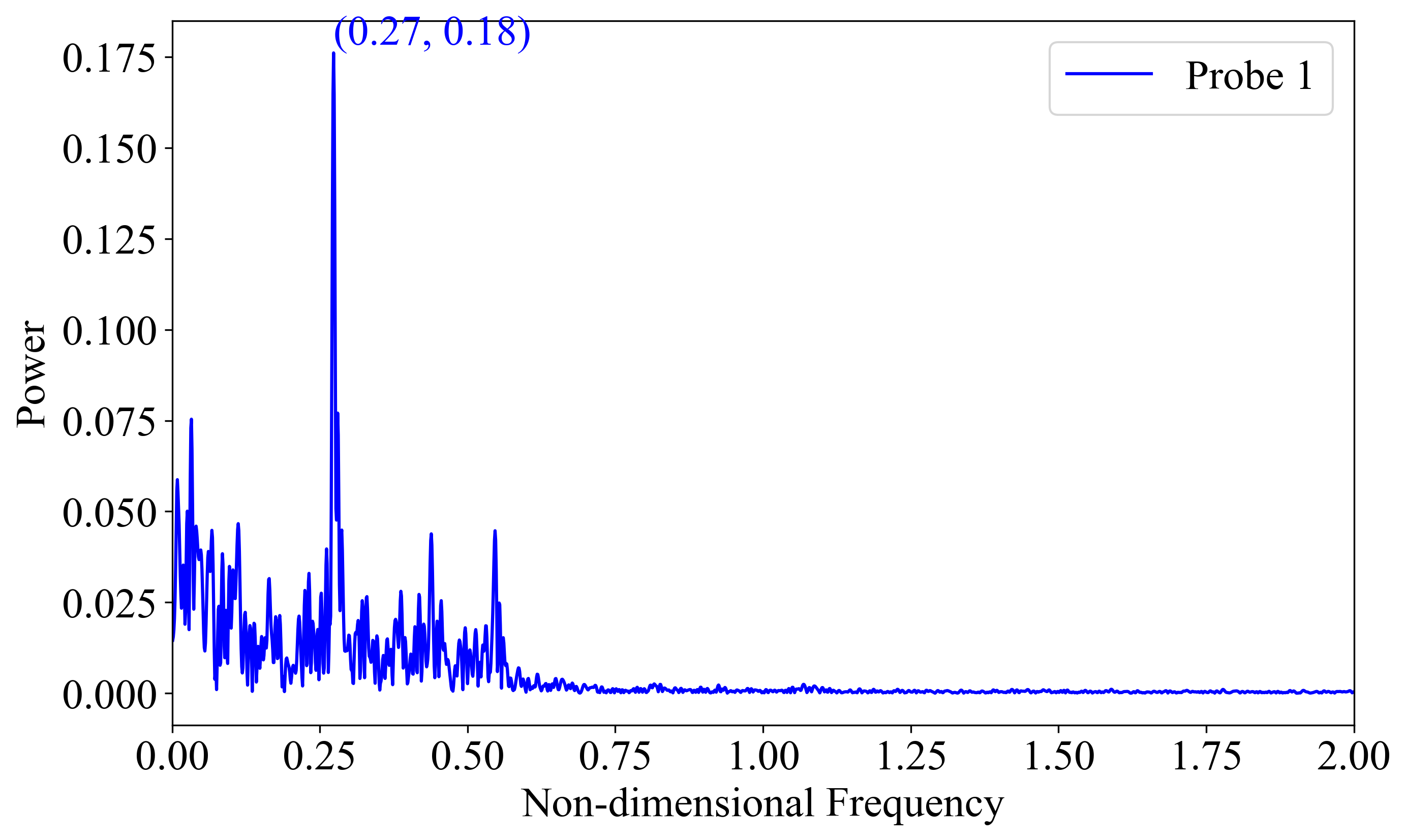} 
 &
  \includegraphics[width=0.45\textwidth]{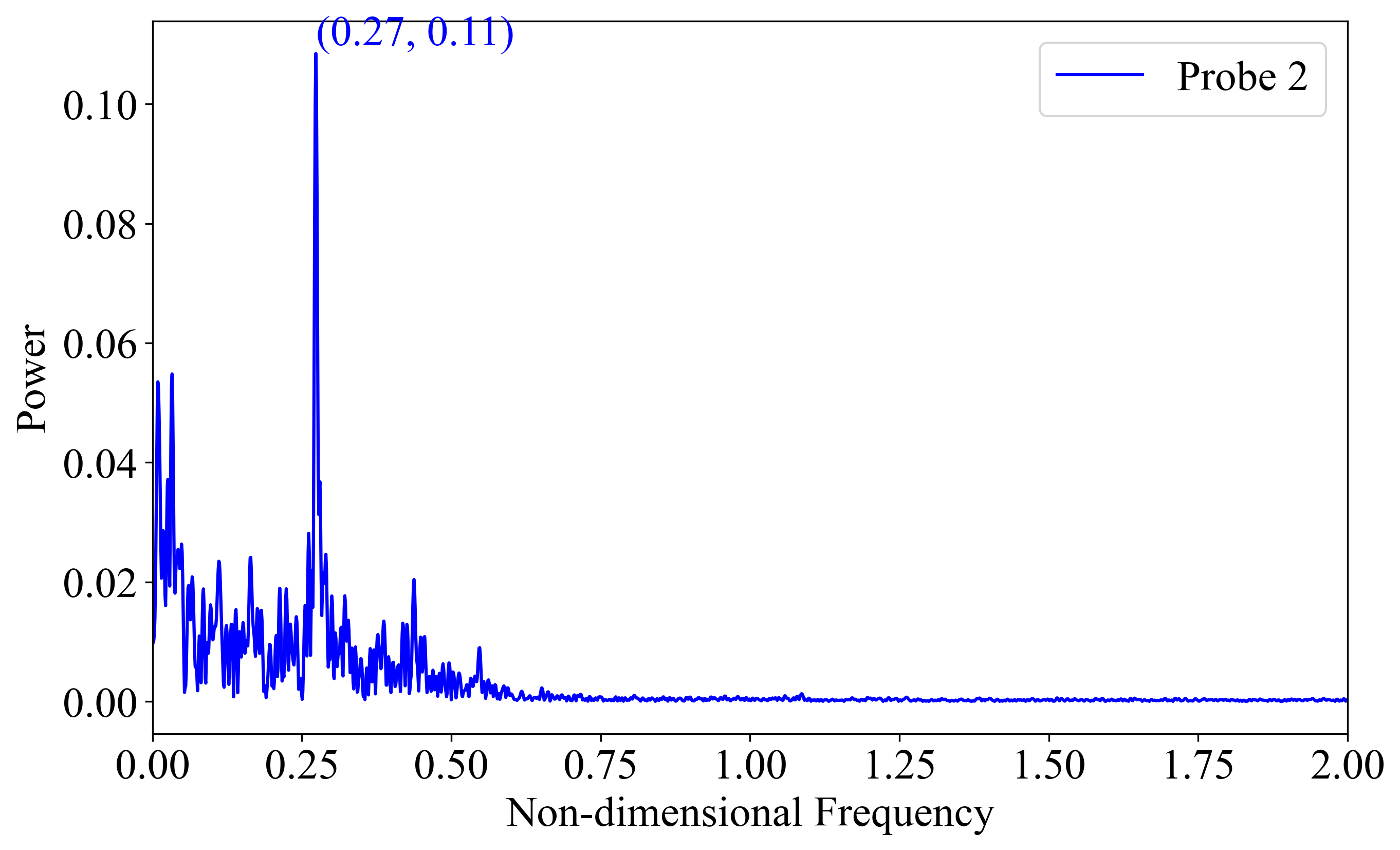} 
  \\
  (a) $p_1=(2d,0)$ &(b) $p_2=(4d,0)$
\end{tabular} 
\end{center}
\caption{Power spectrum for density history collected for probe points located at (a) $p_1=(2d,0)$, and (b)$p_2=(4d,0)$, $f_{nondim}=\frac{f_{dim}\times d}{M}$, where $M$ denotes the Mach number.}
\label{fig:freq_circ_cyl}  
\end{figure}

\begin{figure}[t!]
  \centering
\includegraphics[width=0.9\textwidth]{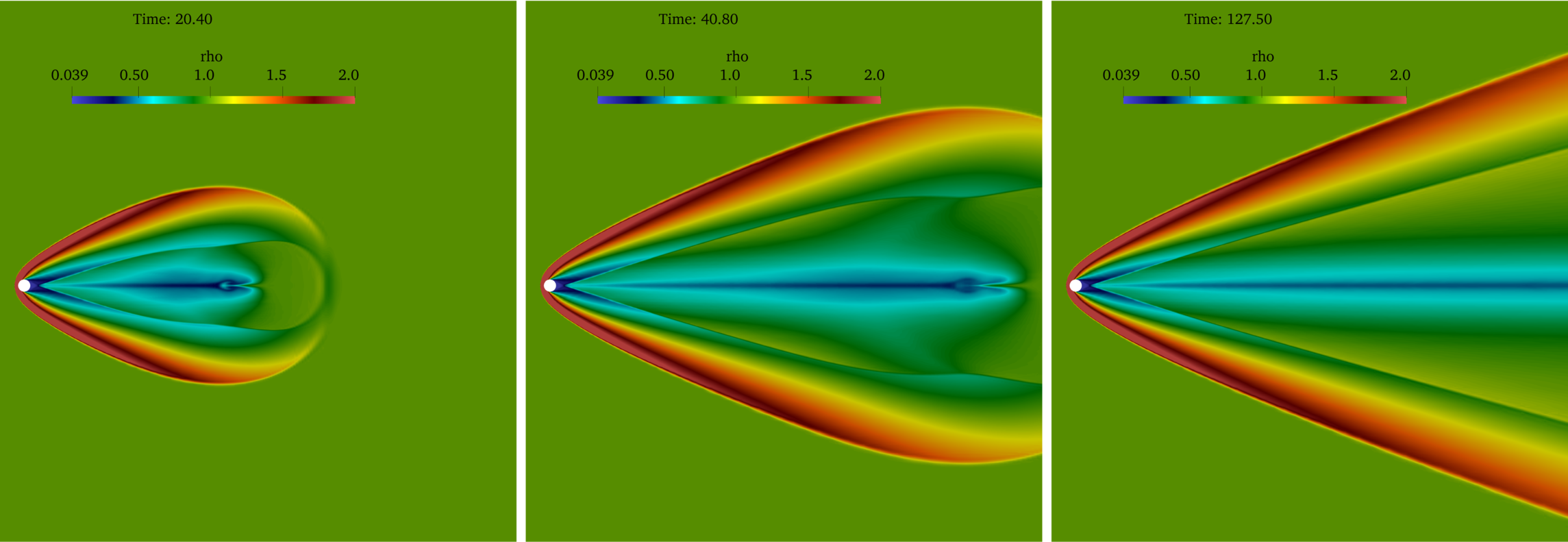} 
\caption{Density contours at T=20.4, 40.8, and 127.5 for circular cylinder in supersonic flows at $M=4.0$ and Re=$2\times10^4$. The solution is obtained using $\mathcal{P}=5$ using 11,452 quadrilateral elements.}
\label{fig:circ_cyl_field}  
\end{figure}

The time series collected at the probe points during the simulation was then analyzed to compute the Strouhal number. We used the Lomb-Scargle periodogram method \cite{lomb1976least, scargle1982studies} to compute the frequency amplitude spectrum for nonuniformly sampled density. This analysis was performed over the time interval $t\in[6.039FT, 10FT]$, where $FT$ indicates flow-through time. According to Figs.~\ref{fig:freq_circ_cyl} (a) and (b), the Strouhal number is computed to be $St=0.27$. This matches the Strouhal number reported in \cite{schmidt2015oscillations} for Reynolds number $2\times10^4$. Figure~\ref{fig:circ_cyl_field} shows three snapshots of density field at different times. According to density contours, it is obvious that there is no formation of vortex street at this Mach and Reynolds numbers; therefore, we can conclude that the very weak oscillations are due to acoustic instabilities.

\subsubsection{Supersonic flow around square cylinder}
\label{sec:cylinder}
\begin{figure}[t!]
  \centering
\includegraphics[width=0.7\textwidth]{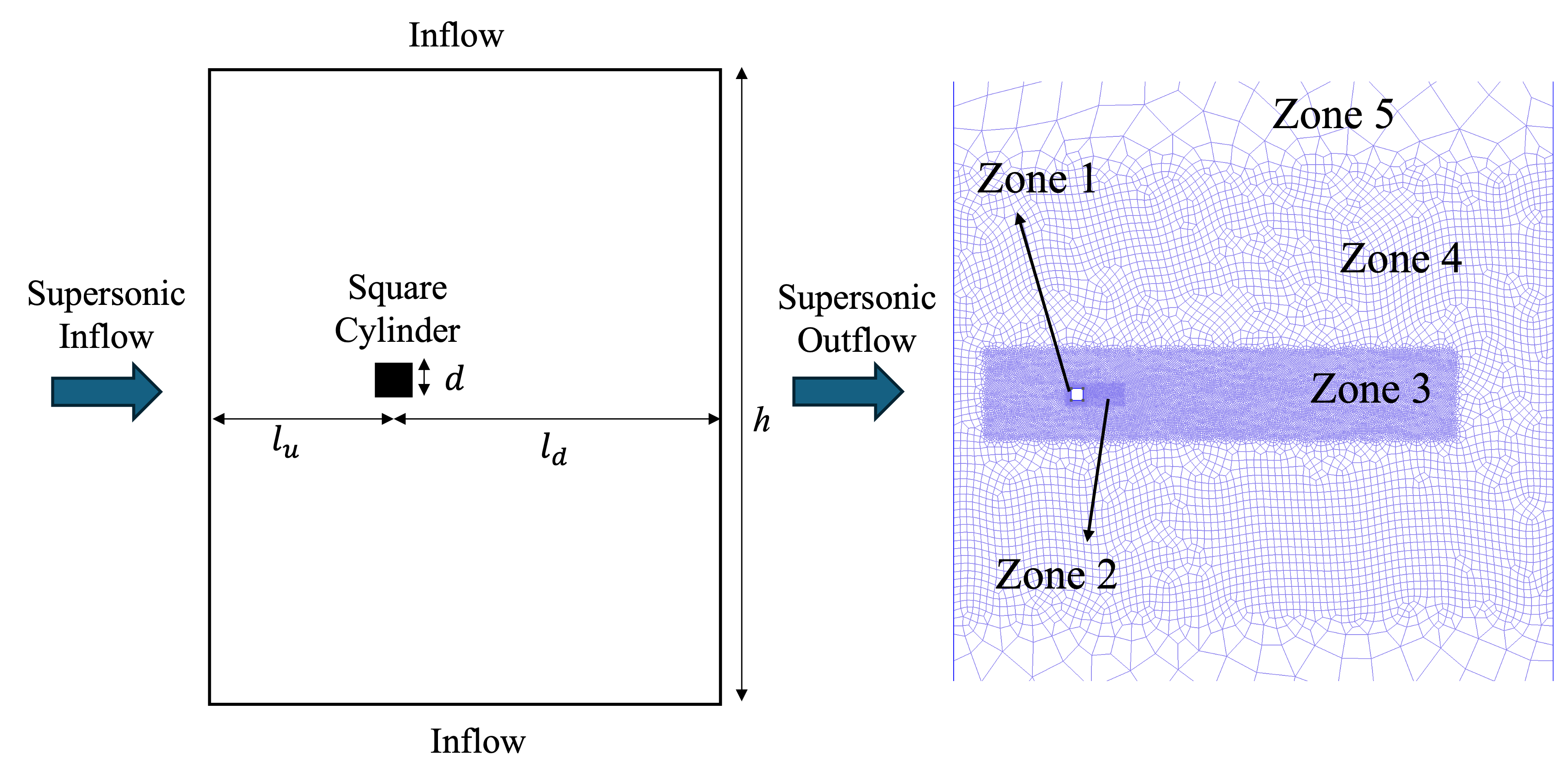} 
\caption{Left: Physical domain and boundary setup for the square cylinder problem. $h$ denotes the width of the domain, $l_u$ is the length of the domain upstream of the cylinder, and $l_d$ denotes the length of the domain downstream of the cylinder. Right: mesh for the small physical domain. Grid is created based on various resolutions for five regions including the vicinity of the cylinder, wake, vortex street, oblique waves, and far field regions.}
\label{fig:square_cyl}  
\end{figure}

We now  solve non-reactive Navier-Stokes equations for a supersonic regime over a square cylinder at $\operatorname{Re}=10^4$ and $\mathrm{M}=1.5$. Our aim is to investigate the frequencies in the wake of a blunt body in supersonic flows; this has been studied experimentally in  \cite{schmidt2015oscillations}. To ensure the reliability of our results for the square cylinder case, we first validated the H$^3$PC solver’s accuracy in predicting the Strouhal number of flow oscillations by applying it to a benchmark problem involving the wake of a circular cylinder subjected to a Mach 4 supersonic flow (see section~\ref{sec:circ_cyl}). The physical domain consists of a rectangular outer boundary that encompasses a unit-length square cylinder, as shown in Fig.~\ref{fig:square_cyl}. The square cylinder is centered vertically within the rectangle. The side length of the square cylinder is $d=1$. The domain extends $l_u=10.5d$ upstream and $l_d=40.5d$ downstream from the cylinder. The coordinate system origin is at the center of the square cylinder. For the simulations, we have chosen two values for the domain height, $h=80d$ and $140d$, to investigate how varying the domain height affects the frequency of vortex shedding events.

The physical domain was meshed using the GMSH software. According to Fig.~\ref{fig:square_cyl}, the grid elements are all quadrilateral. According to Table~\ref{tbl:mesh_res}, we use five element sizing for five zones: cylinder vicinity (zone 1), wake region (zone 2), vortex street region (zone 3), oblique waves region (zone 4), and far-field region (zone 5). The dimensions of zones 1 and 3 to 5 are the same for the two resolution grids. We generate two mesh resolutions which we named mesh (A) for medium resolution and mesh (B) for high resolution grid. For a high-resolution grid, zone 2 is elongated to x=6 to contain the wake region behind the cylinder.  We have used a coarser mesh close to the outer boundaries to dampen reflection artifacts from the boundaries. In general, two mesh resolutions are used, and the element sizes are shown in Table~\ref{tbl:mesh_res}. We investigated the effect of polynomial order on the accuracy of predicting the vortex shedding event. 


\begin{table}[!t]
\caption{Grid size for various zones of the physical domain for supersonic square cylinder.}  
\centering
\begin{tabular}{lcccccc} 
\hline
 & \multicolumn{3}{c}{(A) Medium resolution} & \multicolumn{3}{c}{(B) High resolution}\\
\hline
Zones& x$_{\textrm{range}}$& y$_{\textrm{range}}$& length &x$_{\textrm{range}}$& y$_{\textrm{range}}$& length\\
\hline
Zone 1& $[-0.75d, 0.75d]$& $[-0.75d, 0.75d]$&$d/30$&$[-0.75d, 0.75d]$&$[-0.75d, 0.75d]$&$d/60$\\
Zone 2& $[-d, 4d]$& $[-d, d]$&$2d/25$&$[-1d, 6d]$&$[-d, d]$&$d/25$\\
Zone 3& $[-8d, 32.4d]$& $[-4d, 4d]$&$2d/15$&$[-8d, 32.4d]$& $[-4d, 4d]$&$2d/15$\\
Zone 4&$[-10.5d, 40.5d]$& $[-20d, 20d]$&$d/3$&$[-10.5d, 40.5d]$& $[-20d, 20d]$&$d/3$\\
Zone 5&$[-10.5d, 40.5d]$& $[-70d, 70d]$&$2d$&$[-10.5d, 40.5d]$& $[-70d, 70d]$&$2d$\\
\hline
Elem NO& \multicolumn{3}{c}{27,417}& \multicolumn{3}{c}{38,163}\\
\hline
\end{tabular}
\label{tbl:mesh_res}
\end{table}

For the $\textrm{H}^3\textrm{PC}$ solver, all the boundary conditions are enforced weakly through the advective or viscous fluxes. For the supersonic inflow, an external state of conservative variables is set as the initial conditions. We then assign the advective inflow fluxes as the advective fluxes computed using the external state of the solution. For the viscous fluxes, we use the average of the external state for the internal state of the solution. The supersonic outflow boundary is the same as the supersonic inflow boundary, but we use the internal state of the solution to compute the common advective fluxes at that boundary. No-slip and adiabatic wall boundary conditions are imposed on the cylinder wall. We also utilize the free stream values to define the external state of the solution for the inflow boundaries located at the top and bottom of the domain. We then use the Riemann solver to compute the common flux at the boundaries. The free-stream values are taken to be
\begin{equation}
\rho=1, \quad u_1=1, \quad u_2=0, \quad p=\frac{1}{\mathrm{M}^2 \gamma}.
    \label{init_cond}
\end{equation}
The initial condition and the exterior states on the left, top, and bottom boundaries are set using free-stream values. The physical domain is initialized using the free-stream values defined in Eq.~\eqref{init_cond}. The simulation is performed until a final time of $T=500$. We then record the time history of solution variables over four probe points behind the wake to compare the frequency of vortex shedding within a time interval of $[300, 500]$. The time-stepping scheme for the $\textrm{H}^3\textrm{PC}$ solver is the four-stage-third-order strong stability preserving Runge-Kutta (SSPRK43)~\cite{kraaijevanger1991contractivity,conde2018embedded,ranocha2022optimized} with an adaptive time step size.

\begin{figure}[t!]
  \centering
\includegraphics[width=0.9\textwidth]{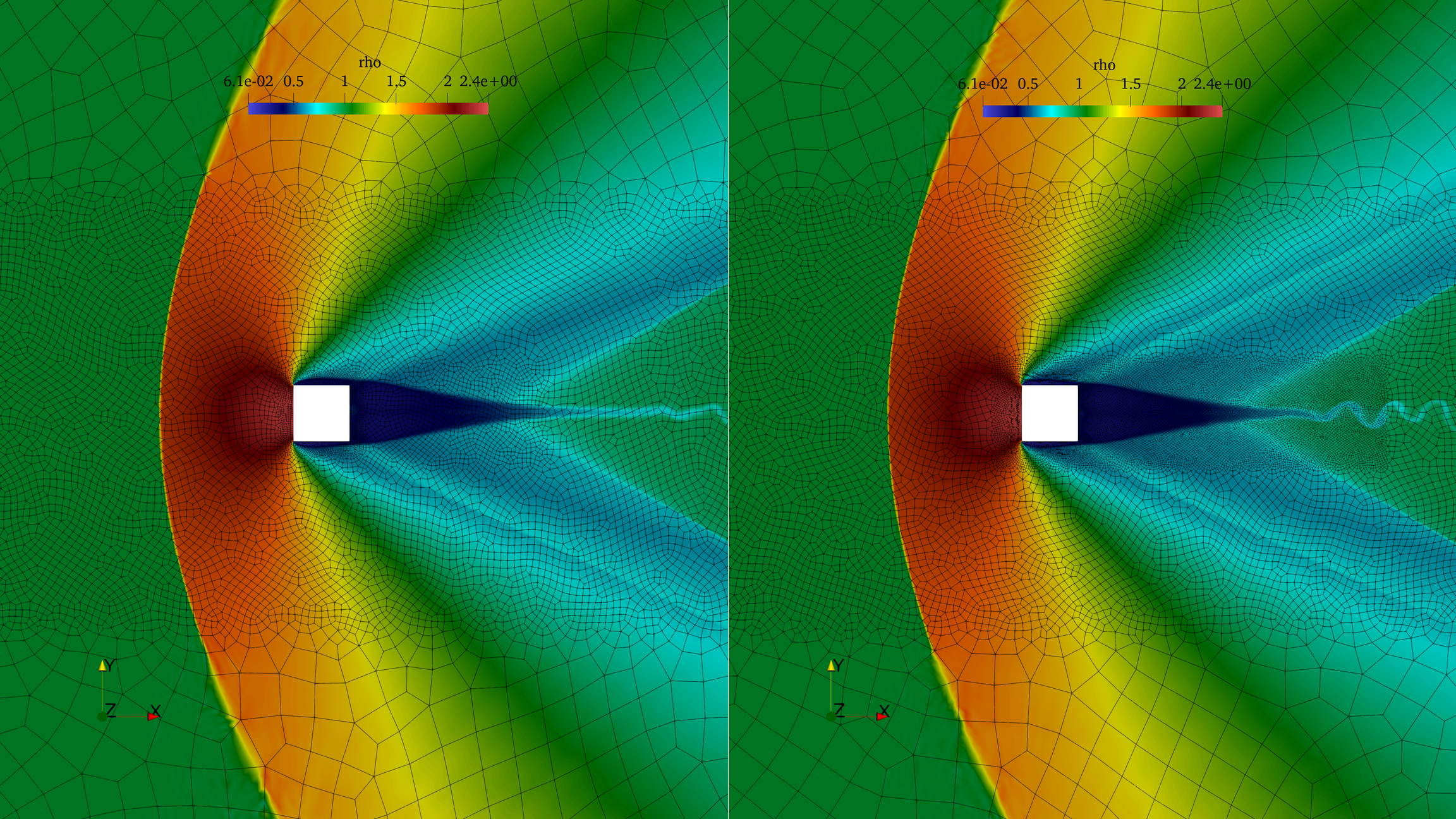} 
\caption{Density contours at $T=500$ and mesh elements outlines for Left: Medium Resolution Mesh (A) $P=5$, Right: High Resolution Mesh (B2) $P=5$. The wake region for high resolution mesh is discretized with higher number of elements and the high resolution mesh is extended to encompass the entire wake region. ($Re=10^4; Ma=1.5$)}
\label{fig:Mesh_res}  
\end{figure}

\begin{figure}[t!]
  \begin{center}
    \begin{tabular}{c}
    \includegraphics[width=0.88\textwidth,trim={0cm 1.5cm 0cm 0cm},clip]{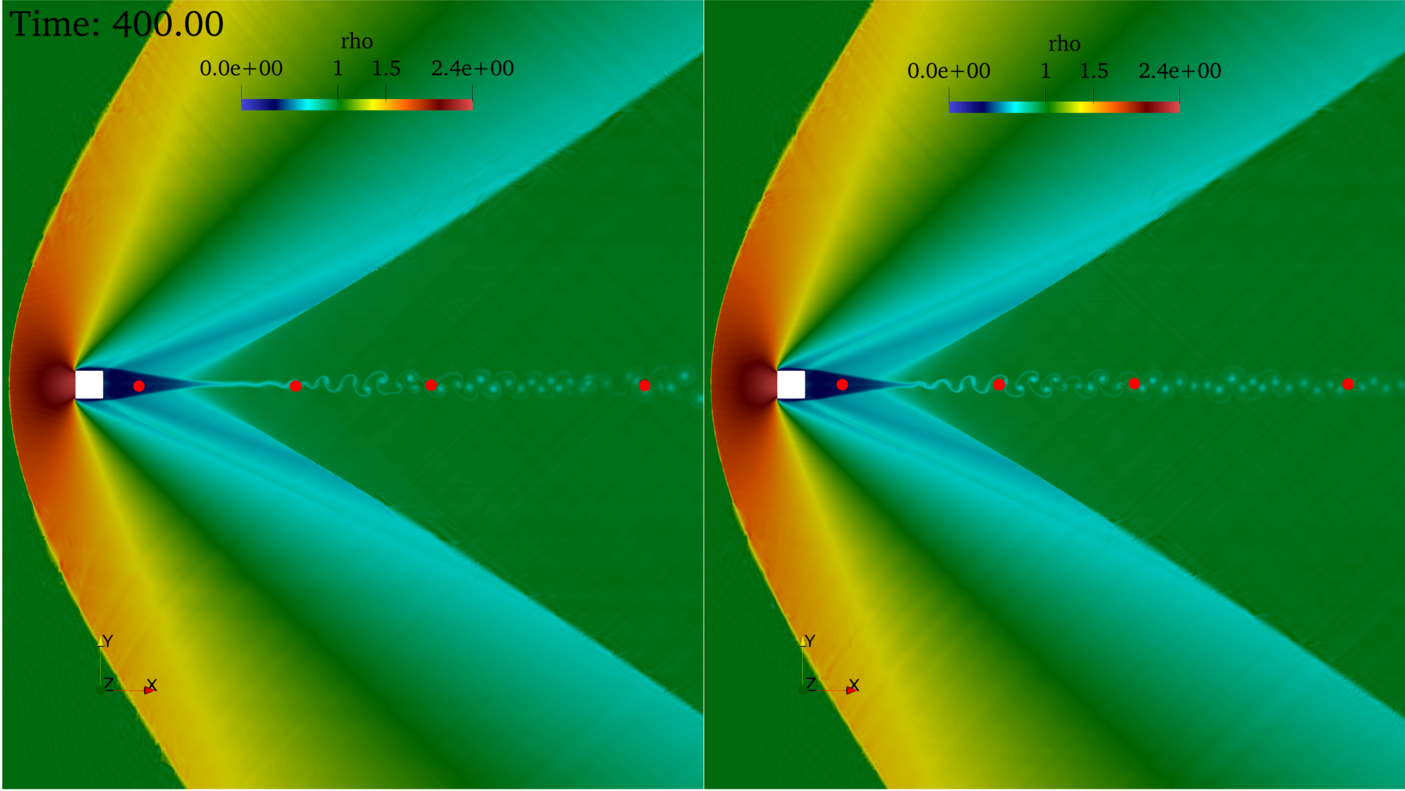}
  \\
  \includegraphics[width=0.88\textwidth,trim={0cm 1.5cm 0cm 0cm},clip]{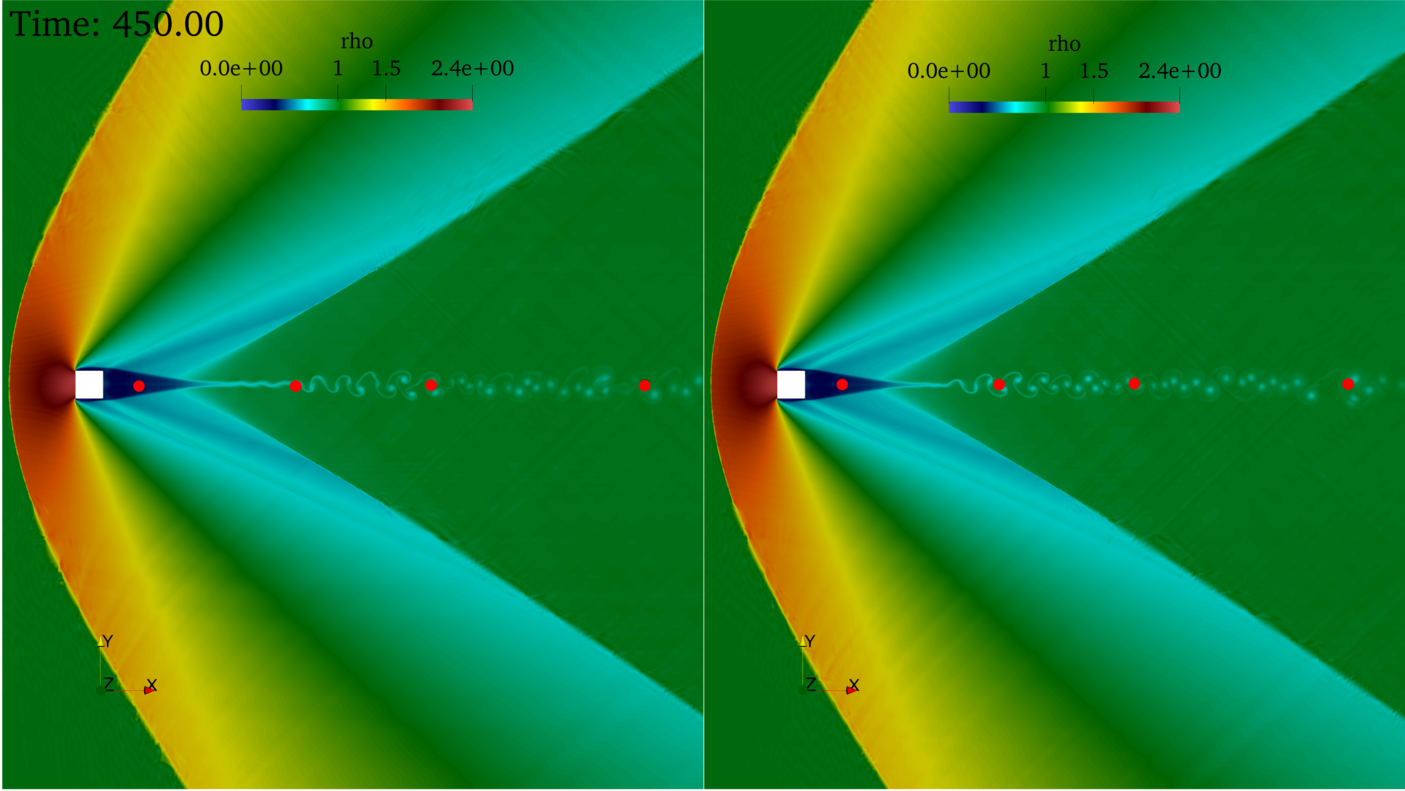}

\\
 \includegraphics[width=0.88\textwidth,trim={0cm 1.5cm 0cm 0cm},clip]{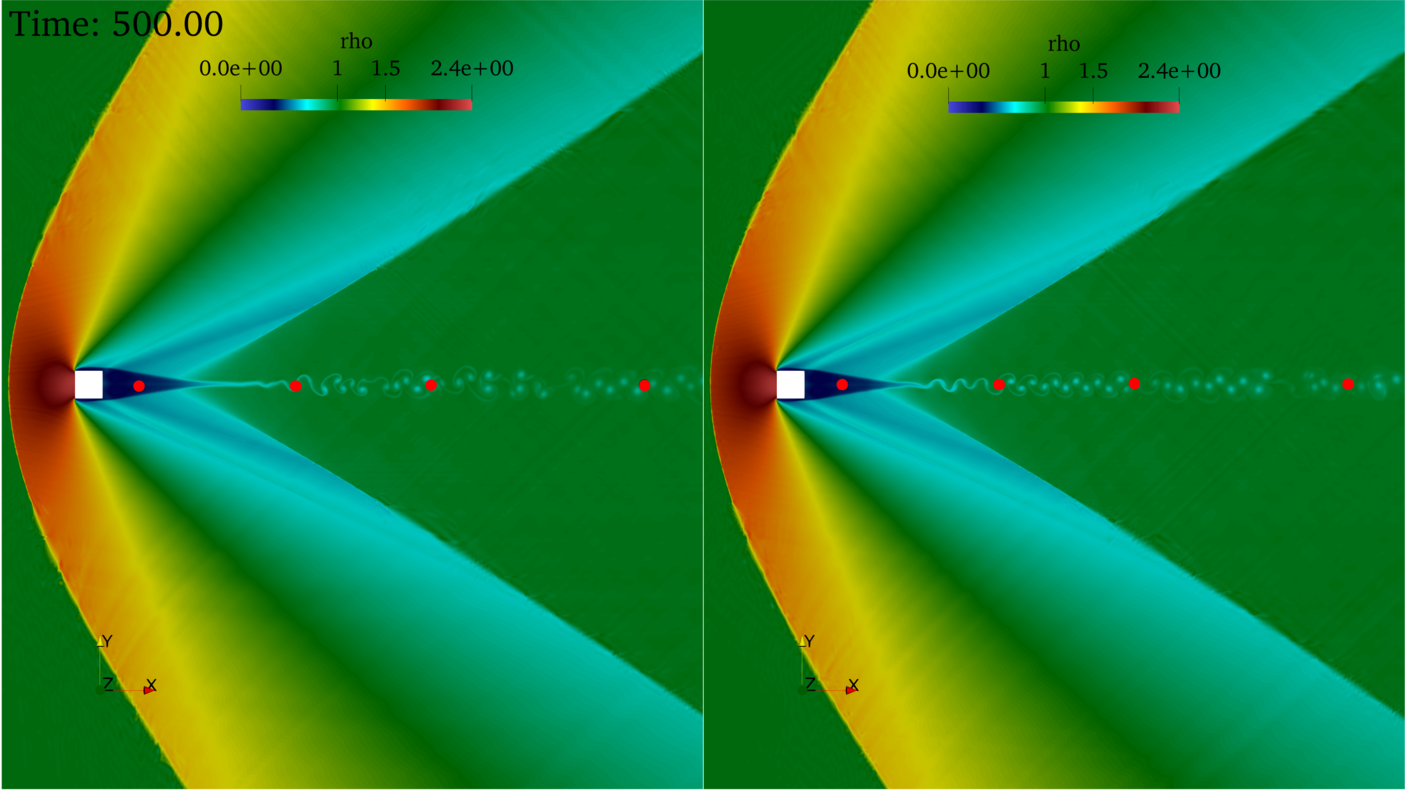}
\\
\end{tabular} 
\caption{Density contours at T=400, 450, 500 obtained using medium mesh resolution (A) on the left and fine mesh resolution (B2) on the right. The probe points are shown using red solid circles. From left to right the red circles indicate probes from P1 to P4. ($Re=10^4; Ma=1.5$)}
\label{fig:time_solution}
\end{center}
\end{figure}

Figure~\ref{fig:Mesh_res} displays density contour plots at $T=500$ for two mesh resolutions according to Table~\ref{tbl:mesh_res}. According to the figure, the wake region in Mesh (B) has higher resolution compare to mesh (A) and it encompasses the entire wake region and therefore the vortex structures generated after the wake are different in mesh (A) and (B). The structure of the bow shock and its trailing edges are the same across various mesh resolutions. According to the boundary condition arrangement, the corner points, which are the intersection of the top and bottom boundaries with the supersonic outflow boundary, can generate solution reflection because of the inconsistencies of the boundary condition setup and the flow regimes. Since the top and bottom boundaries are defined as inflows, it is inconsistent with when the cone-shaped subsonic region reaches the top and bottom boundaries, hence the boundaries reflect numerical artifacts into the domain. Therefore, we selected a large height for the domain to prevent the subsonic region from intersecting with the top and bottom boundaries.

\begin{figure}[t!]
  \begin{center}
    \begin{tabular}{cc}
    \includegraphics[width=0.45\textwidth]{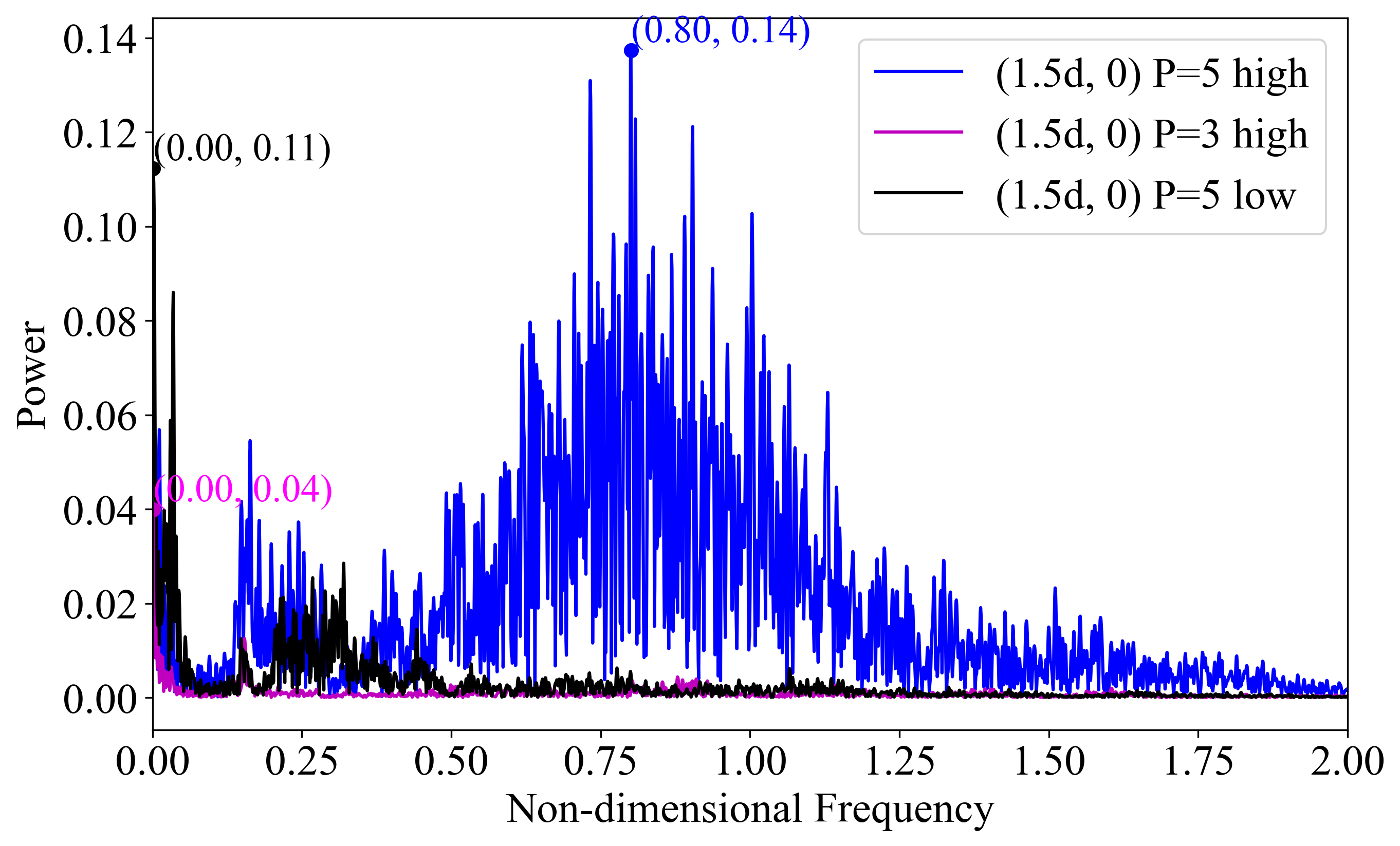}
 &
 \includegraphics[width=0.45\textwidth]{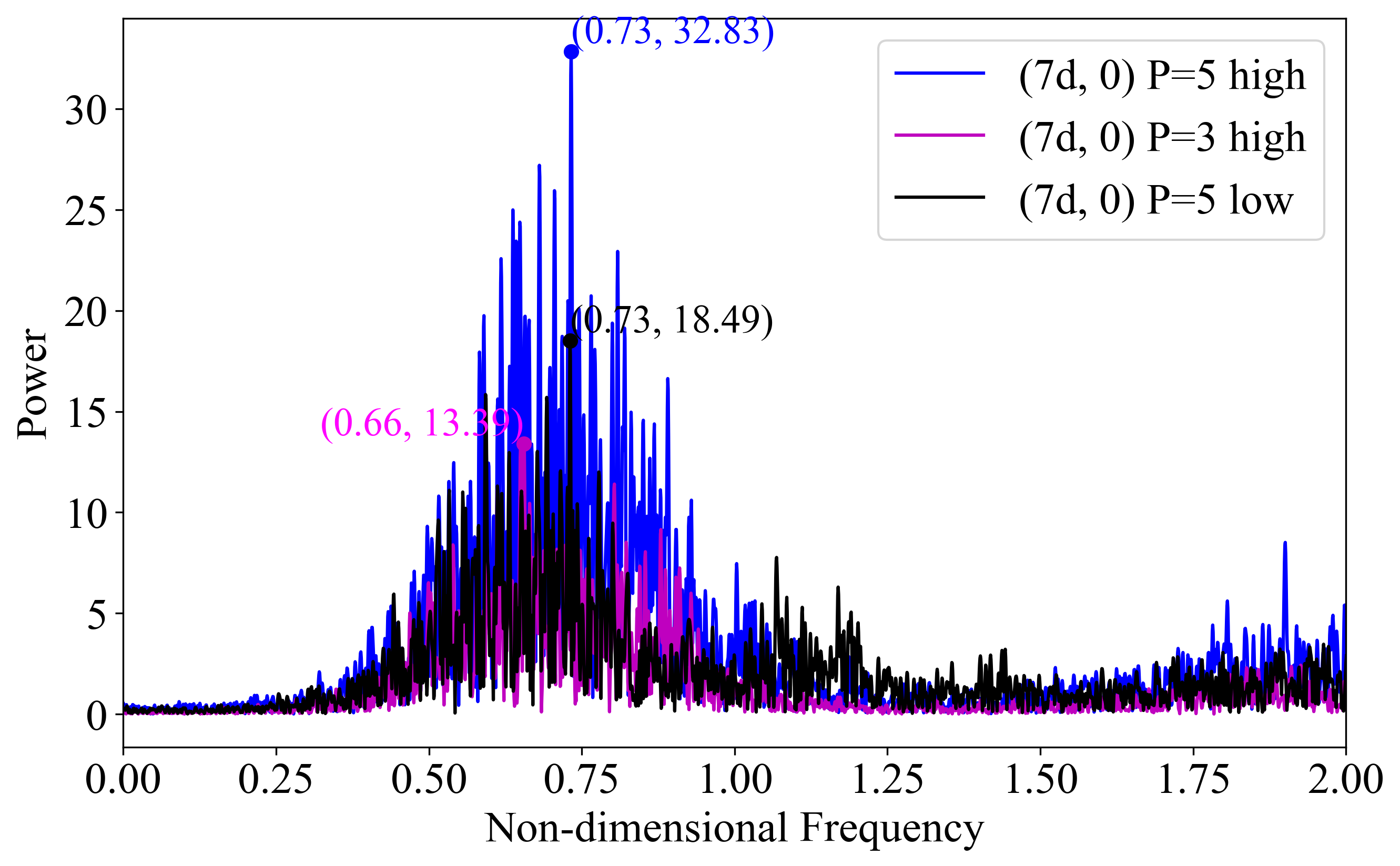}
 \\
    
  (a) Probe at $(1.5d, 0.0)$& (b) Probe at $(7d, 0.0)$ 
\\

 \includegraphics[width=0.45\textwidth]{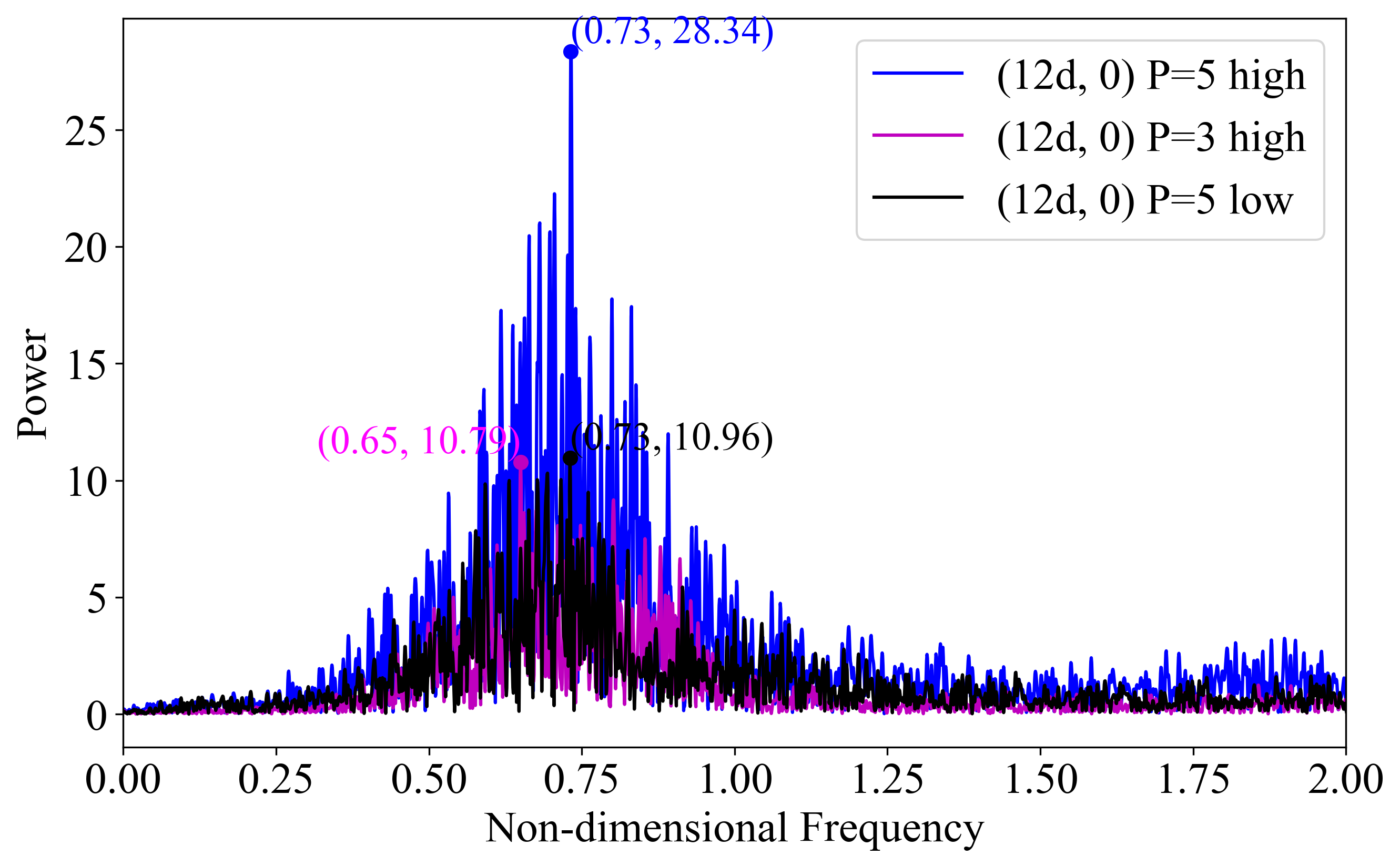}

 &
 \includegraphics[width=0.45\textwidth]{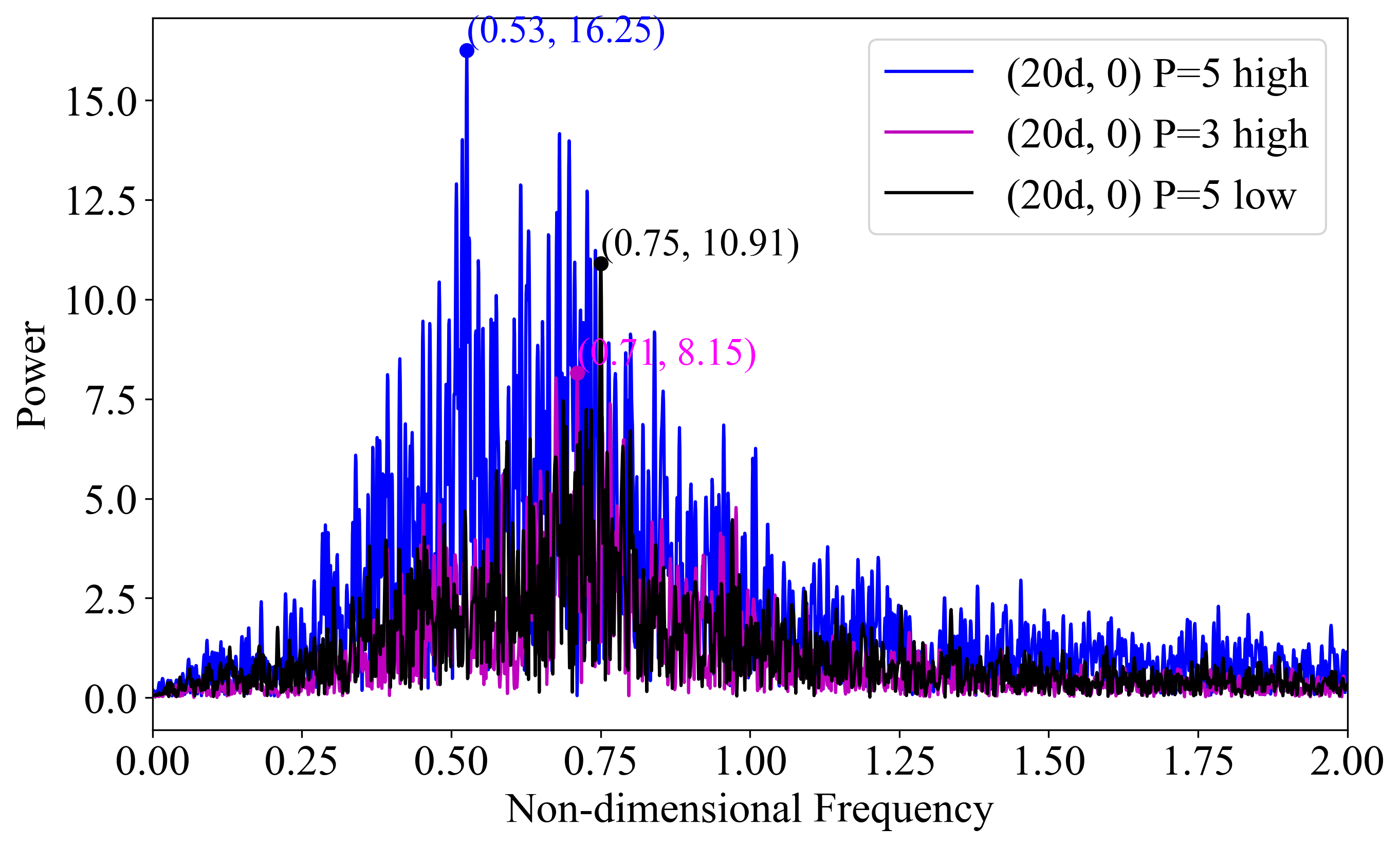}\\
    
  (c) Probe at $(12d, 0.0)$& (d) Probe at $(20d, 0.0)$

\end{tabular} 
\caption{Comparison of frequency spectrum of vertical velocity at four probe points for various grid resolutions. We employed the low resolution (low) and high resolution mesh(high) according to Table~\ref{tbl:mesh_res} with polynomial orders $\mathcal{P}=3$ and $\mathcal{P}=5$, $f_{nondim}=\frac{f_{dim}\times d}{M}$, where $M$ is the Mach number.}
\label{fig:spectrum_sl}
\end{center}
\end{figure}

We investigate the $ \textrm{H}^3 \textrm{PC}$ solver's ability to predict vortex shedding in supersonic flows. For this, we placed four points downstream of the square cylinder and recorded the vertical velocity history at $T\in [300,500]$ for $P_1=(1.5d,0)$, $P_2=(7.0d,0)$, $P_3=(12.0d,0)$, and $P_4=(20.0d,0)$. The history was recorded at nonuniform time samples due to the adaptive time step. We then used the Lomb-Scargle periodogram method \cite{lomb1976least, scargle1982studies} to compute the frequency amplitude spectrum from the nonuniformly sampled time series. Figure~\ref{fig:time_solution} demonstrates time evolution of the density field at $T=400$, $450$, and $500$ for mesh (A) and mesh (B) with $\mathcal{P}=5$. According to this figure the vortex street is sustained at $M=1.5$ and $Re=10^4$. In Fig.~\ref{fig:time_solution}, the probe points are also shown using the red circles.

Figure~\ref{fig:spectrum_sl} compares the supersonic flow simulated using three spatial resolutions. According to Fig.~\ref{fig:spectrum_sl}, the dominant frequency for the second and third probes is 0.73 for low resolution $P=5$ and high resolution $P=5$. At the first probe point, the high-resolution $P=5$ mesh can capture dominant frequencies since it has four times more mesh resolution in the wake of the square cylinder than the low-resolution mesh and predicts the most accurate prediction for the Strouhal number.

\begin{figure}[t!]
  \begin{center}
    \begin{tabular}{c}
    \includegraphics[width=0.7\textwidth,trim={0cm 9cm 0cm 0cm},clip]{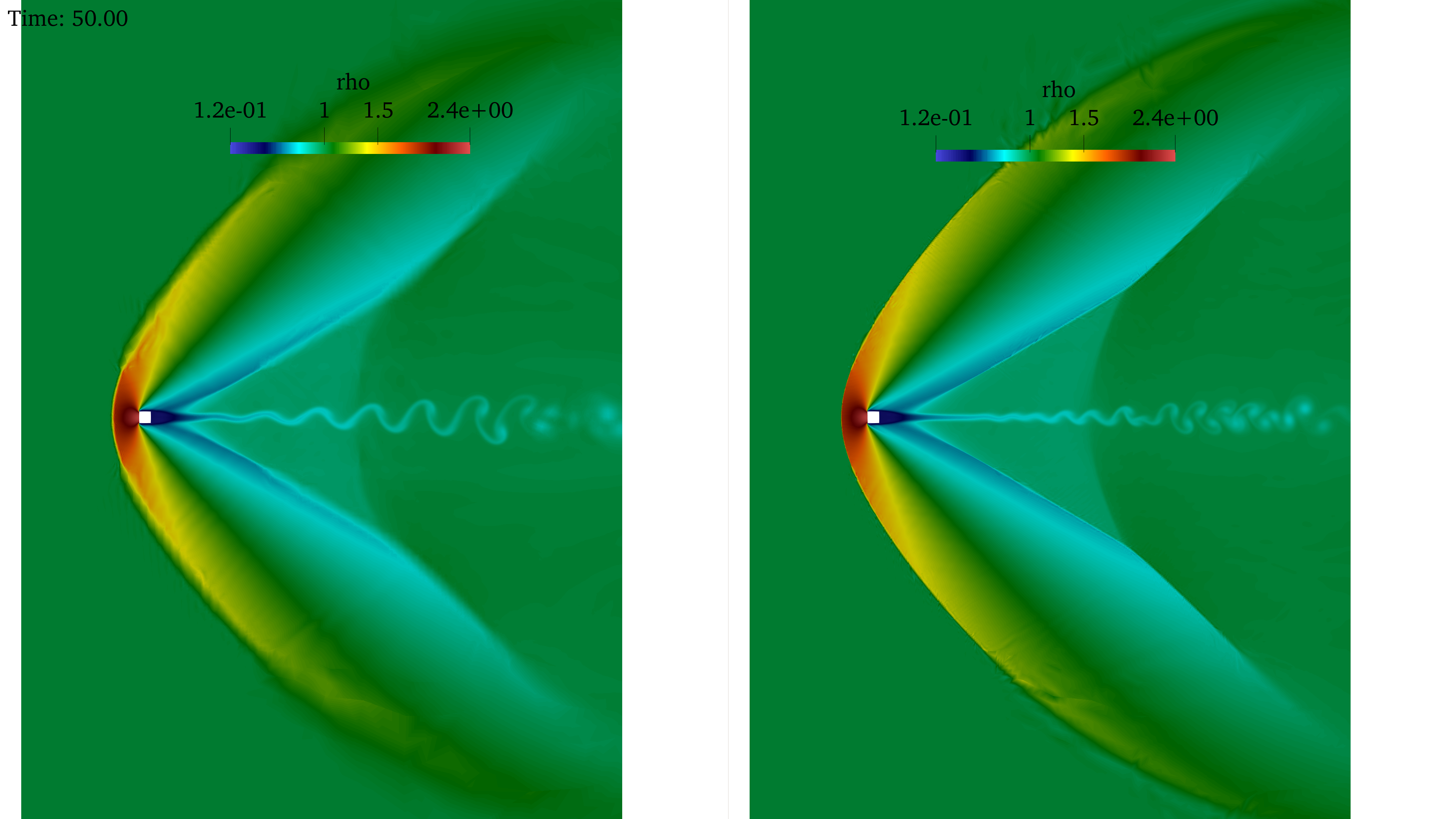}
  \\
  \includegraphics[width=0.7\textwidth,trim={0cm 9cm 0cm 0cm},clip]{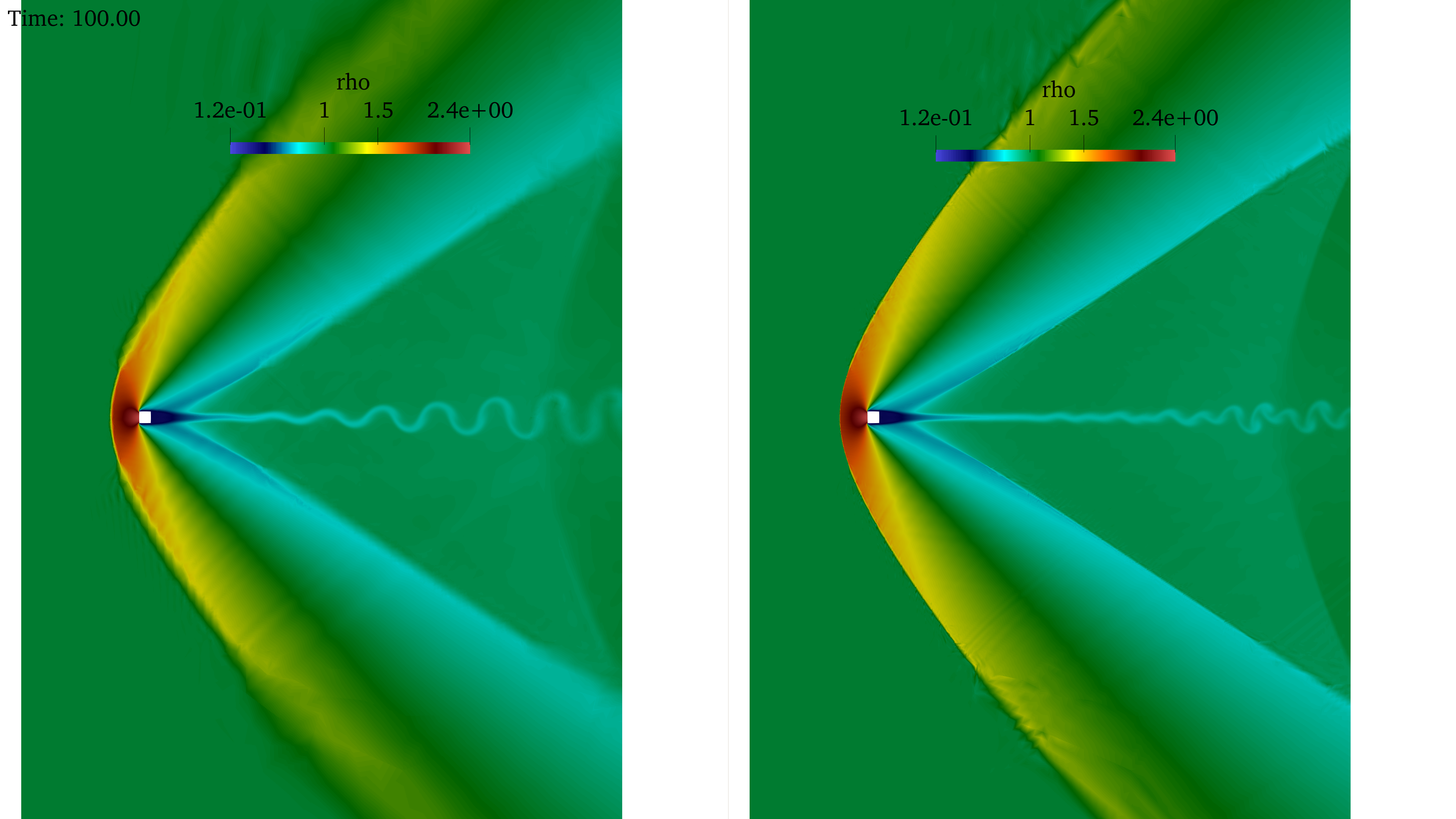}

\\
 \includegraphics[width=0.7\textwidth,trim={0cm 9cm 0cm 0cm},clip]{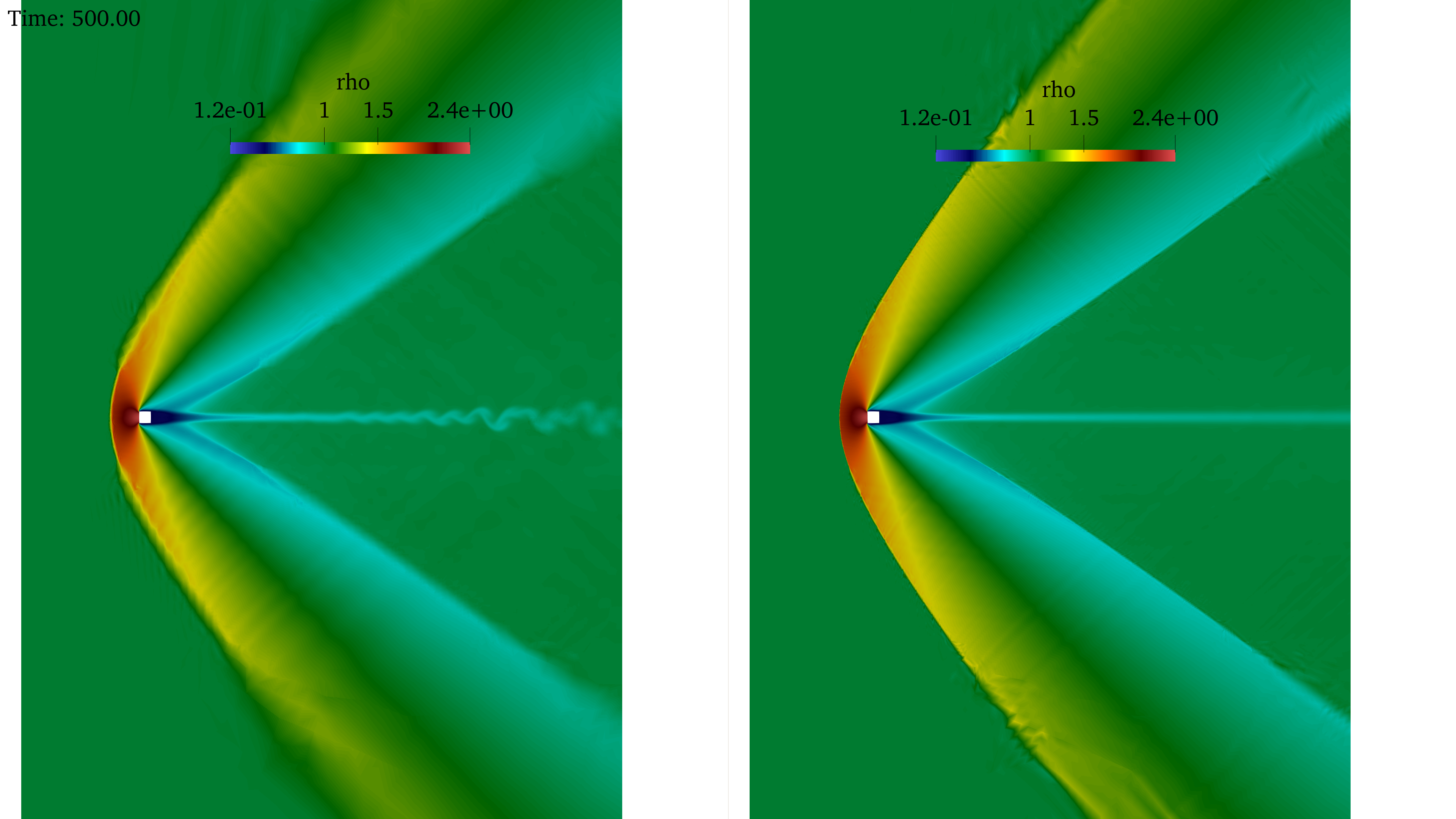}
\\
\end{tabular} 
\caption{Density contours of supersonic flow over square cylinder at $t=50$, $100$, and $500$ obtained using the medium mesh resolution (A) with $\mathcal{P}=2$ on the left and the medium mesh resolution (A) with $\mathcal{P}=5$ on the right. $Re=1000$ and $M=1.5$.
Despite the early emergence of the vortex street, the flow becomes steady eventually at the highest resolution.}
\label{fig:re1000_comp}
\end{center}
\end{figure}

We further investigate vortex shedding in supersonic flows by reducing the Reynolds number to ensure the solution field is sufficiently resolved. Therefore, we solve the NS equations using mesh (A) with two polynomial orders to demonstrate the accuracy order's effectiveness in resolving transient flows. Figure~\ref{fig:re1000_comp} shows solution snapshots at $t=50$, $100$, and $500$. The left column shows the density field obtained with mesh (A) and $\mathcal{P}=2$, and the right column shows the solution obtained with mesh (A) and $\mathcal{P}=5$. We can observe that the vortex street is sustained when we use a coarse resolution ($\mathcal{P}=2$) while it dies out when a high resolution grid ($\mathcal{P}=5$) is employed. This phenomenon indicates that, with sufficient resolution for these blunt bodies, the vortex street may not even form. This phenomenon could happen in the $Re=10^4$ case, meaning that if we further increase the mesh resolution, the vortex street may disappear. We have also used the UCN3D solver to resolve the supersonic flow over a square cylinder at $Re=1000$ and $M=1.5$, and to compare the results with those obtained with the H$^3$PC solver. They both show vortex streets because the mesh resolution of both solvers is not sufficient, see Appendix \ref{sec:compare_solvers}. 

This phenomenon can be explained as follows. For
incompressible wakes, the vortex street emerges due to {\em absolute instability} that can be triggered due to random arithmetic noise. However, the effect of density stratification is to alter the instability to convective 
type, which is sustained only by the continuous feeding of disturbances, which are larger at low resolution. In the physical experiment, it is plausible that free stream turbulence in the wind tunnel can trigger and sustain this convective instability and possibly sustain a vortex street in the far wake. At higher Mach numbers, the density stratification is even stronger and thus high resolution simulations do not produce a vortex street, as it was the case for the circular cylinder presented in the previous section. However, stability analysis presented in \cite{rodriguez2023compressibility} suggests that the transition happens at Ma=1.2, which is very close to the Mach number used in the square cylinder simulations.

\subsection{P8-Inlet}
\begin{figure}[t!]
  \centering
\includegraphics[width=0.7\textwidth]{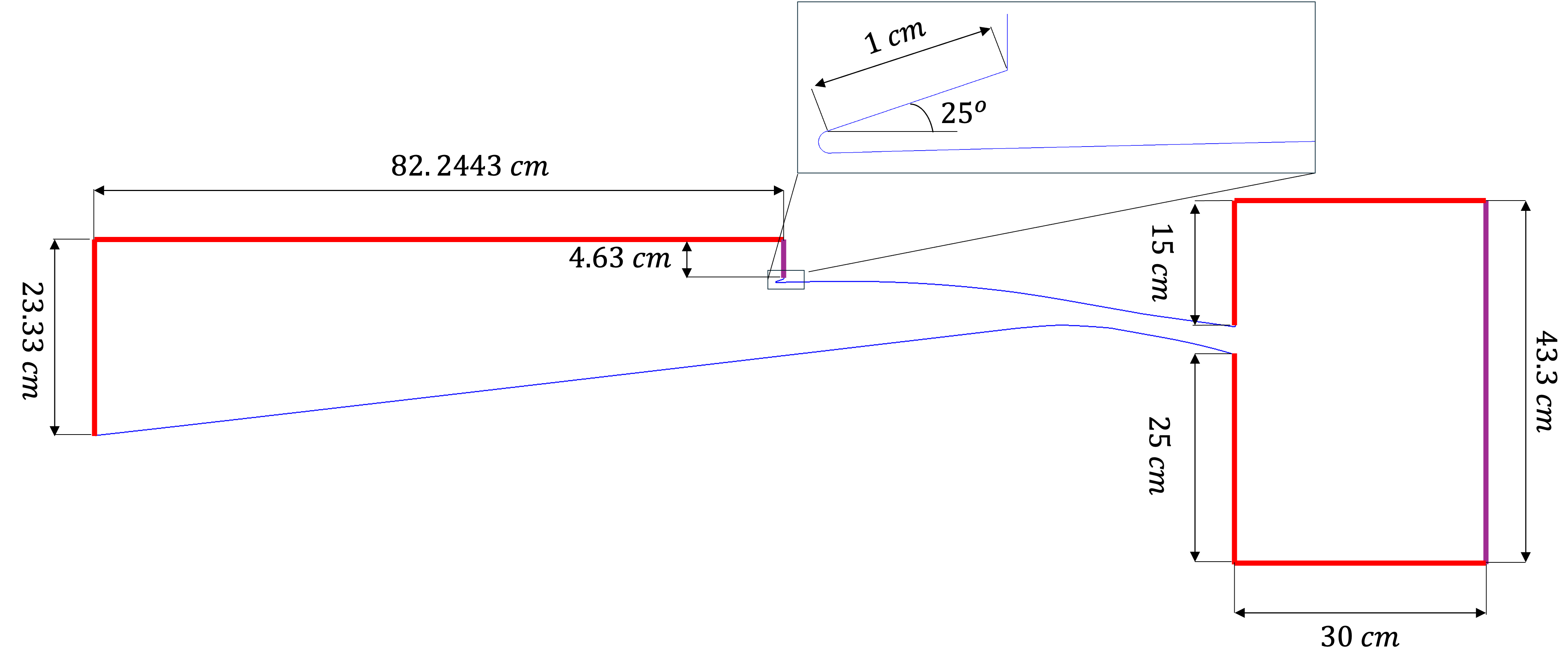} 
\caption{Left: Physical domain and boundary setup for the P8-inlet problem. At the red boundaries, supersonic inflow boundary condition is imposed. The magenta lines indicate the supersonic outflow boundary conditions. We impose isothermal wall boundary condition at the the blue lines, which are the walls of P8-inlet configuration.}
\label{fig:p8inlet}  
\end{figure}

In this section, we tackle a challenging problem that investigates the aerodynamic performance of an air-induction system of a hypersonic vehicle. Gnos and Watson \cite{gnos1973investigation} designed three large-scale inlet models, each providing different compression ratios. Specifically, they designed P2, P8, and P12 inlet configurations corresponding to the internal compression ratios of 2, 8, and 12, respectively. In this context, we focus on the P8-inlet configurations. The geometry of the P8-inlet consists of a wedge forebody contour, inlet centerbody (bottom wall), and cowl (top wall). To construct the inlet geometry, we use the coordinate points reported in reference \cite{gnos1973investigation}, applying Splines for smooth contouring. Figure~\ref{fig:p8inlet} demonstrates the geometric configuration of the physical domain. Several numerical studies \cite{gnos1973investigation, knight1977numerical,anderson1980numerical,ng1989turbulence,kapoor1994numerical} have attempted to solve this problem and match experimental results. These studies commonly solve Reynolds-averaged Navier-Stokes equations along with various turbulence models. For example, Knight \cite{knight1977numerical} used two overlapping regions to solve the equations separately. However, that study modeled the circular shape of the cowl's leading edge by extending the cowl wall and using a sharp leading edge. The author claims that this geometric simplification marginally affects the flow field. In contrast, this study reveals that such simplifications significantly alter the flow structure inside the inlet duct. Similarly, Kapoor \emph{et al.} \cite{kapoor1992comparative} also use a sharp edge for the cowl leading edge. In our current study, however, we use the actual geometry of the cowl leading edge, as shown in the zoomed-in box of Fig.~\ref{fig:p8inlet}. Here, the circular arc has a radius of 0.114 cm and a center at $P_c=(81.394, 18.33)$.

We solve the non-reactive compressible Navier-Stokes equations on the physical domain shown in Fig.~\ref{fig:p8inlet}. According to Fig.~\ref{fig:p8inlet}, the red lines indicate the supersonic inflow boundary condition, while the magenta lines denote the supersonic outflow boundary condition. The blue lines are the isothermal wall boundary conditions with a wall temperature of 304.125 K. The free-stream mach number is at $M=7.4$ and $Re=8.86\times 10^6 \textrm{ m}^{-1}$. The inflow conditions are set such that the total temperature $T_o=811$ K, total pressure is $P_o=4.14\times 10^6$ Pa. As a result, using the ideal gas law, we can compute the free-stream pressure and density as $p_\infty=701.3879$ Pa, and $\rho_\infty=0.03601\textrm{ kg}/m^3$. The sponge region is designed to avoid unnecessary reflections from the outflow boundary. The dynamic viscosity is computed employing Sutherland's law as

\begin{equation}
    \label{eq:sutherland}
    \mu = \mu_0 \left(\frac{T}{T_0}\right)^{\frac{3}{2}}\frac{T_0+S}{T+S}
\end{equation}
where $S=110$, and $T_0$ and $\mu_0$ are the reference temperature and viscosity. 

\begin{figure}[t!]
  \begin{center}
    \begin{tabular}{c}
    \includegraphics[width=0.7\textwidth]{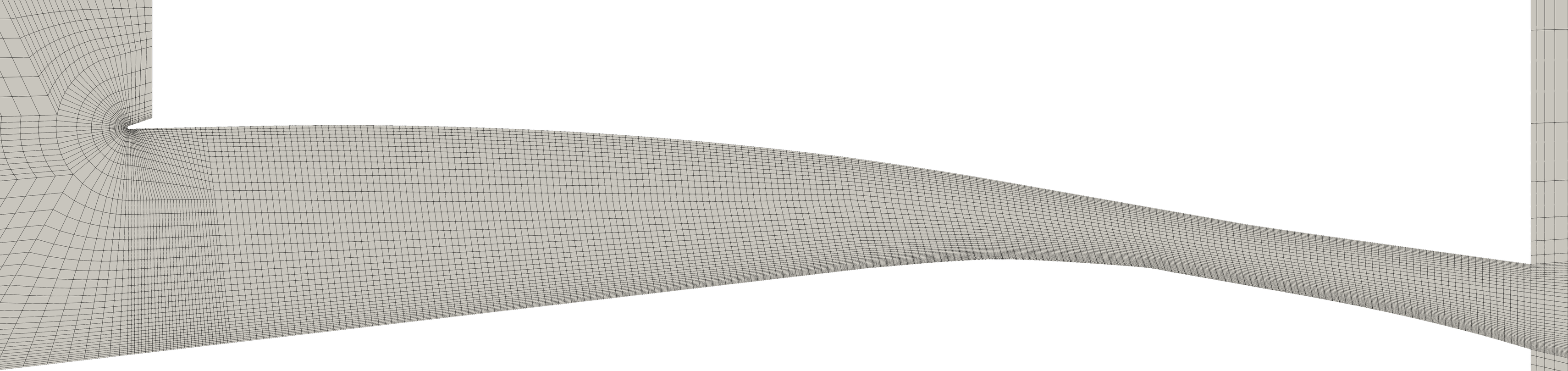}

  \\
  (a) Coarse
  \\
  \includegraphics[width=0.7\textwidth]{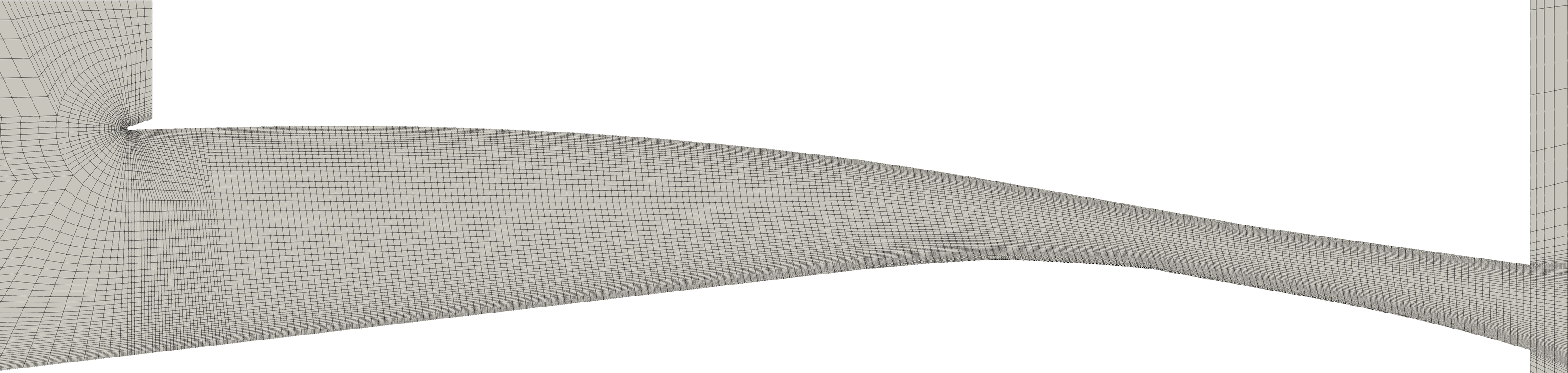}
    
  \\
  (b)Medium
  \\
  \includegraphics[width=0.7\textwidth]{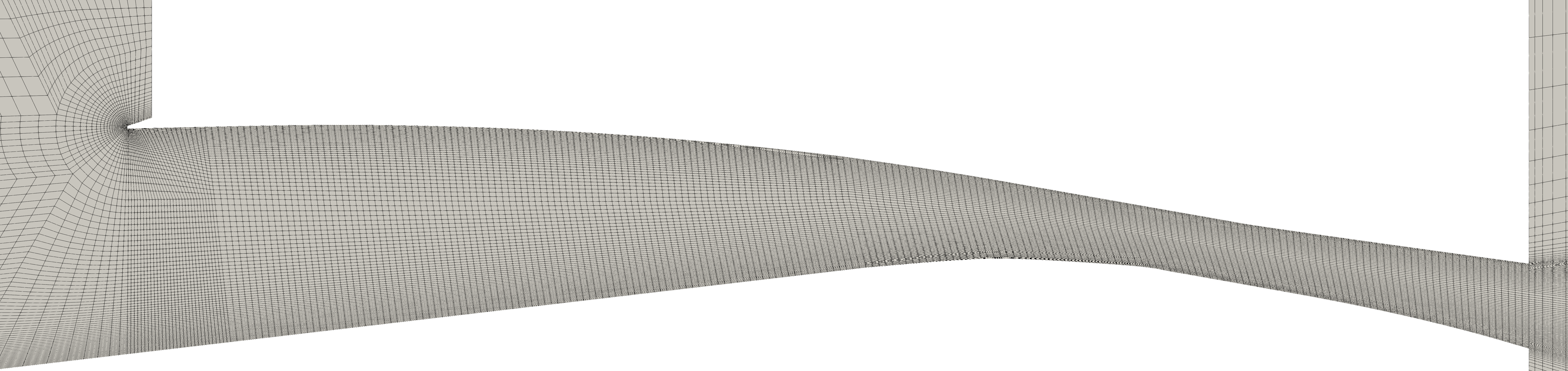}
    
  \\
  (c)fine

\end{tabular} 

\caption{Grid resolution for the P8-inlet geometry. The mesh resolution in x direction is the same across cases while the resolution in wall normal varies as 42, 64, and 99 elements. In x direction, we placed 229 elements. We employed GMSH to generate the mesh. We used a progression ratio of 1.09, 1.07, and 104 near the bottom wall and progression ratio of 1.1, 1.09, and 1.06 near the top wall for coarse, medium, and fine resolution grids, respectively.}
    
    \label{fig:p8-inlet mesh}
  \end{center}
  
\end{figure}

Three mesh resolutions are selected to discretize the physical domain shown in Fig.~\ref{fig:p8inlet}. The grid resolution is refined only inside the inlet geometry, since we observe the most change in flow structure; see Fig.~\ref{fig:p8-inlet mesh}. In the x-direction, we maintain a constant resolution while refining in the normal wall directions. In total, we have $12,784$, $19,334$, and $29,611$ quad elements for coarse, medium, and fine resolution grids. The grid resolution inside the sponge region is very coarse, and the elements progressively become larger toward the outflow boundary. We created a supersonic outflow boundary on top of the cowl wall to allow the incident shock wave to exit the domain without any reflection.

The simulation is performed for 20 flow-through times. Each flow through ($FT$) time is computed using the entire length of the physical domain, excluding the sponge region, divided by the free-stream Mach number. We then collect the average statistics in time by computing the following integral
\begin{equation}
\bar{\phi}= \frac{1}{\Delta t}\int_{t_1}^{t_2}\phi dt,
    \label{eq:ave_int}
\end{equation}
using the trapezoidal rule. In Eq.~\eqref{eq:ave_int}, $\phi$ represents any primitive variable. For this simulation, the averaging is performed from $t_1=10 FT$ to $t_2=20FT$. The time stepping scheme is SSPRK43, and the maximum blending factor for the shock capturing scheme is $0.5$. We also employ the positivity preservation scheme of Zhange and Shu \cite{zhang2011positivity} for density and pressure. The polynomial order is selected at $\mathcal{P}=3$ for all three cases.

\begin{figure}[t!]
  \centering
\includegraphics[width=0.7\textwidth]{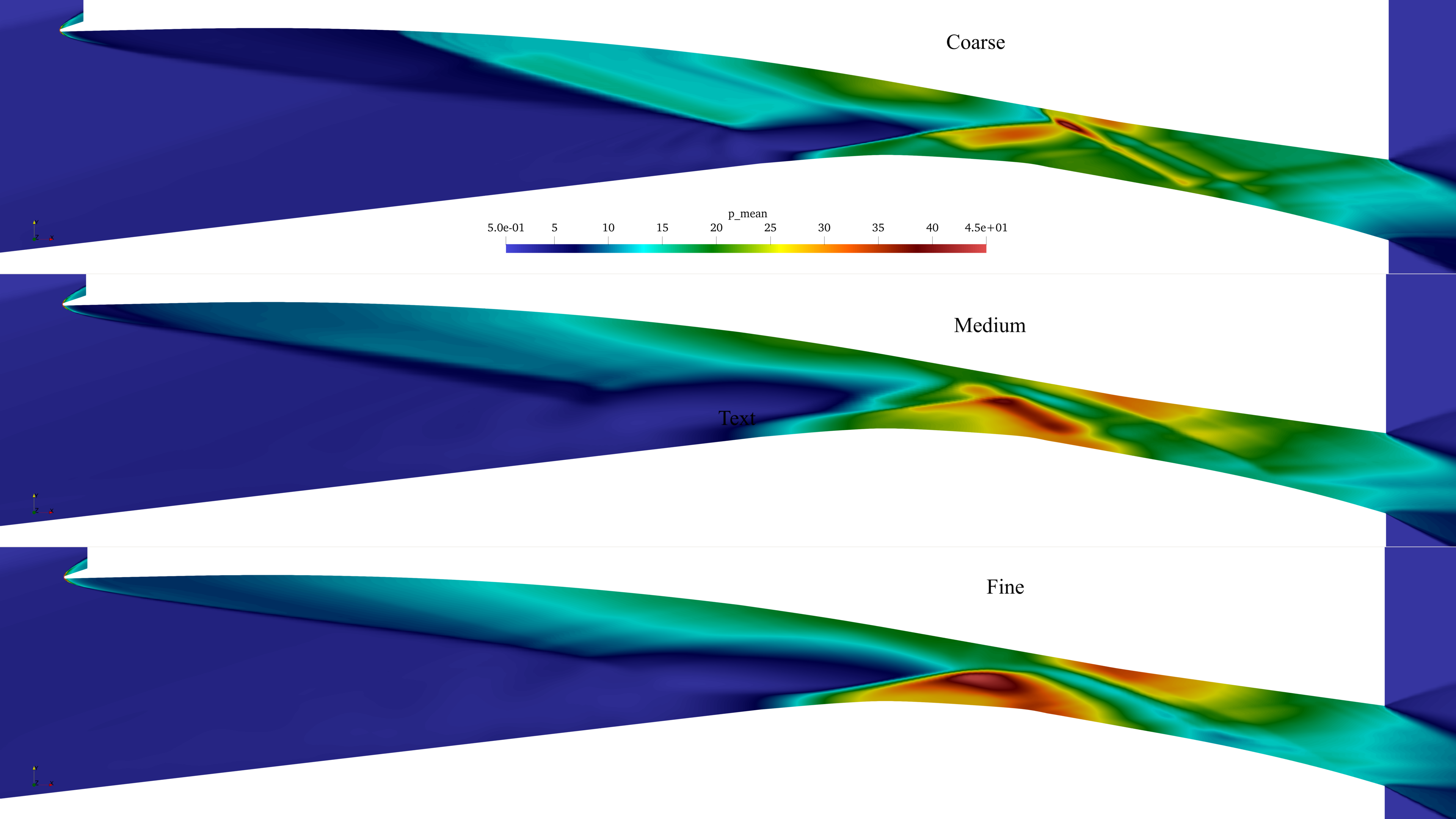} 
\caption{Average pressure for coarse to fine resolution mesh.}
\label{fig:p8inlet_pressure}  
\end{figure}
\begin{figure}[t!]
  \centering
\includegraphics[width=0.7\textwidth]{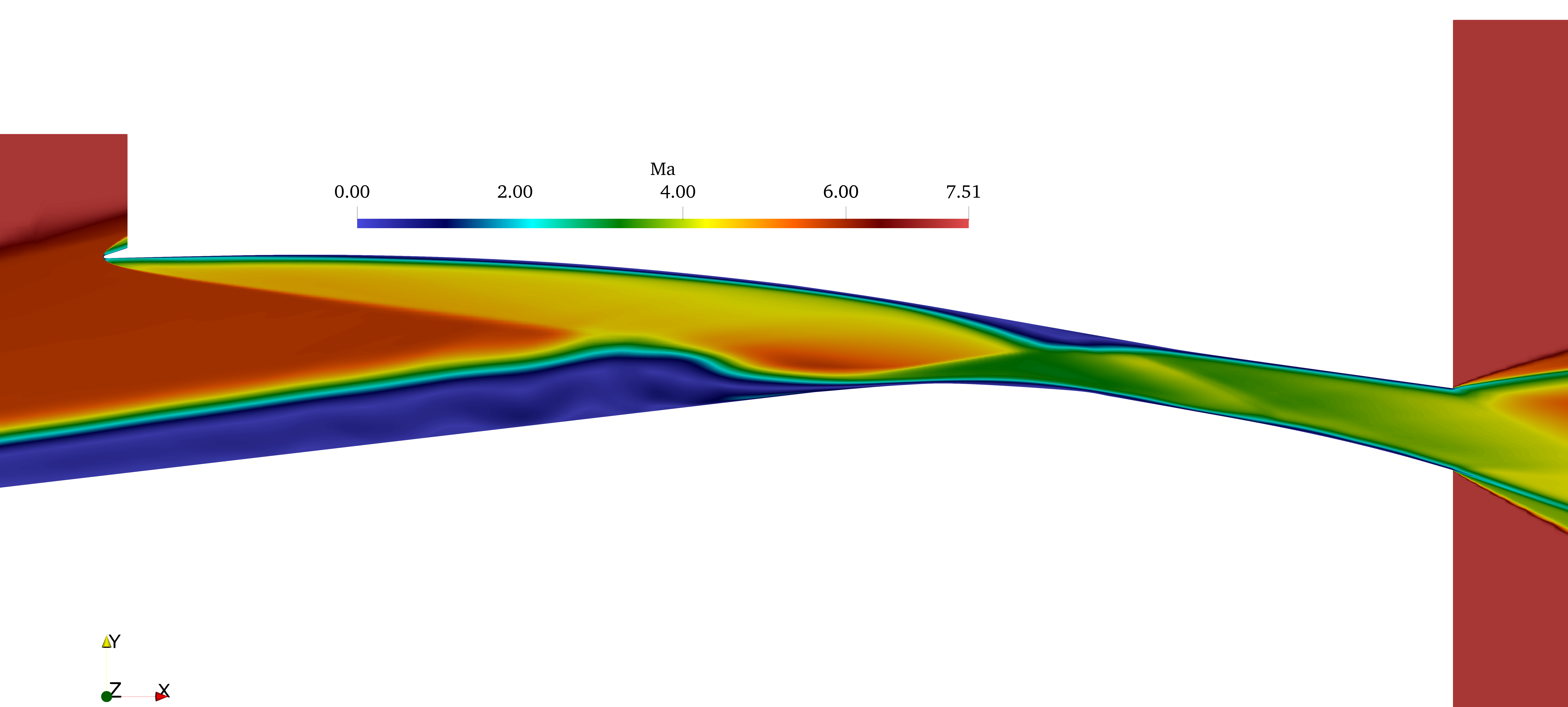} 
\caption{Average Mach contours for fine resolution mesh.}
\label{fig:p8inlet_mach}  
\end{figure}
\begin{figure}[t!]
  \begin{center}
    \begin{tabular}{cc}
    \includegraphics[width=0.45\textwidth]{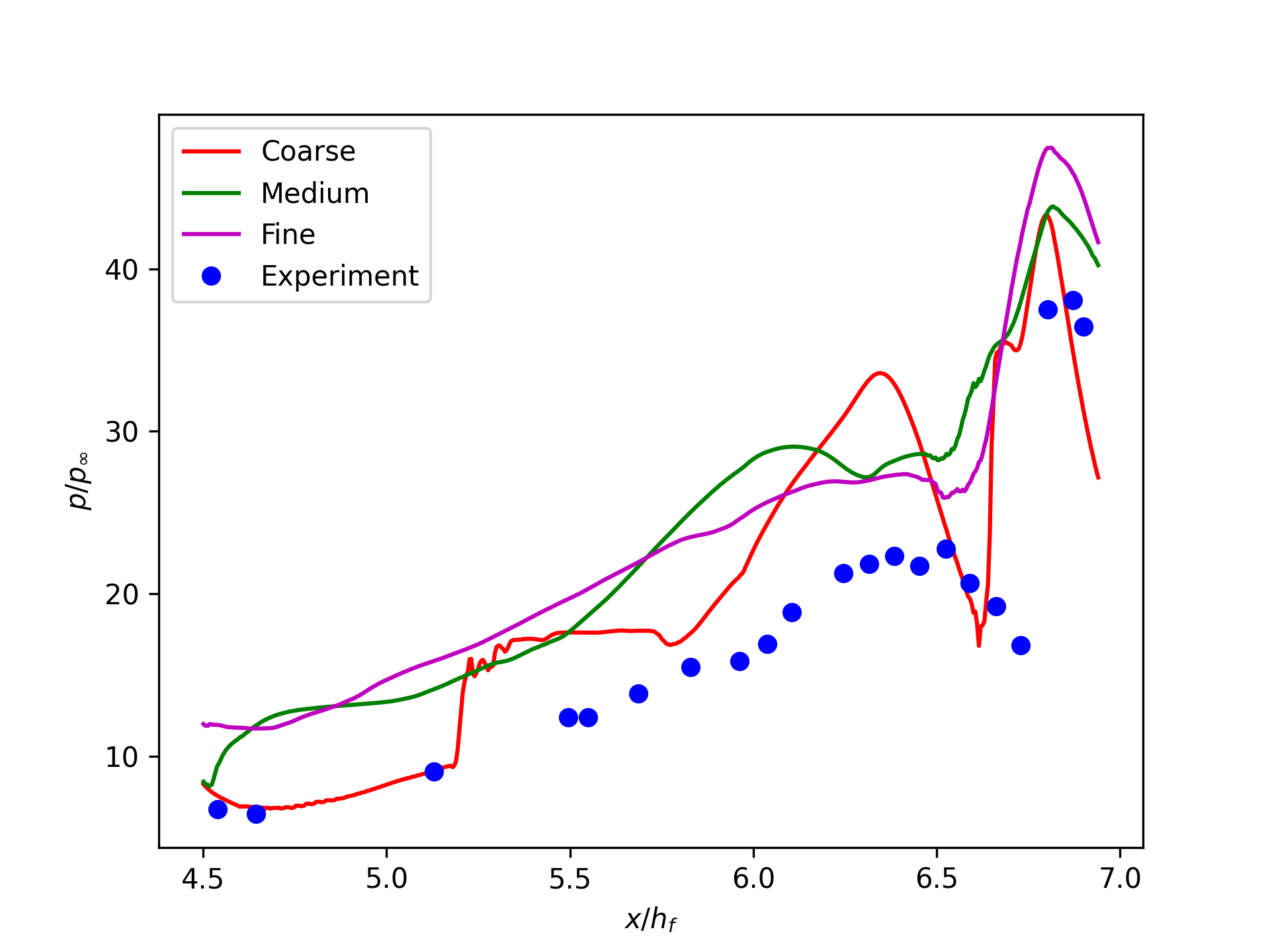}
    &
    \includegraphics[width=0.45\textwidth]{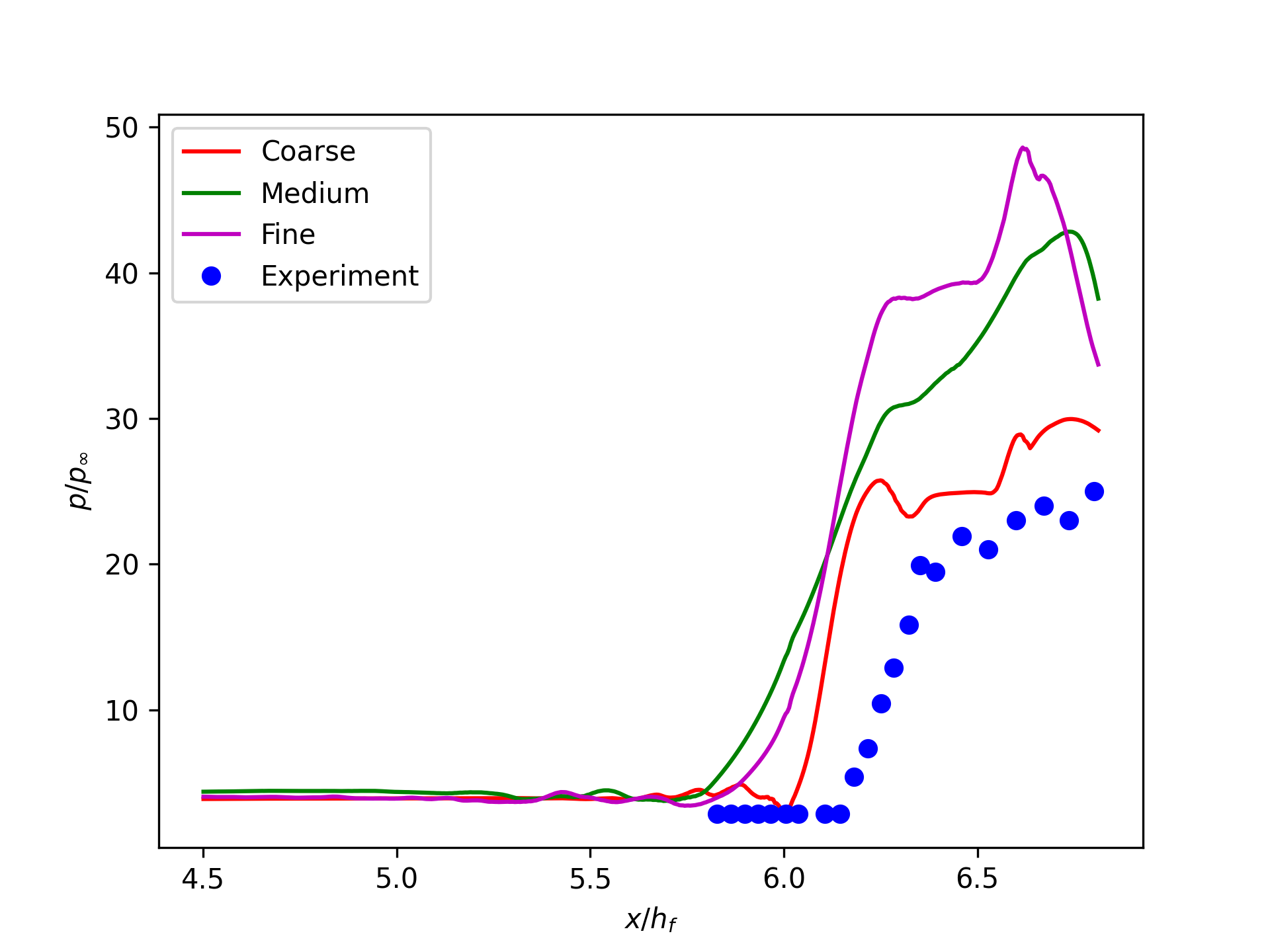}

  \\
  (a) Pressure ratio profile on the cowl& (b) Pressure ratio profile on the bottom wall

\end{tabular} 

\caption{Pressure ratio profile on the (a) cowl wall, and (b) on the bottom wall of the P8-inlet. $h_f=18.33$ cm.}
    
    \label{fig:p8inlet_pressure_prof}
  \end{center}
  
\end{figure}

Figure~\ref{fig:p8inlet_pressure} shows a comparison of the average pressure contours for three grid resolutions. It is clear that the solution converges at the fine grid level. The shapes of high-pressure regions for the fine grid are similar to the contour plots of pressure in reference \cite{kapoor1994numerical} except for a separation region that exists at $x=108$ cm. This separated region can be observed in the Mach contour plots of Fig.~\ref{fig:p8inlet_mach} before the throat area. The separation of flow inside the P8-inlet area has not been reported in the experimental results. We attribute this dependency to the inaccurate condition of the inflow in our simulation. The inflow in the experiment must have some level of turbulence that could affect the downstream flow. Figure~\ref{fig:p8inlet_pressure_prof} (a) and (b) compare the pressure ratio ($p/p_{\infty}$) profile on the cowl and bottom walls predicted by the H$^3$PC solver with experiment \cite{gnos1973investigation}. Focusing on the fine grid resolution profile (Fig.~\ref{fig:p8inlet_pressure_prof} (a)), we can observe that the pressure on the cowl wall shape is very similar to the experiment with a shift of about 7 units according to the vertical axis. We attribute this shift to the existence of a separated flow region that causes a contraction of flow and an increase in pressure on the cowl wall. We can observe the effect of separation in the prediction of pressure ratio profile on the bottom wall (Fig.~\ref {fig:p8inlet_pressure_prof}(b)) in the fact that the pressure rise in the simulation occurs upstream of the location in the experiment. 

Because of this discrepancy, we decided to try several modifications in the geometry of the physical domain and boundary conditions. First, the modification in geometry is accomplished based on the suggestion of Knight's study \cite{knight1977numerical}, where he extends the leading edge of the cowl and uses an edge instead of a circular leading edge. Knight extends the leading edge to a virtual origin, which is at the intersection of the tangent to the cowl surface and the experimental shock location downstream of the inlet entrance. Knight suggests extending the leading edge by 2.794 CM. However, in our experience, this extension is too much as it prevents the incident oblique wave from emanating out of the physical domain. Therefore, we decided to extend the leading edge by the radius of the cowl leading edge for 0.11 cm. For the second modification, we use the real geometry of the cowl and then employ a different boundary condition for the inflow boundaries of the sponge box at the outflow of the P8-inlet. For the original case, the incoming flux from the inlet boundaries of the sponge box was computed using the free stream conditions. For the current condition, we employed the pressure reported by the experimental results. For the upper boundary in the input, we used $p/p_\infty=36.44$, and for the lower boundary in the input, we used $p/p_\infty=25$. Other primitive variables were computed from the simulation of fine grid resolution, such as $\rho/\rho_f=6$ and $M=4$. 

Figure~\ref{fig:p8inlet_pressure_prof} compares the pressure prediction on the medium resolution grid to impose back pressure at the exit of the P8-inlet duct. We can observe that the setup with the back pressure produces a better prediction; however, it does not remove the separation region that causes the mismatch of the numerical and experimental results. Using a straight, sharp leading edge completely changes the dynamics of the flow. Using the setup with the straight edge creates an unstart phenomenon where the supersonic flow transitions into subsonic regimes due to the separation of the flow. (The results are not shown here.)

\begin{figure}[t!]
  \begin{center}
    \begin{tabular}{cc}
    \multicolumn{2}{c}{\includegraphics[width=0.55\textwidth]{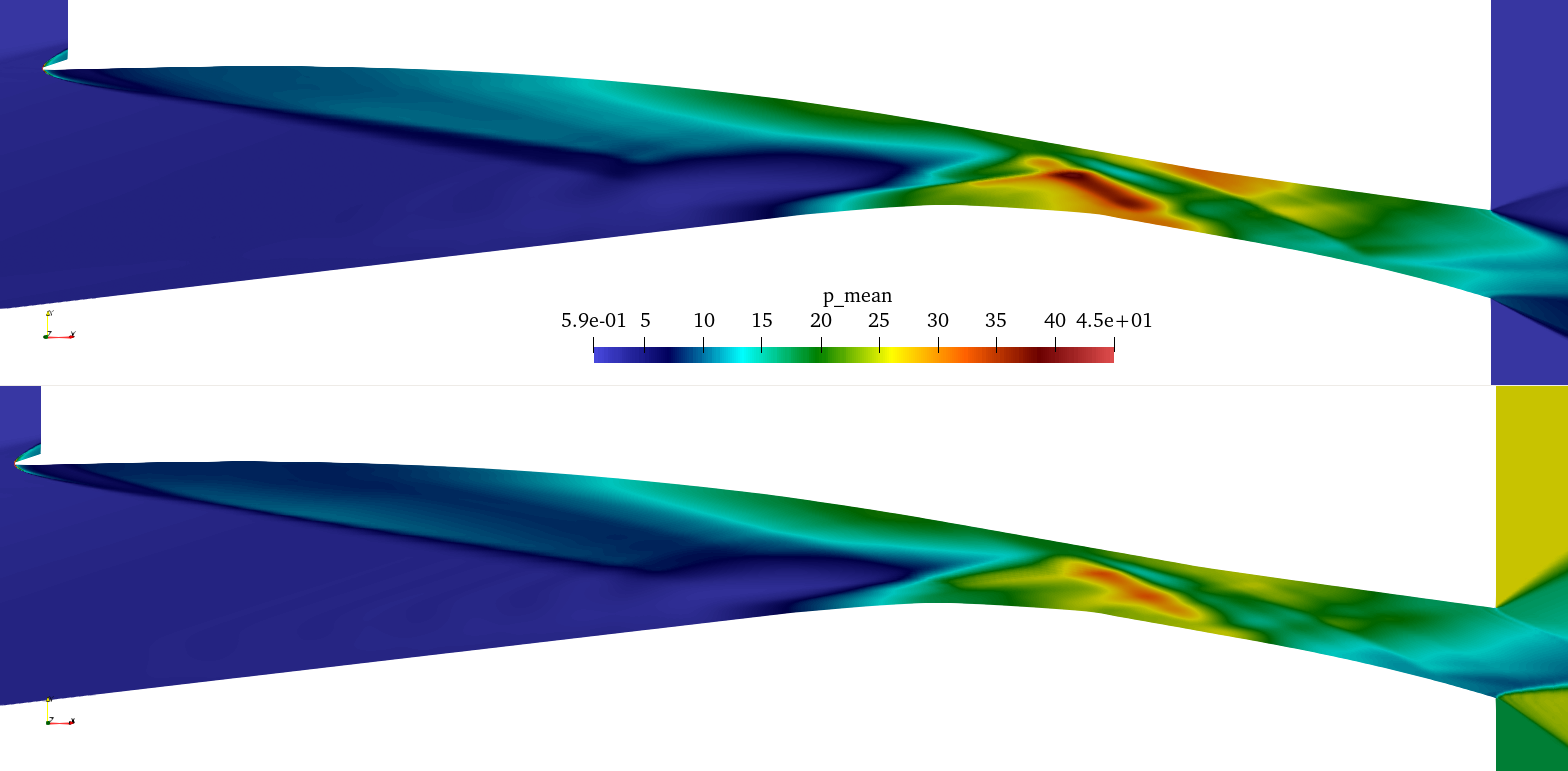}}    
\\\multicolumn{2}{c}{(a) Contour plots of average pressure. }
\\
\includegraphics[width=0.45\textwidth]{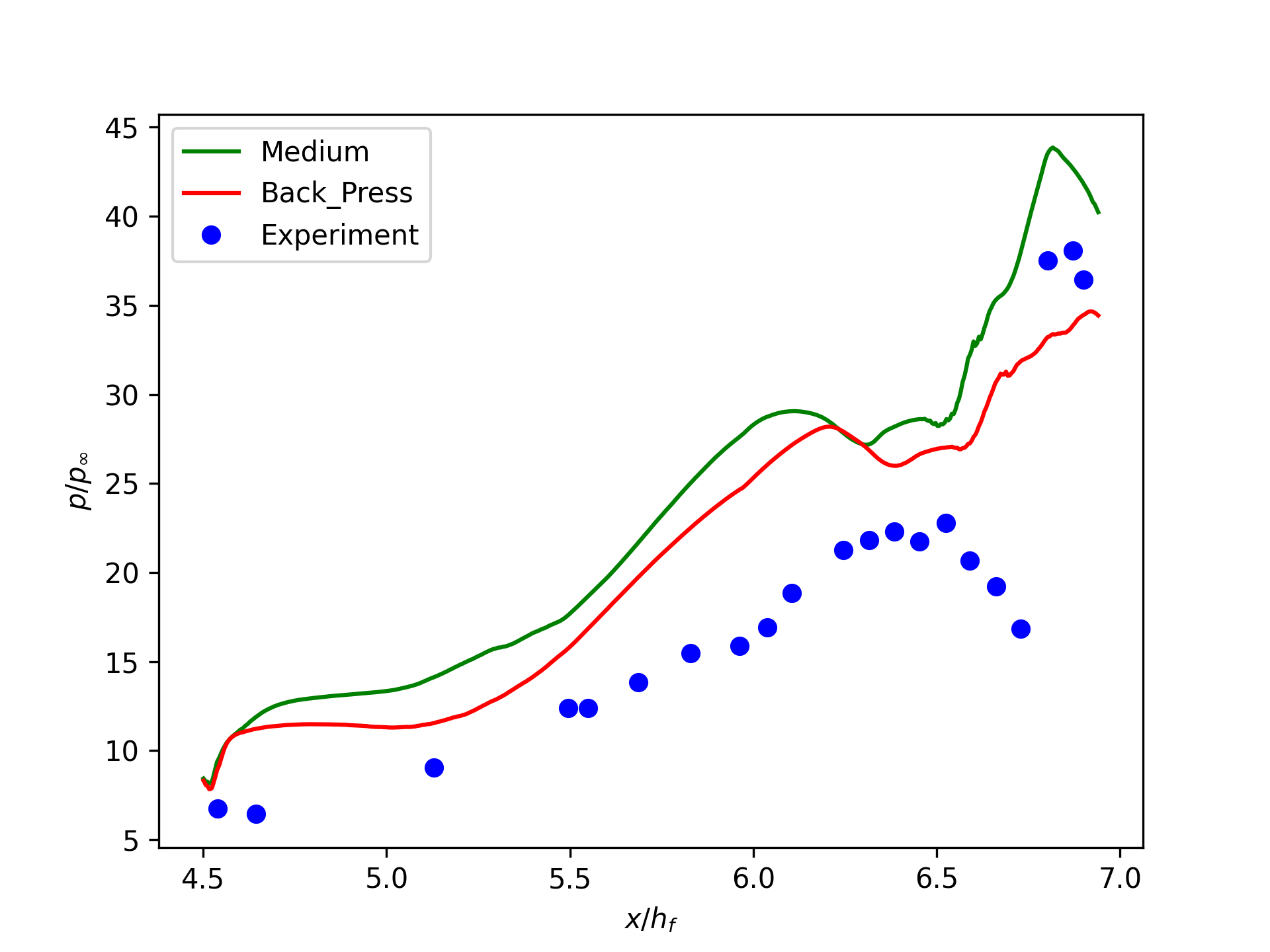} & \includegraphics[width=0.45\textwidth]{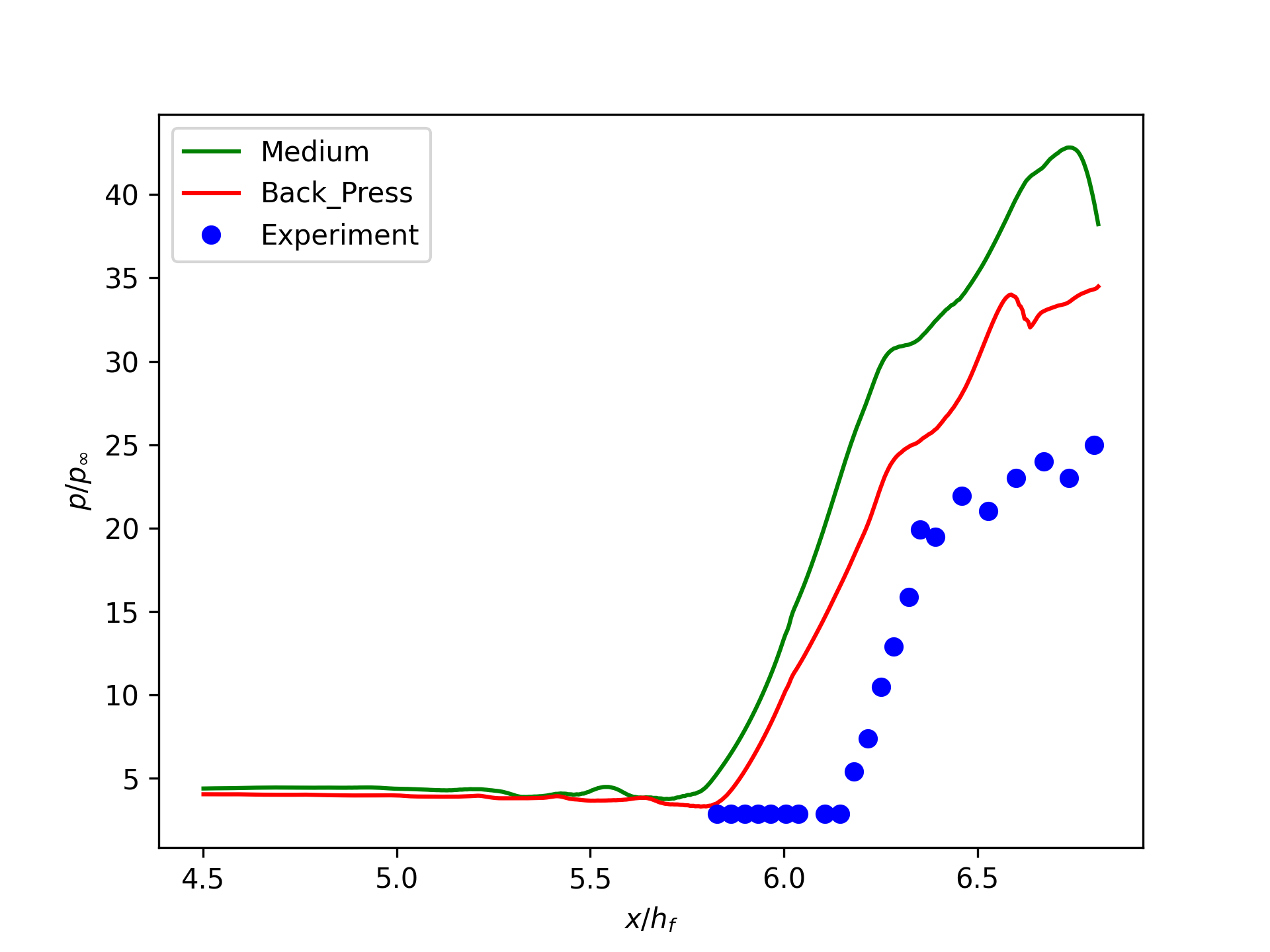} 
  \\
  (a) Pressure ratio profile on the cowl& (b) Pressure ratio profile on the bottom wall
\end{tabular} 
\caption{(a) Top plot shows average pressure for medium resolution grid with free stream boundary condition for the inflow boundaries of sponge region, and the bottom plot shows the average pressure for medium resolution grid with experimental pressure boundary conditions for the inflow boundaries of the sponge region, (b) Pressure profile on the cowl wall, (c) Pressure profile comparison on the bottom wall}    
    \label{fig:p8inlet_bcs}
  \end{center}
 
\end{figure}

%% file: conclusion.tex
\section{Summary}
We have developed a versatile numerical framework capable of simulating subsonic to hypersonic regimes, with realistic modeling of thermo-chemical events at hypersonic speeds. The H$^3$PC solver is developed by implementing four main features into Trixi.jl framework. A parabolic solver is implemented for the P4est mesh type. The Parabolic solver can handle static and adaptive meshes using flux patching across interfaces and mortars. We then use the MPI library to parallelize the parabolic solver and couple it with Trixi's existing hyperbolic solver.jl code. As a result, the H$^3$PC solver employs adaptive mesh refinement for solving Euler and Navier-Stokes equations in parallel mode. The code uses the Mutation++ library for non-equilibrium chemistry modeling with minimum computational overhead. The implementation of the Mutation++ in the H$^3$PC solver was validated through a 1D Sod problem for a gas mixture with a non-equilibrium chemistry assumption. In this assumption, air is modeled as a gas mixture of five species, and the internal energy is partitioned into translational, vibrational, rotational, and electronic modes. The high-order method and entropy stability improve the accuracy of the under-resolved simulations of turbulent flows. For instance, the H$^3$PC solver achieves higher accuracy than the UCNS3D framework in predicting energy dissipation evolution. Because the H$^3$PC solver features entropy stability, we can employ very high polynomial orders while preserving numerical stability. We have also compared the results from our framework with experimental data to predict the vortex shedding frequency in a supersonic flow over a circular cylinder. We used the H$^3$PC solver to predict the vortex shedding of a square cylinder in supersonic flow. We demonstrated that, with sufficient spatial resolution and a high order of accuracy, these transient flows can be resolved accurately using the H$^3$PC solver.

The H$^3$PC solver provides a modular platform for robust, large-scale simulation of hypersonic flows on CPU architectures. In the future, we will extend this framework to support multi-GPU  computation. We will also incorporate other gas dissociation models, including plasma modeling that requires consideration of the ionization of gas-mixture species. The \textit{Mutation.jl} library supports 5- and 11-species gas-mixture models; therefore, extending it to simulate hypersonic flows with ionization will be straightforward.

%% file: appendix.tex
\section{Appendix}

\subsection{Thermodynamics}
\label{termo}
In this study, we employ a simple Rigid-Rotor and harmonic Oscillator model (RRHO) thermodynamic database \cite{Scoggins2020}. Here, the assumption is that the internal energy consists of four modes: translational, rotational, vibrational, and electronic. This study treats translational energy as a continuous quantity, while the other three modes are treated using quantum mechanic modeling, where we assume discrete energy bins for computations. The gas mixture species can be molecules or atoms. We consider three modes of internal energy, including translation and electronic energies. For molecules, we consider four modes: translation, rotation, vibration, and electronic. The total internal energy for each species can be computed as

\begin{equation}
e_s(T)= 
\begin{cases}
     e_s^t+e_s^r+e_s^v+e_s^e+e_s^0,& \text{if } s \in \mathcal{H}\\
     e_s^t+e_s^e+e_s^0,              &\text{if } s \notin \mathcal{H}
\end{cases},
    \label{Internal_energy}
\end{equation}
where $\mathcal{H}$ denotes sets of all molecule species in the gas mixture ans superscripts $t$, $r$, $v$, and $e$ indicate translational, rotational, vibrational, and electronic modes. The definition of each mode can be illustrated as
\begin{equation}
e_s^t = \frac{3}{2}R_s T\quad \textrm{for } s=1,\cdots,N_s,
    \label{trans_energ}
\end{equation}

\begin{equation}
e_s^r = \frac{L_s}{2}R_s T\quad\textrm{for } s\in\mathcal{H},
    \label{rot_energ}
\end{equation}

\begin{equation}
e_s^v = R_s\sum_{k=1}^{N_v}\frac{\theta^v_{ks}}{\exp{\left(\frac{\theta^v_{ks}}{T}\right)-1}}\quad  \textrm{for } s\in\mathcal{H},
    \label{vib_energ}
\end{equation}

\begin{equation}
e_s^e = \frac{R_s}{Q^{e}_s}\sum_{k=1}^{N_e}a_{ks}^e\theta^e_{ks}\exp{\left(-\frac{\theta^e_{ks}}{T}\right)}\quad \textrm{for } s=1,\cdots,N_s,,
    \label{elec_energ}
\end{equation}
where $Q^{e}_s$ is an electronic internal partition function that is defined as 
\begin{equation}
Q^{e}_s = \sum_{k=1}^{N_e}a_{ks}^e\exp{\left(-\frac{\theta^e_{ks}}{T}\right)}\quad \textrm{for } s=1,\cdots,N_s.
    \label{elec_partition}
\end{equation}
In Eq.~\eqref{rot_energ}, $L_s$ denotes the molecule's linearity and takes a value of two for linear and 3 for non-linear molecules. The terms $\theta^v_{ks}$ are the characteristic temperatures for each vibrational mode $k$ and $\theta^e_{ks}=E^e_{ks}/k_B$ are the characteristic temperature for each electronic level $k$ with energy $E^e_{ks}$ and degeneracy $a^e_{ks}$. The gas constant for each species can be computed as $R_s=R_u/W_s$ with universal gas constant $R_u$ and molecular weight $W_s$.

\subsection{Transport}
\label{Transport}
The transport fluxes consist of the mass diffusion velocity, stress tensor, and heat flux that must be computed employing the primitive variables. The species diffusion velocity can be calculated by solving the following system of equations derived from the solution of generalized Stefan-Maxwell equations \cite{scoggins2017development}.

\begin{equation}
\sum_{s=1}^{N_s} G^V_{is}V^k_s = d^k_i,\quad \textrm{for } i=1,\cdots,N_s, \textrm{ and } k=1,2,3,
    \label{diffusion}
\end{equation}
where $d_i^k$ are the driving force in the $k^\textrm{th}$-direction of the coordinates system.
\begin{equation}
d_i^k=\frac{1}{nk_BT}\frac{\partial p_i}{\partial x_k}-\frac{y_ip}{nk_BT}\frac{\partial \ln{p}}{\partial x_k}+\mathcal{X}_i \frac{\partial \ln{T}}{\partial x_k}.
\label{driving_force}
\end{equation}
The transport system matrix $G^V_{is}$ depends on the species binary collision integrals and compositions. The exact formulation for the transport matrix is given in \cite{scoggins2017development}. In Eq~\eqref{visc_fluxes}, the stress tensor $\sigma_{ij}$ is expressed as 

\begin{equation}
\sigma_{ij} = \mu \left(\frac{\partial u_i}{\partial x_j}+\frac{\partial u_j}{\partial x_i}-\frac{2}{3}\frac{\partial u_m}{\partial x_m}\delta_{ij}\right)+\lambda \frac{\partial u_m}{\partial x_m}\delta_{ij},
    \label{stress}
\end{equation}
where $\mu$ and $\lambda$ are the shear dynamic and bulk mixture viscosities. In hypersonic regimes, the bulk viscosity is assumed to be negligible compared to dynamic viscosity \cite{scoggins2017development}. The dynamic viscosity $\mu$ is derived from the first Laguerre-Sonine polynomial approximation of Chapman-Enskog expansion. It can be computed by solving a linear system of equations defined as

\begin{equation}
\sum_{s=1}^{N_s} G^\mu_{is}\beta_s^\mu = x_i,\quad \textrm{for } i=1,\cdots,N_s, 
    \label{dynamic_visc}
\end{equation}
with unknown $\beta^\mu_i$. After computing unknowns, the dynamic viscosity is determined as
\begin{equation}
\mu = \sum_{s=1}^{N_s} \beta_s^\mu x_i,\quad \textrm{for } i=1,\cdots,N_s.
    \label{dynamic_visc}
\end{equation}
The viscosity transport matrix $G^\mu_{is}$ is a function of species mole fractions and binary collision integrals described in \cite{scoggins2017development}. 

The component of heat flux vector $q^k$ can be expressed as 
\begin{equation}
q^k = \sum_{s=1}^{N_s}\rho_sh_s V^k_s+nk_BT\sum_{s=1}^{N_s}\mathcal{X}_s V^k_s-\kappa \frac{\partial T}{\partial x_k},\quad \textrm{k=1,2,3,}
    \label{heat_flux}
\end{equation}
where $\mathcal{X}_s$ represent thermal diffusion ratios of species. The term $\kappa$ denotes the gas mixture thermal conductivity consisting of translational and internal conductivities defined as
\begin{equation}
\kappa=\kappa^t+\kappa^{int}.
    \label{thermal_cond}
\end{equation}
The translational conductivity can be computed by solving a linear system of equations for unknowns $\beta^{\kappa^t_s}$ expressed as
\begin{equation}
\sum_{s=1}^{N_s}G^{\kappa^t_{is}}\beta^{\kappa^t_s}=x_i,\quad \textrm{for }i=1,\cdots,N_s,
    \label{thermal_cond}
\end{equation}
with transport matrix $G^{\kappa^t_{is}}$ given in \cite{scoggins2017development}. After determining the unknowns, the thermal conductivity of the gas mixture is computed as

\begin{equation}
\kappa^t = \sum_{s=1}^{N_s} \beta_s^{\kappa^t} x_i,\quad \textrm{for } i=1,\cdots,N_s.
    \label{thermal_cond1}
\end{equation}
In Eq.~\eqref{heat_flux}, the thermal diffusion ratio is calculated via 
\begin{equation}
\mathcal{X}_i = \frac{5}{2}\sum_{s=1}^{N_s}\Gamma^{01}_{is} \beta_s^\kappa,\quad \textrm{for } i=1,\cdots,N_s,
    \label{thermal_cond1}
\end{equation}
with another transport matrix $\Gamma^{01}_{is}$ based on collision integrals.

The internal conductivity consists of three terms, each corresponding to an internal energy mode expressed as 

\begin{equation}
\kappa^{int}=\sum_m\kappa^m\quad \textrm{for }m=r,v,e,
    \label{intrnal_cond}
\end{equation}
where the superscripts $r$, $v$, and $e$ denote rotational, vibrational and electronic modes. The internal heat conductivity can be expressed as 
\begin{equation}
\kappa^m=\sum_{i=1}^{N_s}\frac{\rho_ic_{pi}^{m,int}}{\sum_{j=1}^{N_s}x_j/\mathcal{D}_{ij}},\quad \textrm{for } m=r,v,e,
    \label{internal_cond}
\end{equation}
where $\mathcal{D}_{ij}$ is the binary diffusion coefficient for species $i$ and $j$. In Eq.~\eqref{internal_cond}, the internal specific heat is defined as
\begin{equation}
c_{pi}^{m,int} = \frac{\partial h_i^r}{\partial T}+\frac{\partial h_i^v}{\partial T}+\frac{\partial h_i^e}{\partial T}.
    \label{heatint}
\end{equation}

\subsection{Chemical Kinetics}
\label{chem_kinetics}
The production and destruction rates of species due to the elementary chemical reactions are estimated according to the reaction mechanism that is considered. Here, since we model air as a mixture of five species, we consider a five-reaction mechanism introduced by Park\cite{park1993review}. All the chemical reactions can be cast into common formulations as 

\begin{equation}
\sum_{s=1}^{N_s} \nu_{sr}B_s\rightleftarrows\sum_{s=1}^{N_s} \nu^{'}_{sr}B_s,\quad r=1,\cdots, N_r,
    \label{chemreact}
\end{equation}
where $\nu_{sr}$ and $\nu^{'}_{sr}$ denote the forward and reversed stoichiometric coefficients for species $B_s$ in chemical reaction $r$. According to the law of action, the production rate of a reaction product is proportional to the product of reactant densities raised to their stoichiometric coefficients. Since  the reactions are reversible, we consider the reversed reaction if the calculation of the molar rate of progress for the $r^\textrm{th}$ reaction can be expressed as 

\begin{equation}
\mathcal{R}_r= 
\begin{cases}
\left[k_{f,r}\prod_{s=1}^{N_s}\hat{\rho}^{\nu_{sr}}_s-k_{b,r}\prod_{s=1}^{N_s}\hat{\rho}^{\nu^{'}_{ir}}_s\right]\sum_{s=1}^{N_s}\alpha_{sr}\hat{\rho}_i& \text{for third-body reaction} \\
k_{f,r}\prod_{s=1}^{N_s}\hat{\rho}^{\nu_{sr}}_s-k_{b,r}\prod_{s=1}^{N_s}\hat{\rho}^{\nu^{'}_{ir}}_s,            &\text{Otherwise,}
\end{cases}
    \label{nrp}
\end{equation}
where $k_{f,r}$ and $k_{b,r}$ denote forward reaction and backward rate coefficients. The forward rate coefficient is defined as 

\begin{equation}
k_{f,r}= A_r T^{\xi_r}\exp{\left(-\frac{E_r}{R_u T}\right)},
    \label{forward_rate}
\end{equation}
where $A_r$, $E_r$, and $\xi_r$ are the rate constants that are determined by experiments. The backward rate coefficient is determined using the following formula
\begin{equation}
k_{f,r} = K_{eq,r} k_{b,r},
    \label{backward}
\end{equation}
where the equilibrium constant $K_{eq,r}(T)$ is defined with a reference pressure, $p_{eq}=1 \,Pa$, as

\begin{equation}
\ln K_{eq,r}=-\sum_{s=1}^{N_s}\left(\nu_{sr}^{'}-\nu_{sr}\right)\frac{G^0_{s}}{R_uT}+\sum_{s=1}^{N_s}\left(\nu_{sr}^{'}-\nu_{sr}\right)\ln{\left(\frac{p^0}{R_uT}\right)}\quad \textrm{for }r=1,\cdots,N_r,
    \label{kedef}
\end{equation}
Where $G^0_s=W_s g_s$ is the standard state Gibbs free energy of species $s$.
\begin{equation}
g_s(T,p_{eq})= 
\begin{cases}
     g^t_{s}+g^r_{s}+g^v_{s}+g^e_{s},& \text{if } s \in \mathcal{H}\\
     g^t_{s}+g^e_{s},              &\text{if } s \notin \mathcal{H}
\end{cases},
    \label{gibbs}
\end{equation}
The translational Gibbs free energy is defined as 

\begin{equation}
g^t_{s}(T,p_{eq})=-\frac{R_u}{W_s}T\ln{\left(\frac{R_uT}{N_A p_{eq}}\left[\frac{2\pi W_s R_u T}{N_A^2 h_p^{2}}\right]^{3/2}\right)},\quad s=1,\cdots,N_s,
    \label{trans_gibbs}
\end{equation}
where $N_A$ is Avogadro's number, and $h_p$ is Planck's constant. The specific rotational Gibbs free energy and specific vibrational Gibbs free energy are also defined as
\begin{equation}
g^r_{s}(T)=\frac{R_uT}{W_s}\left(\frac{L_s}{2}\ln{\left(\frac{T}{\theta^r_s}\right)}-\ln{\sigma_s}\right),
\label{rot}
\end{equation}
where $\theta_s^r$ indicates the characteristic rotational temperature. The vibrational specific Gibbs energy is formulated as 
\begin{equation}
g^v_{s}(T)=-\frac{R_uT}{W_s}\sum_{k=1}^{N_v}\ln{\left(1-\exp{\left(\frac{-\theta^v_{sk}}{T}\right)}\right)}.
\label{vib}
\end{equation}
In Eq.~\eqref{rot} $\sigma_s$ indicates the steric factor for species $s$. The electronic-specific Gibbs energy can be defined as 
\begin{equation}
g^e_{s}(T)=\frac{R_uT}{W_s}\ln{\sum_{k=1}^{N_e}a_{ks}^e\exp{\left(-\frac{\theta_{ks}^e}{T}\right)}}.
\label{elec}
\end{equation}
The set of reactions includes three third-body and two regular chemical reactions shown as 

\begin{equation}
\begin{aligned}
&N_2+M \rightleftarrows 2N+M \quad &A_r = 3\times10^{16},\quad \xi_r= -1.6, \quad E_r=113200 \\
&M=N_2,NO,O_2&\Hat{\alpha}_{N_2}=0.2333,\Hat{\alpha}_{NO}=0.2333,\Hat{\alpha}_{O_2}=0.2333\\
&O_2+M \rightleftarrows 2O+M \quad &A_r = 1\times10^{16},\quad A_r= -1.5, \quad E_r=59360 \\
&M=N_2,NO,O_2&\Hat{\alpha}_{N_2}=0.2,\Hat{\alpha}_{NO}=0.2,\Hat{\alpha}_{O_2}=0.2\\
&NO+M \rightleftarrows N+O+M \quad &A_r = 5\times10^{9},\quad \xi_r= 0.0, \quad E_r=75500 
\\
&M=NO,N,O&\Hat{\alpha}_{NO}=22.0,\Hat{\alpha}_{N}=22.0,\Hat{\alpha}_{O}=22.0\\
&N_2+O \rightleftarrows NO+N \quad &A_r = 5.7\times10^{6},\quad \xi_r= 0.42, \quad E_r=42938 \\
&NO+O \rightleftarrows O_2+N \quad &A_r = 8.4\times10^{6},\quad \xi_r= 0.0, \quad E_r=19400. \\
\end{aligned}
    \label{reactions}
\end{equation}
In the end, we can compute the rate of production of each species employing the following formula

\begin{equation}
\dot{\omega}_s = W_s \sum_{r=1}^{N_r}\left(\nu^{'}_{sr}-\nu_{sr}\right)\mathcal{R}_r,\quad\textrm{for }s=1,\cdots,N_r.
    \label{calc_prod}
\end{equation}

\subsection{Non-equilibrium Chemistry Validation}
\begin{figure}[t!]
  \begin{center}
    \begin{tabular}{cc}
    \includegraphics[width=0.4\textwidth,height=0.3\textwidth]{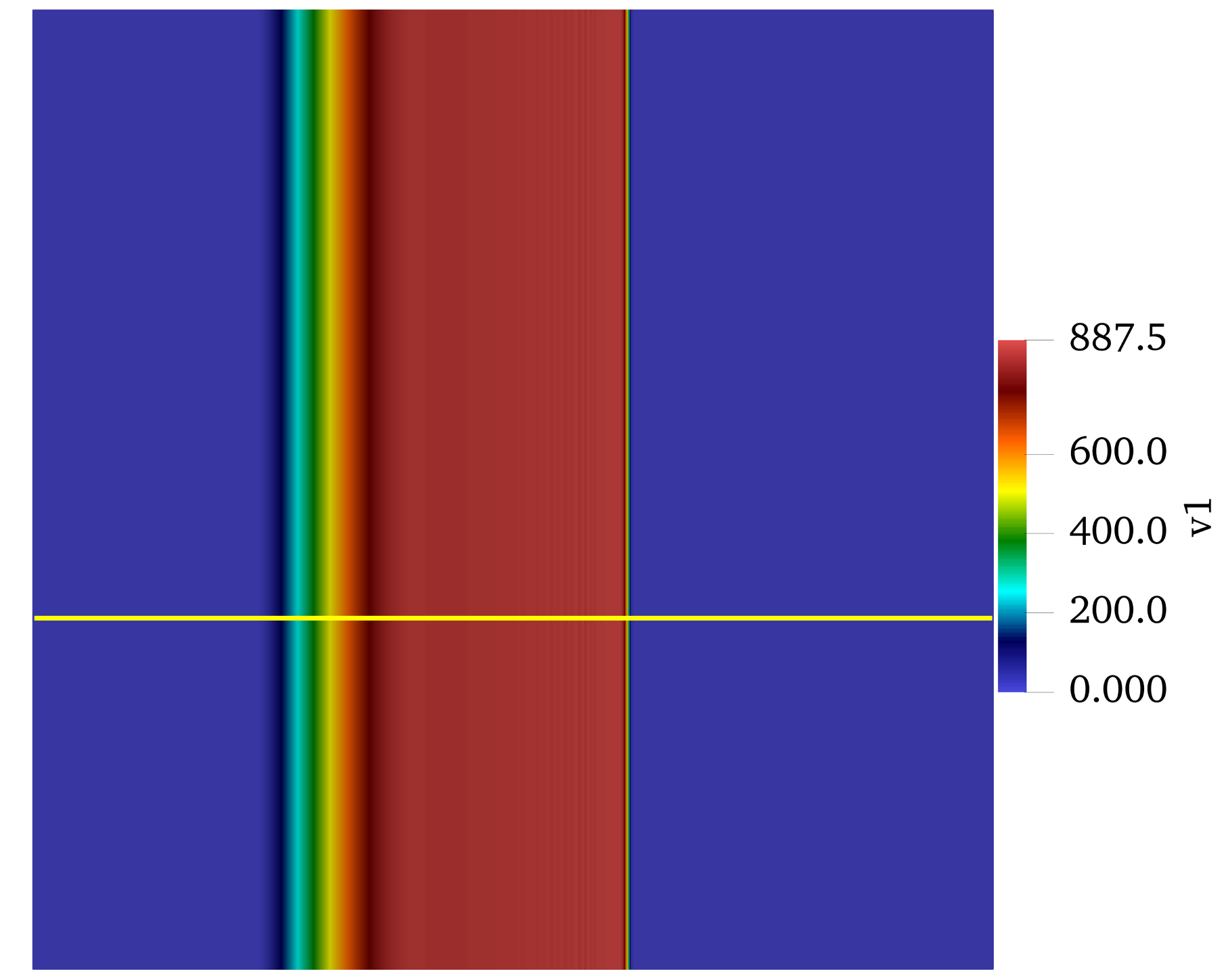}
 &
 \includegraphics[width=0.4\textwidth,height=0.3\textwidth]{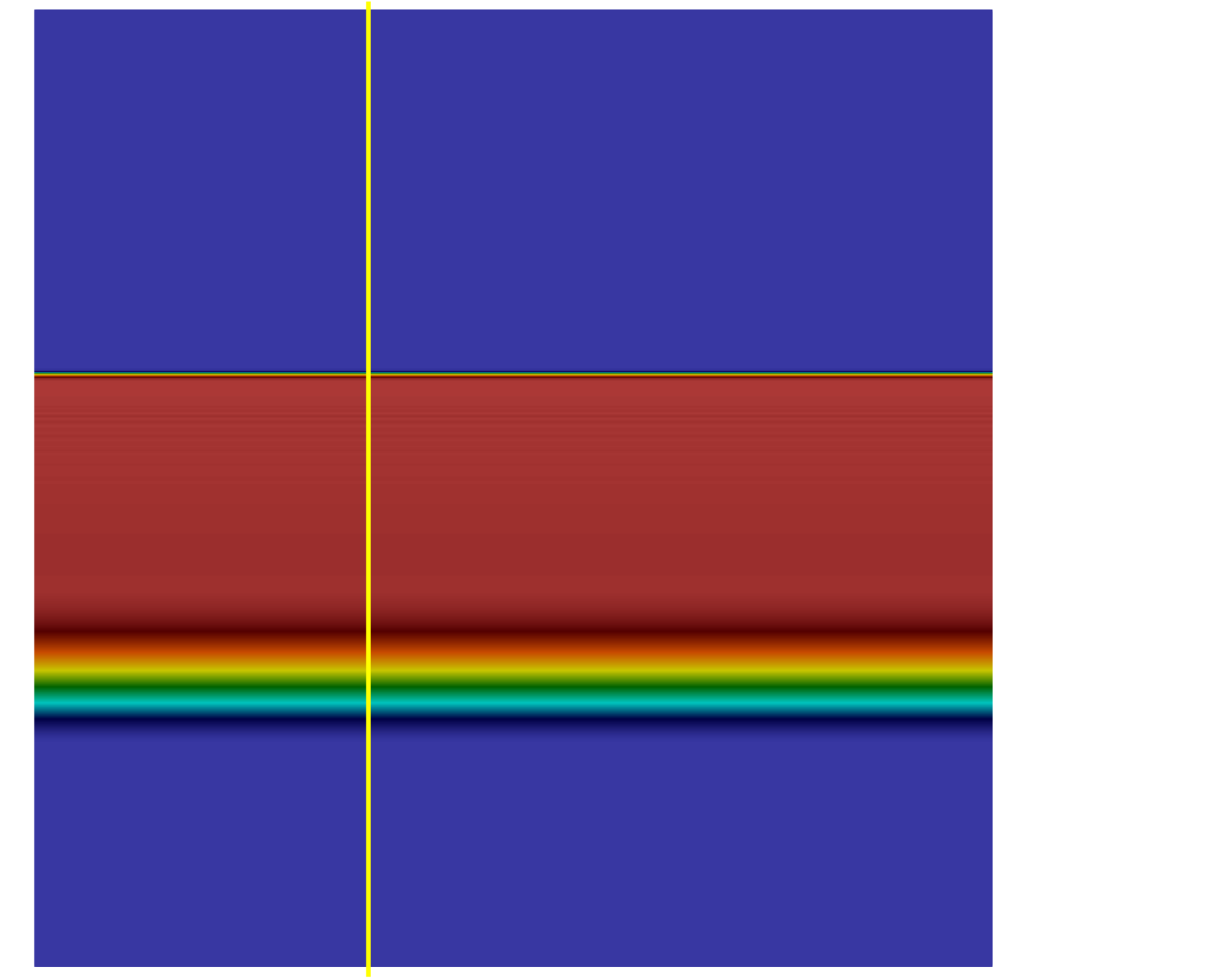}
    
  \\
  (a) $v_1$ velocity 
 & (b) $v_2$ velocity 

\end{tabular} 
  \end{center}
  \caption{1D Grossman Sod problem solved on a 2D domain. The discontinuous solutions are initialized in (a) x-direction and (b)y-direction.}
  \label{fig:grossman_lines}
\end{figure}

\begin{figure}[t!]
\begin{center}
   \begin{tabular}{c}
\includegraphics[width=0.7\textwidth]{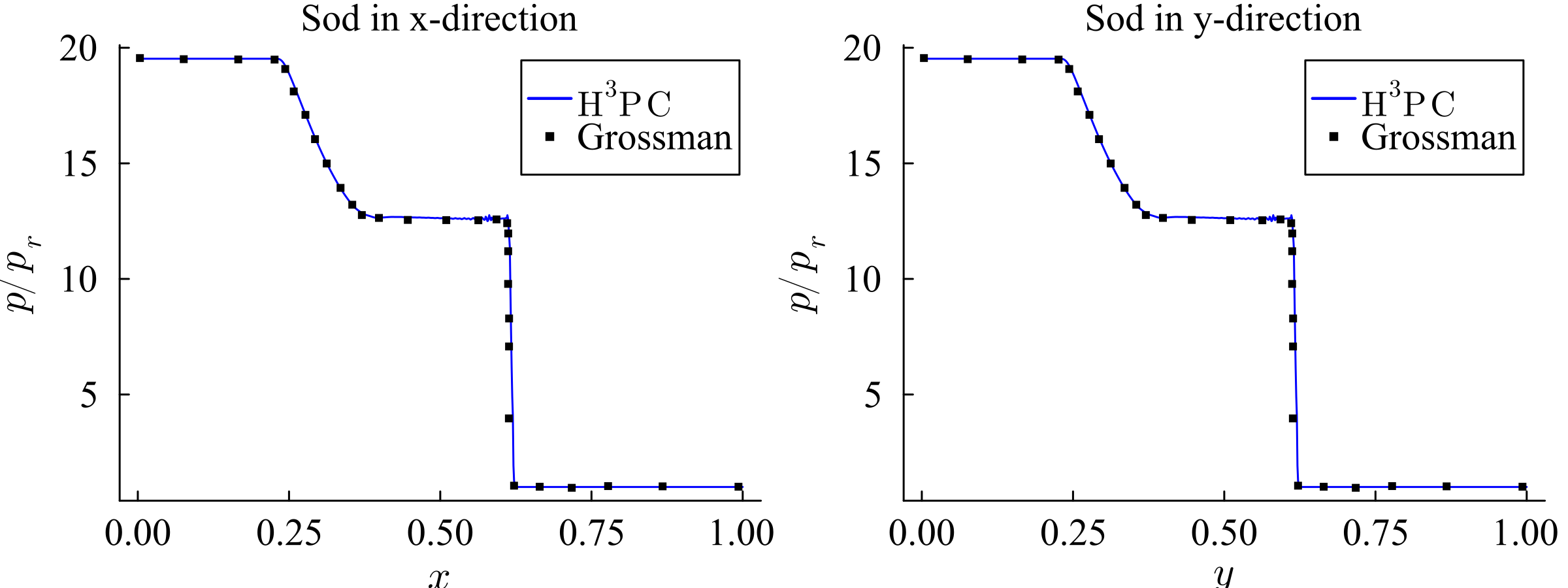} 
  \\
  (a) $\frac{p}{p_r}$
  \\
  \includegraphics[width=0.7\textwidth]{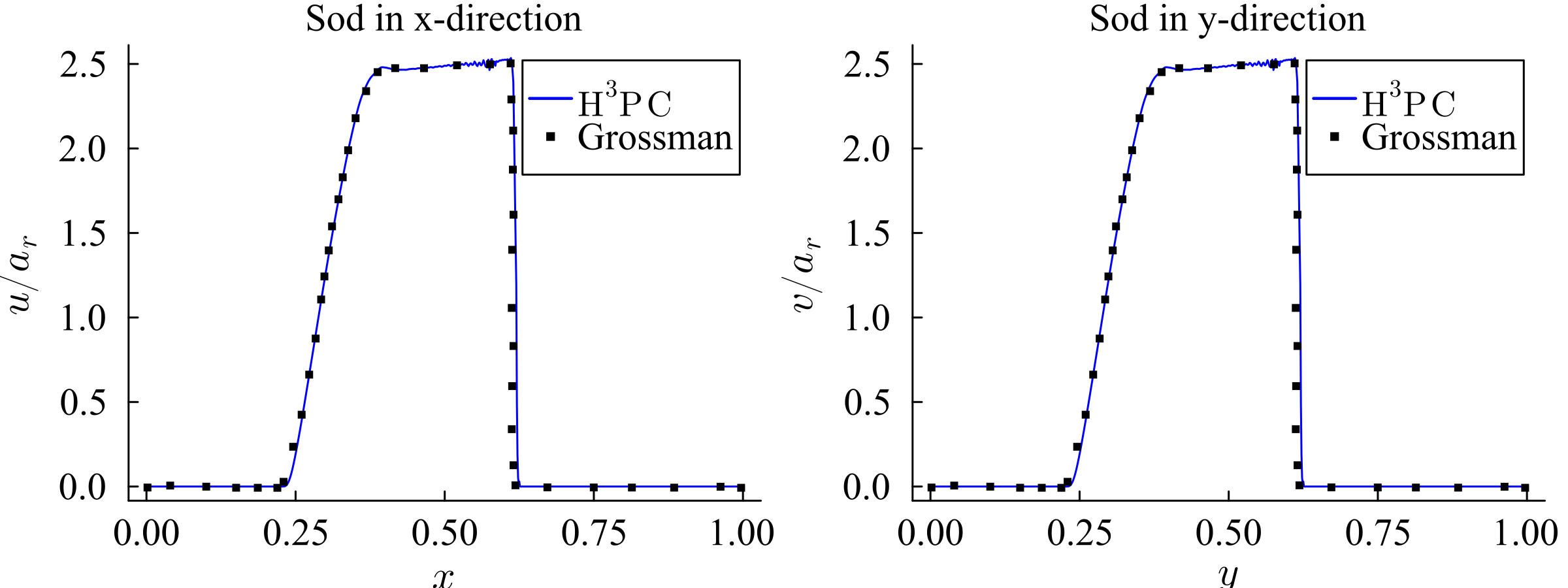} 
  \\
  (b) $\frac{u}{a_r}$
 \\
  \includegraphics[width=0.7\textwidth]{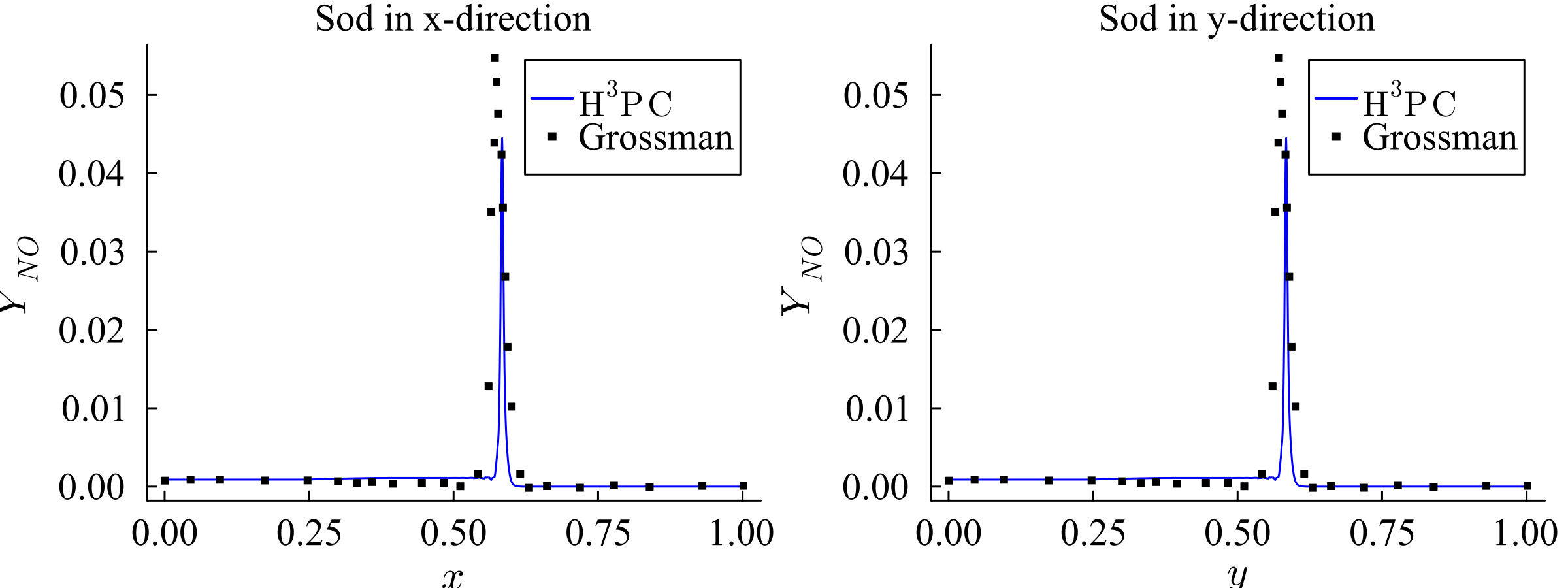} 
    
  \\
  (c)$Y_{NO}$

\end{tabular} 
\end{center}

\caption{Validation for coupling \texttt{Mutation.jl} library with H$^3$PC solver. Grossman's \cite{grossman1990flux} sod problem with non-quilibrium chemistry assumption. $200$ elements are considered in the desired direction and a single element is considered for the other Cartesian direction with $\mathcal{P}=3$.}
\label{fig:grossman}  
\end{figure}

To validate the coupling of the Mutation++ library with our H$^3$PC solver, we initialized the 1D Grossmann \cite{grossman1990flux} Sod problem in a 2D domain for which the discontinuous states of initialization are imposed first in x-direction and then in y-directions. We impose periodic boundary conditions in a perpendicular direction to avoid solution gradients. The physical domain is a 2D square with $x,y \in [0,1]^2$. For the initialization of two test problems, we have employed the following formula:
\begin{align}
\textrm{Case1: }\left(T,u,P\right)= 
\begin{cases}
     \left(9000\textrm{ K}, 0\textrm{ }\frac{\textrm{m}}{\textrm{s}}, 195,256 \textrm{ Pa}\right),& \text{if } x<0.5\\
     \left(300 \textrm{ K}, 0\textrm{ }\frac{\textrm{m}}{\textrm{s}}, 10,000 \textrm{ Pa}\right),              &\text{if } x>0.5
\end{cases}\\
\textrm{Case2: }\left(T,u,P\right)= 
\begin{cases}
     \left(9000\textrm{ K}, 0\textrm{ }\frac{\textrm{m}}{\textrm{s}}, 195,256 \textrm{ Pa}\right),& \text{if } y<0.5\\
     \left(300 \textrm{ K}, 0\textrm{ }\frac{\textrm{m}}{\textrm{s}}, 10,000 \textrm{ Pa}\right),              &\text{if } y>0.5
\end{cases}
    \label{eq:grossman}
\end{align}
The simulation time is $t=0.0001 \textrm{ s}$. Figure~\ref{fig:grossman_lines} demonstrates velocity profiles for two cases where the 1D sod problem is initialized in x and y directions for a rectangular domain discretized with 200 elements with $\mathcal{P}=3$. We then plot profiles of primitive variables over the yellow line depicted in Figure~\ref{fig:grossman_lines}(a) and (b) and compare it with the results of Grossman and Cinnella \cite{grossman1990flux}. Figure~\ref{fig:grossman}(a)-(c) depicts the pressure and velocity ratios and $NO$ mass fraction profile at $T=0.0001$ s. We can observe that the non-equilibrium chemistry solver generates the correct results as it agrees well with the reference solution.

\subsection{Comparison of H$^3$PC and UCNS3D solvers for supersonic flow around square cylinder for Re=1000 and Ma=1.5} \label{sec:compare_solvers}
\begin{figure}[t!]
  \begin{center}
    \begin{tabular}{c}
    \includegraphics[width=0.55\textwidth,trim={0cm 11cm 0cm 0cm},clip]{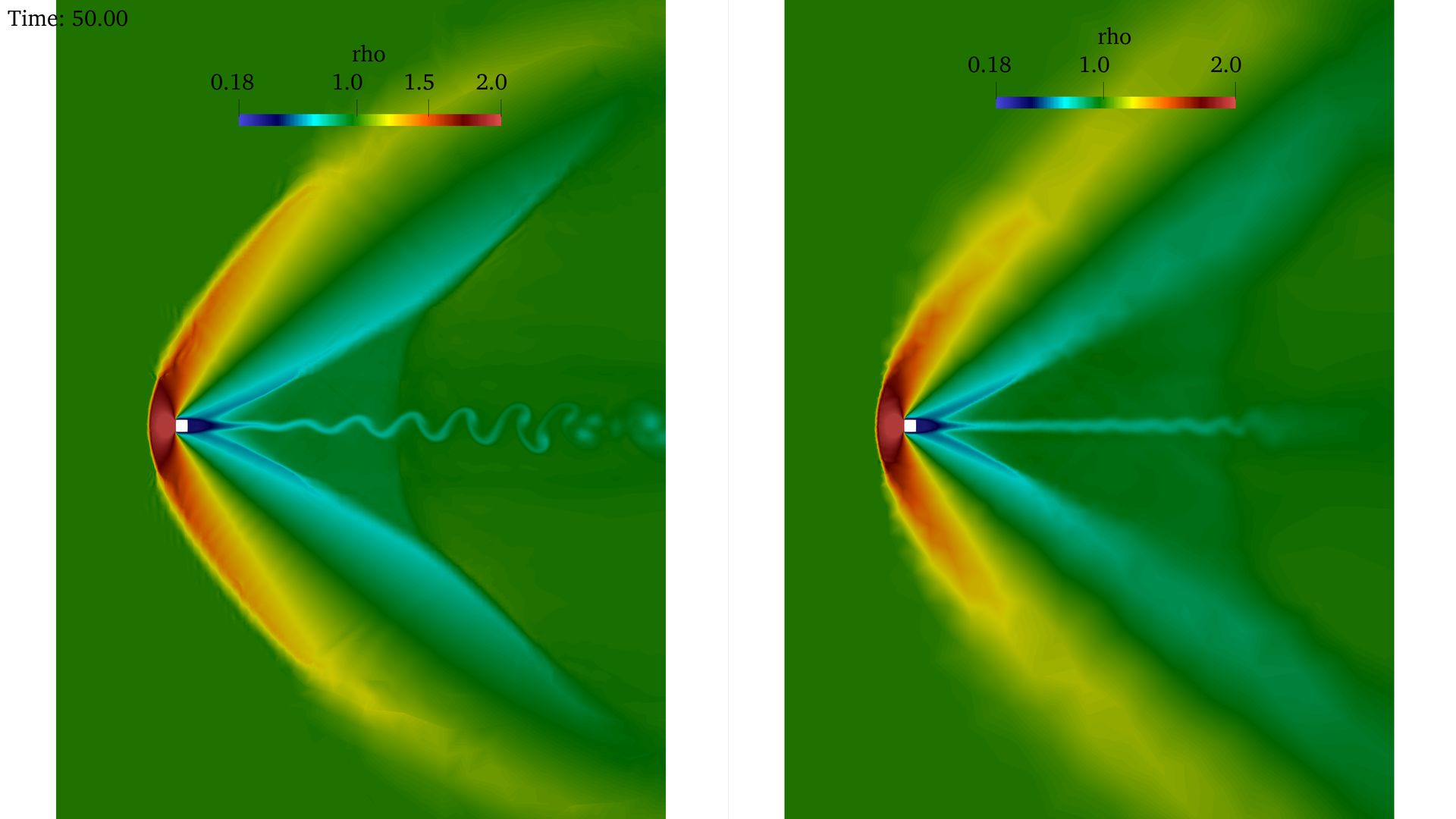}
  \\
  \includegraphics[width=0.55\textwidth,trim={0cm 11cm 0cm 0cm},clip]{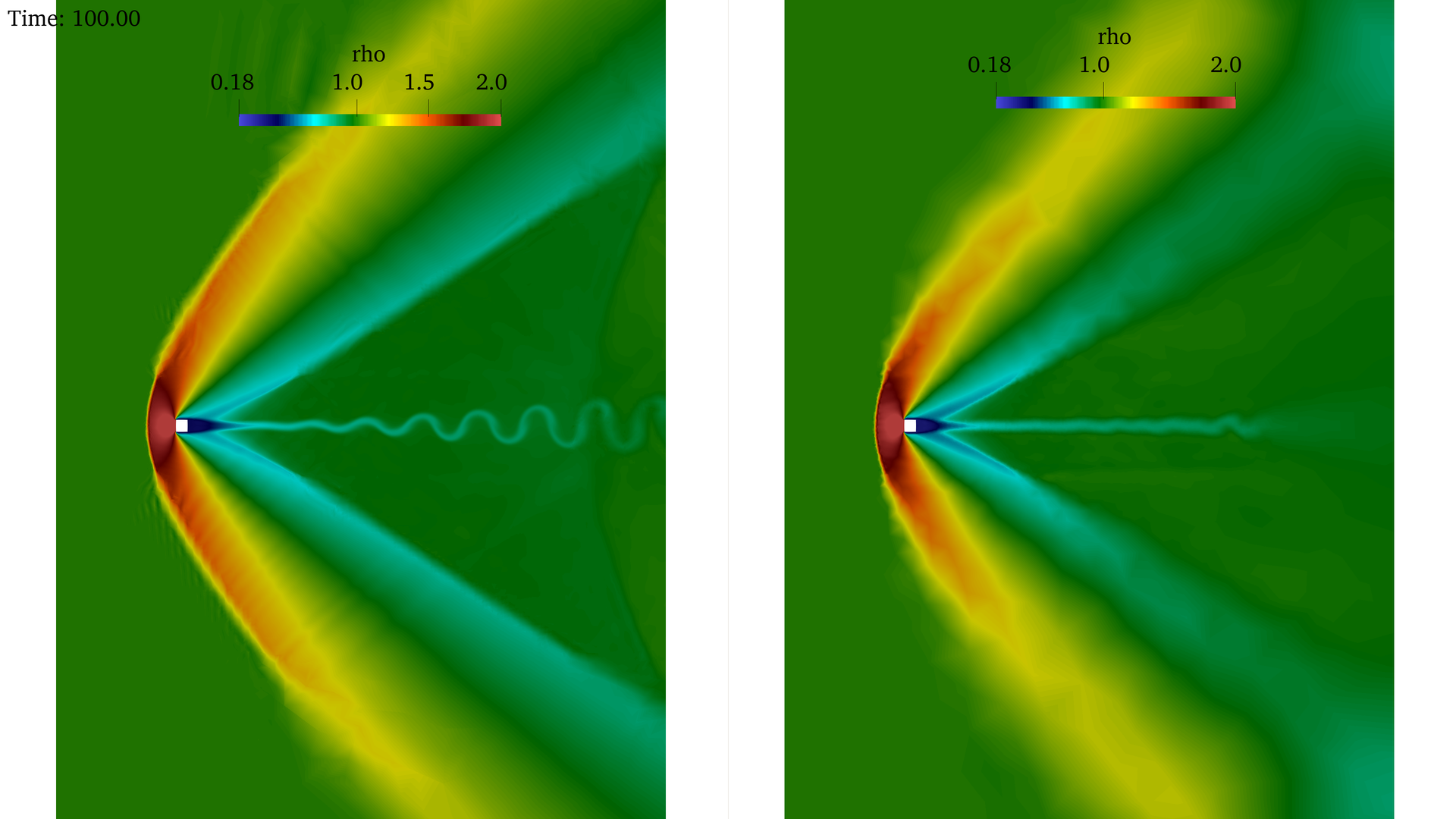}

\\
 \includegraphics[width=0.55\textwidth,trim={0cm 11cm 0cm 0cm},clip]{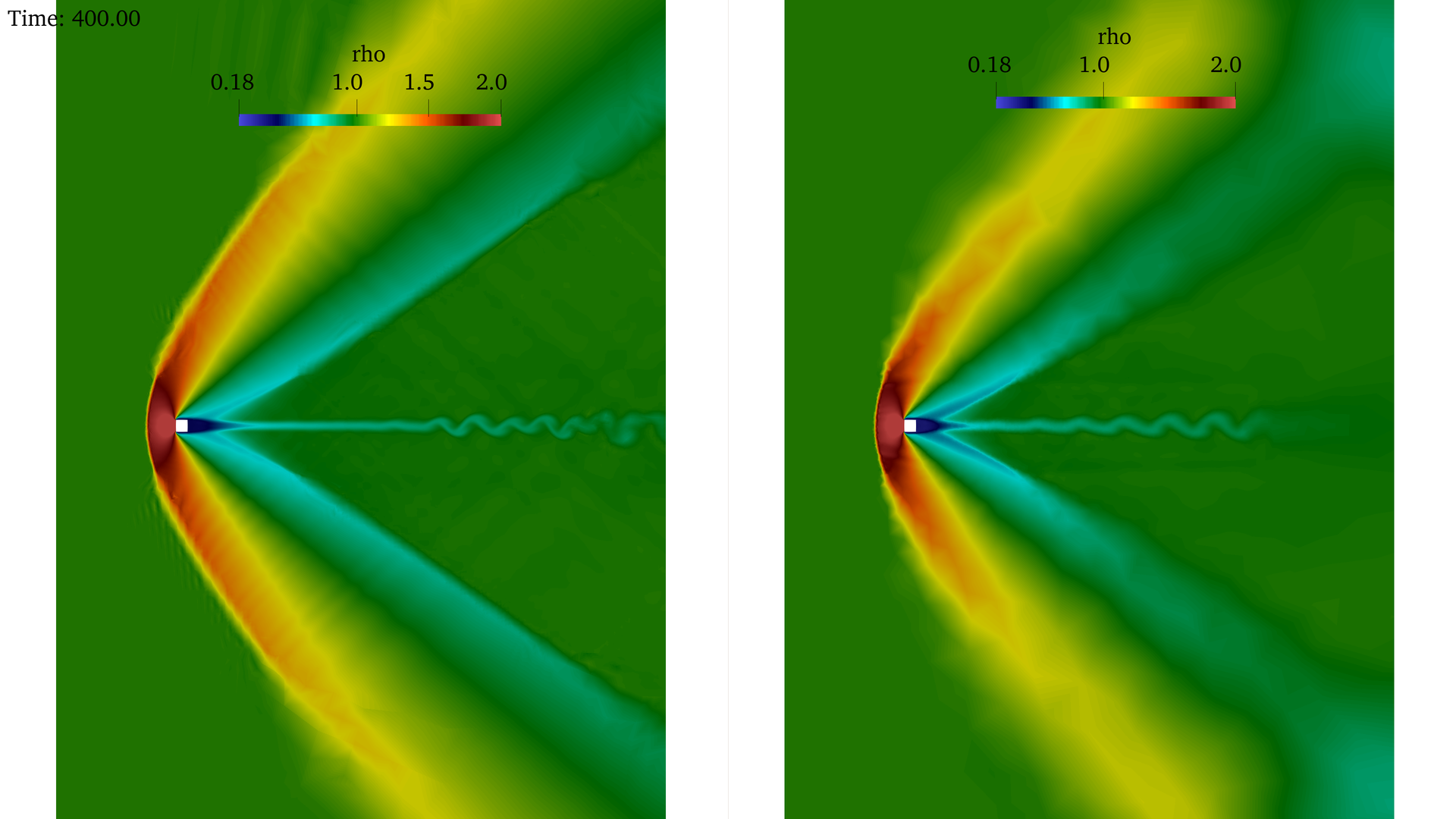}
\\
 \includegraphics[width=0.55\textwidth,trim={0cm 11cm 0cm 0cm},clip]{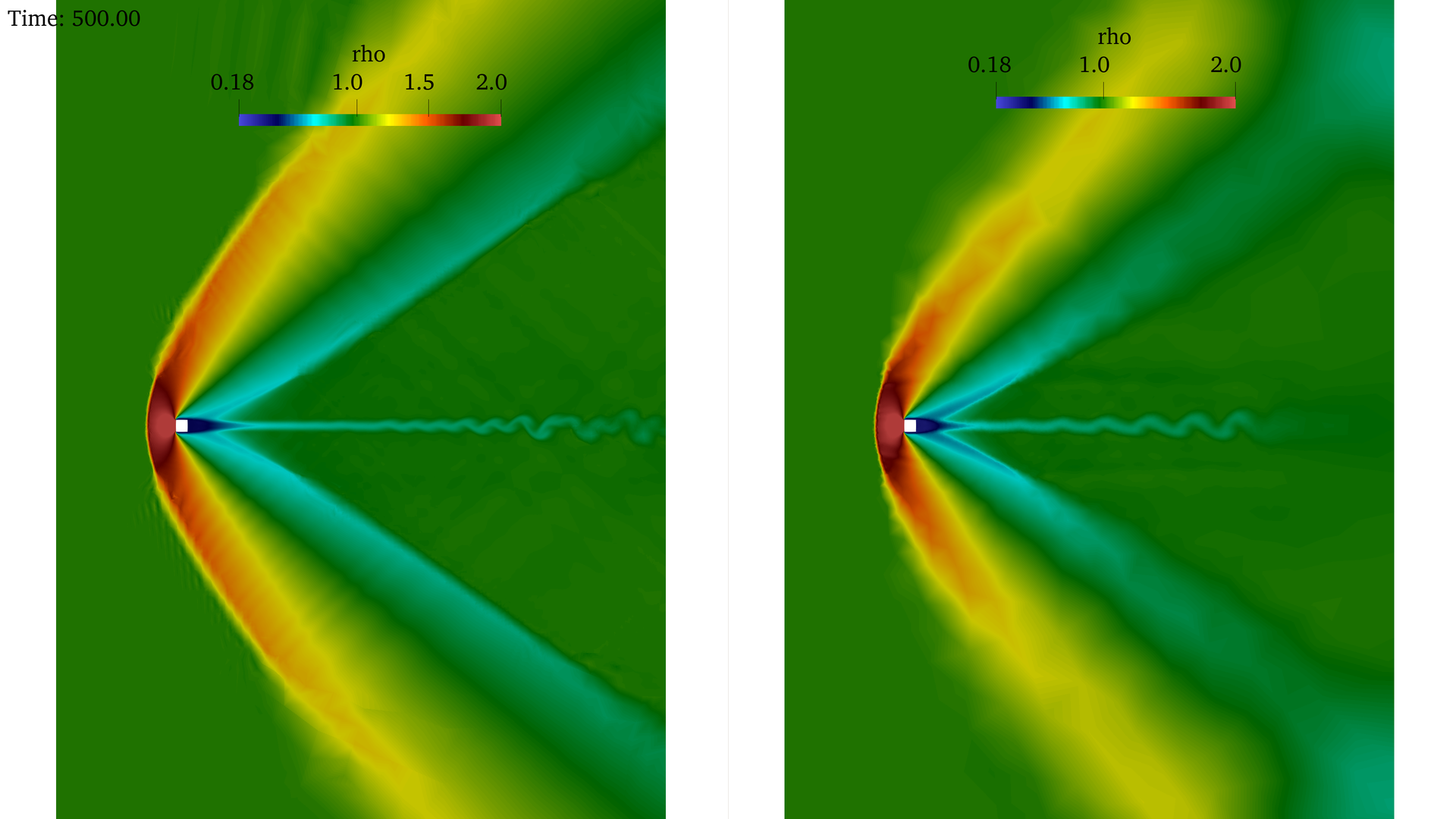}
\\
\end{tabular} 
\caption{Density contours at T=50, 100, 400, and 500 obtained using H$^3$PC solver on the left column and UCNS3D right column for $Re=1000$ and $M=1.5$ flow over square cylinder. For the H$^3$PC solver, we use medium resolution mesh with $\mathcal{P}=2$.}
\label{fig:re1000_UCNS3D}
\end{center}
\end{figure}

\noindent To validate our results, we used the UCNS3D solver to simulate the same flow past a square cylinder described in Section~\ref{sec:cylinder}. The computational domain was discretized using unstructured triangular elements, resulting in 324,216 elements. The mesh was refined near the cylinder and progressively coarsened toward the far-field boundaries. The compressible Navier–Stokes equations were spatially discretized using the Conservative WENO (CWENO) scheme~\cite{tsoutsanis2011weno} with a sixth-order polynomial basis. To maintain numerical stability and suppress spurious oscillations, the Venkatakrishnan limiter~\cite{venkatakrishnan1995convergence} was employed. Time integration was performed using a second-order implicit dual time-stepping scheme, and the simulation was advanced to ( t = 500 ) to compute the Strouhal number and corresponding frequency. The Reynolds number was reduced to ( Re = 1000 ) to ensure sufficient spatial resolution for accurately capturing the vortex street downstream of the cylinder. The Strouhal number computed at four probe locations is approximately 0.24 while the H$^3$PC solver computes a Strouhal number of $0.27$ using a coarse resolution mesh with $\mathcal{P}=2$. 
Figure~\ref{fig:re1000_UCNS3D} shows a comparison of density field obtained by the H$^3$PC and UCNS3D solvers at $t=50$, $100$, $400$, and $500$. The solutions obtained by both solvers match in predicting large structures, while the vortex-street region differs slightly. The comparison shows the different characteristics of the H$^3$PC solver and the UCNS3D solver in predicting fine details of the transient flow with less spatial resolution.

